\documentclass[11pt]{article}
\usepackage[utf8]{inputenc}

\usepackage{latexsym}
\usepackage{amssymb,amsmath}
\usepackage{amsthm}
\usepackage{authblk}
\usepackage{graphicx}
\usepackage{sgame}
\usepackage{color}
\usepackage{graphicx}
\usepackage{chngcntr}  
\counterwithin{figure}{section}  
\usepackage[hidelinks]{hyperref}
\usepackage{empheq}
\usepackage{blkarray}
\usepackage{cancel}
\usepackage{breqn}
\usepackage{comment}
\usepackage{makecell}
\usepackage{tikz} 
\usepackage{setspace}
\usepackage{appendix}
\usepackage{enumerate}

\usepackage{caption}
\usepackage{subcaption}

\numberwithin{equation}{section}

\newcommand{\ds}{\displaystyle}
\newcommand{\mc}{\mathcal}

\newcommand{\bbm}{\begin{bmatrix}}
\newcommand{\bpm}{\begin{pmatrix}}
\newcommand{\ebm}{\end{bmatrix}}
\newcommand{\epm}{\end{pmatrix}}

 \newcommand{\dsdel}[2]{\displaystyle\frac{\partial #1}{\partial #2}}

 \newcommand{\doubledelsame}[2]{\displaystyle\frac{\partial^2 #1}{\partial #2^2}}

\newcommand{\dsddt}[1]{\displaystyle\frac{d #1}{dt}}

\usepackage[numbers,sort&compress]{natbib}
\usepackage{algorithm}
\usepackage{algpseudocode}

\makeatletter
\renewcommand*{\@fnsymbol}[1]{\ensuremath{\ifcase#1  \mathsection\or  *\or \dagger\or \ddagger\or
  \mathparagraph\or \|\or **\or \dagger\dagger
   \or \ddagger\ddagger \else\@ctrerr\fi}}
\makeatother

\title{Pattern Formation in Agent-Based and PDE Models for Evolutionary Games with Payoff-Driven Motion}
\author[1,2,$\mathsection$]{Tianyong Yao}
\author[1,3,$\mathsection$]{Chenning Xu}
\author[1,6,$\mathparagraph$]{Daniel B. Cooney}
\affil[1]{\small{Department of Mathematics, University of Illinois Urbana-Champaign, Urbana, IL, USA}}
\affil[2]{\small{Department of Mathematics, University of Michigan, Ann Arbor, MI, USA}}
\affil[3]{\small{Department of Mathematics and Department of Computer Science, California Institute of Technology, Pasadena, CA, USA}}
\affil[4]{\small{Carl R. Woese Institute for Genomic Biology, University of Illinois Urbana-Champaign, Urbana, IL, USA}}
\affil[$\mathsection$]{\small{These authors contributed equally and should be considered joint first authors. \thanks{The order among the co-first authors was determined by the draw of a random number via the random.random() function in Python.}}}
\affil[$\mathparagraph$]{\small{Correspondence to @dbcoone2@illinois.edu.}}

\date{\today}

\begin{document}

\newtheorem{definition}{Definition}[section]
\newtheorem{theorem}{Theorem}[section]
\newtheorem{lemma}[theorem]{Lemma}
\newtheorem{corollary}[theorem]{Corollary}
\newtheorem{claim}[theorem]{Claim}
\newtheorem{fact}[theorem]{Fact}
\newtheorem{proposition}[theorem]{Proposition}
\newtheorem{remark}[theorem]{Remark}
\newtheorem{observation}[theorem]{Observation}
\newtheorem{example}[theorem]{Example}

\def\danielnote#1{{\color{blue}{\bf(}#1 [Daniel]{\bf)}}}
\def\seokhwannote#1{{\color{magenta}{\bf}#1 [Seokhwan]{\bf)}}}
\def\chenningnote#1{{\color{red}{\bf}#1 [Chenning]{\bf)}}}
\def\Tianyongnote#1{{\color{teal}{\bf}#1 [Tianyong]{\bf)}}}

\maketitle

\begin{abstract}
Spatial structure can play an important role in the evolution of cooperative behavior and the achievement of collective success of a population. In this paper, we explore the role of random and directed motion on spatial pattern formation and the payoff achieved by populations in both stochastic and deterministic models of spatial populations who engage in social interactions following a hawk-dove game. For the case of purely diffusive motion, both a stochastic spatial model and a partial differential equation model show that Turing patterns can emerge when hawks have a greater movement rate than doves, and in both models hawks and doves see an increase in population size and average payoff as hawk mobility increases. For the case of the payoff-driven motion, the stochastic model shows an overall decrease in population size and average payoff, but the PDE model displays more subtle behavior in this setting and will depend on the relative diffusivities of the two strategies. The PDE model also displays a biologically infeasible short-wave instability in the case of payoff-driven motion and equal diffusivities, indicating that we need to be careful about the mathematical properties of PDE models with payoff-driven directed motion and indicating potential use for nonlocal PDE models for spatial patterns in evolutionary games with directed motion. 
\end{abstract}

\begin{spacing}{0.01}
\renewcommand{\baselinestretch}{0.1}\normalsize
\tableofcontents
\addtocontents{toc}{\protect\setcounter{tocdepth}{2}}
\end{spacing}
\singlespacing

\section{Introduction}

Social dilemmas are a common feature that arise in a range of biological and social systems, with evolutionary forces like natural selection or cultural imitation producing a tension between an individual incentive to cheat and a collective incentive to sustain cooperation within a population. Evolutionary game theory provides a tractable mathematical framework for describing cooperative behavior and social interactions between individuals, with games such as the Prisoners' Dilemma, Hawk-Dove game, and Stag-Hunt game providing examples of resulting evolutionary dynamics featuring dominance of cheaters, coexistence of cooperators and cheaters, and alternative stable outcomes of all-cheater or all-cooperator populations \cite{smith1973logic,hofbauer1998evolutionary,nowak2006evolutionary,sandholm2010population}. The role of spatial interactions and spatial motion has often been proposed as a mechanism for promoting the evolution of cooperative behavior beyond the level achievable in a well-mixed population \cite{nowak1994spatial,durrett1994importance,sicardi2009random}, with clusters of cooperators forming in spatial or network neighborhoods allowing cooperators to achieve higher payoffs and invade regions previously occupied by defectors \cite{ohtsuki2006evolutionary,ohtsuki2006simple,ohtsuki2006replicator,allen2017evolutionary,szabo2007evolutionary}.

The spatial motion of individuals within a population is one mechanism that can impact the distribution of strategies, with increased birth rates due to greater payoff potentially allowing clusters of cooperators to increase in size and increase the population size and cooperation level across a spatial domain \cite{vickers1989spatial,vickers1993spatial,hutson1992travelling,wakano2009spatial,wakano2011pattern}. This has been particularly explored by Wakano, Hauert, and coauthors, who have used reaction-diffusion equations and ecological public goods games to show how rapid undirected motion for defectors can produce Turing patterns and promote cooperation \cite{hauert2008ecological,wakano2009spatial,wakano2011pattern,funk2019directed,gerlee2019persistence}. Further work on PDE models in evolutionary games has explored how directed motion can help to shape spatial patterns of cooperation, either assuming that individuals can perform directed motion %
towards cooperators and away from defectors \cite{kimmel2019time,funk2019directed}, assuming that individuals perform payoff-driven motion in which they climb payoff gradients to find regions with increasing payoff for their respective strategies \cite{helbing2008migration,helbing2009pattern,deforest2013spatial,deforest2018game,xu2017strong,young2018fast}, or that individuals perform directed motion based on a quantity like environmental quality that serves as a proxy for payoff \cite{yao2025spatial}.

In addition to the PDE modeling approach for describing diffusive or payoff-driven motion, it is also possible to formulate agent-based models that describes the rules by which individuals choose to move to neighboring spatial locations in a metapopulation lattice. One comparison between the individual-based and PDE models for spatial evolutionary games was provided by Durrett and Levin \cite{durrett1994importance,pacala2020importance}, who showed the key roles played by the choices of spatially explicit or well-mixed domains and the role of discrete or continuous modeling of population states in determining the long-term survival or coexistence of competing strategies. Their analysis of various game-theoretic scenarios and their range of spatial and population states highlighted the importance of comparing across different modeling frameworks, and emphasized approaches to careful derivation of continuum models from the discrete interaction and movement rules of individuals in a population \cite{durrett1994importance,seri2012sustainability,cantrell2004deriving}. This comparative approach of using agent-based and mean-field models to study spatial behavior of populations has also been applied to study complex ecological systems \cite{lewis1993modelling,white1996model}, and this approach has been applied to explore human social systems and collective spatial phenomena from the emergence of economic aggregation through labor migration to the formation of heterogeneous patterns within cities \cite{lindstrom2020qualitative,hasan2020transport}. The formulation of such spatial models have been particularly helpful in describing assumptions about the rules of interactions between individuals and biased random walks taken by individuals through spatial lattices, with these individual-based rules also allowing for the derivation of a range of emergent PDE models incorporating features incorporating chemotaxis-type phenomena \cite{short2008statistical,alsenafi2018convection,painter2019mathematical,alsenafi2021multispecies,codling2008random,plank2025random}.

We find that, in the case of purely undirected spatial motion, both the stochastic and PDE models can display Turing pattern formation when hawks have a higher diffusivity than doves, and we observe that this pattern-forming mechanism helps to increase the total population size and average payoff for both strategies across the spatial domain. When we incorporate payoff-driven motion, we see that patterns can form in the stochastic model when doves are more effective than hawks at performing directed motion towards patches with higher payoff, but that the resulting patterns feature lower overall payoff than the level of payoff achieved in spatially well-mixed populations. The behavior of payoff-driven motion is more difficult to study in the PDE model, as we see that, in line with prior observations by Helbing \cite{helbing2009pattern} and by Funk and Hauert \cite{funk2019directed}, a short-wave / infinite-wavenumber instability will arise when the directed motion of doves can produce spatial patterns in the case of equal diffusivities of the two strategies. To avoid this issue of short-wave instability, we extend our analysis of the PDE model to show that allowing the possibility of increased hawk diffusivity or a nonlocal evaluation of payoff gradients can result in biologically feasible finite-wavenumber patterns through sufficiently strong payoff-driven motion by doves.

The remainder of the paper is organized in the following manner. In Section \ref{sec:ModelDesign}, we summarize the game-theoretic background for our models and formulate our baseline stochastic process and PDE models for evolutionary games with payoff-driven motion. We then present results for the simulations of our stochastic spatial model in Section \ref{sec:StochasticResults}, discussing the case of purely diffusive motion and the case of patterns due solely to differences in the rules of payoff-driven motion. In Section \ref{sec:PDEResults}, we provide analytical results and numerical simulations for spatial patterns in the PDE model, highlighting the behavior of Turing patterns in the purely diffusive case and the short-wave instability in the case of payoff-driven motion and equal diffusivities. We then provide a comparison between the behavior of the stochastic and PDE models in Section \ref{sec:mixedeffectsresults} for the case of faster diffusion by hawks and greater ability of payoff-driven motion for doves, and we formulate nonlocal PDE models of payoff-driven motion and demonstrate the presence of finite-wavenumber instabilities in Section \ref{sec:nonlocalPDEmodels}. We summarize our results and discuss our outlook for future work in Section \ref{sec:discussion}, and we include additional details about the stochastic simulations and the derivation of our local and nonlocal PDE models in the appendix. 

\section{Formulation of the Stochastic Process and PDE Models for Spatial Evolutionary Dynamics}
\label{sec:ModelDesign}

In this section, we present our baseline models for game-theoretic interactions, the population dynamics arising due to payoff and density-dependent regulation, as well as the rules for spatial movement due to diffusive and payoff-driven motion. We first present our model of two-player, two-strategy games in Section \ref{sec:game-background}, and then use these game-theoretic ideas to formulate our stochastic spatial model and our deterministic PDE model 
in Sections \ref{sec:StochasticModel} and \ref{sec:PDEModel}.

\subsection{Game-Theoretic Background: Frequency-Dependent Evolutionary Games with Density-Dependent Population Regulation}
\label{sec:game-background}

In this paper, we will consider spatial evolutionary dynamics based on underlying hawk-dove or snowdrift games played at each spatial location. In particular, we will consider a two-player symmetric game in which individuals can play one of two strategies: a cheater/defector strategy called Hawk ($H$) and a cooperative strategy called Dove ($D$). We represent the payoffs received when playing each of these strategies through the following payoff matrix
\begin{equation} \label{eq:generalpayoff}
\begin{blockarray}{ccc}
& H & D \\
\begin{block}{c(cc)}
H & P &  T \\
D & S & R\\
\end{block}
\end{blockarray},
\end{equation}
where $P$ is a punishment for mutual defection (hawkish or aggresive behavior), $R$ is the reward for mutual cooperation (dove-like or peaceful behavior), $T$ is the temptation to play hawk against a dove, and $S$ is the sucker payoff for playing dove against a hawk. For this payoff matrix to represent a hawk-dove or snowdrift game, we require that the payoffs have the following ranking
\begin{equation}
T > R > S > P,
\end{equation}
which ensures that individuals receive higher payoffs by cooperation (playing dove) against hawks (as $S > P$) and by defecting (playing hawk) against doves (as $T > R$). %
This ranking of payoffs typically ensures that evolutionary dynamics supports long-time coexistence of hawks and doves in an infinite population, with the anti-coordination structure of the payoffs favoring individuals who are currently rare in a population.

For our analysis, we will typically focus on a special case of the hawk-dove game that is used to model a pairwise contest over a resource of total value $V$ and a cost $\frac{C}{2}$ for fighting over a resource. Assuming that two doves split the resource evenly, a hawk takes the entire resource when interacting with a dove, and that hawks doves split the resource evenly after fighting over the resource, we can express the payoffs for such a hawk-dove game with the following matrix

\begin{equation} \label{eq:CVpayoff}
\begin{blockarray}{ccc}
& H & D \\
\begin{block}{c(cc)}
H & \frac{V-C}{2} &  V \\
D & 0 & \frac{V}{2} \\
\end{block}
\end{blockarray}.
\end{equation}
We will assume that $C > V > 0$, so the cost of fighting over a resource is greater than the value of sharing the resource, which allows us to see that this payoff matrix will feature the ranking of payoffs associated with hawk-dove games. 

Now that we have formulated the payoffs obtained through pairwise interactions, we can describe how individuals receive payoff by playing games in a well-mixed population featuring a population with densities $u$ of cooperators and $v$ of defectors. If individuals interact with all members of the population, then the average payoffs received by doves and hawks are given by
\begin{subequations} \label{eq:payoff_equations}
\begin{align}
p_H\left(u,v\right) &= P \left(\frac{u}{u+v} \right) + T \left( 
\frac{v}{u+v} \right) \\
p_D\left(u,v\right) &= S \left(\frac{u}{u+v} \right) + R \left( 
\frac{v}{u+v} \right).
\end{align} 
\end{subequations}

Following the approach used by Brown and Hansell \cite{brown1987convergence} and Durrett and Levin \cite{durrett1994importance}, we consider population dynamics for the number of hawks $u$ and doves $v$ with a combination of a baseline frequency-dependent net birth rate and a density-dependent term featuring logistic regulation based on the total population density $u+v$. This yields the following system of ODEs 
\begin{subequations} \label{eq:densityODEsystem}
\begin{align}
\dsddt{u} &= u \left[ p_H(u,v) - \kappa \left(u+v \right) \right] \\
\dsddt{v} &= v \left[ p_D(u,v) - \kappa \left(u+v \right) \right].
\end{align}
\end{subequations}
Because the payoff functions $p_H(u,v)$ and $p_D(u,v)$ depend only on the fraction of hawks $s := \frac{u}{u+v}$ and doves $1-s = \frac{v}{u+v}$, we can also rewrite the payoff functions in the form
\begin{subequations}
\begin{align}
p_D(s) &= P s + T (1-s) \\
p_H(s) &= S s + R (1-s),
\end{align}
\end{subequations}
and we can use a change of variables to represent the population in terms of $s = \frac{u}{u+v}$ and the total population size $q = u+v$, which allows us to rewrite the dynamics of Equation \eqref{eq:densityODEsystem} as the following system of ODEs:
\begin{subequations}
\label{eq:ODEsq}
\begin{align}
\dsddt{s} &= s (1-s) \left[ p_H(s) - p_D(s) \right]  \\
\dsddt{q} &= q \left[ s p_H(s) + (1-s) p_D(s) - \kappa q \right].
\end{align}
\end{subequations}
The first equation is the replicator equation typically used to study frequency-dependent selection in evolutionary game theory, which is typically used to model the evolution of strategy frequency \cite{hofbauer1998evolutionary,weibull1997evolutionary,sandholm2010population}. Notably, the first equation is independent of the total population size $q = u + v$, so we can use the first equation alone to determine the fraction of cooperators $s$ achieved in the long-time limit of the ODE system. In particular, for the payoff rankings corresponding to the Hawk-Dove game, we see that the stable equilibrium of Equation \eqref{eq:ODEsq} is given by the coexistence equilibrium
\begin{equation}
 \left(s_0,q_0 \right) = \left( \frac{S-P}{S+T-R-P}, \frac{ST - RP}{\kappa \left( S + T - R - P \right)}  \right),   
\end{equation}
and we can correspondingly represent this stable equilibrium in terms of the original variables $u$ and $v$ as the equilibrium point
\begin{equation}
\label{eq:uv_equilibium}
\left(u_0,v_0 \right) = \left( \frac{(T-R)(ST-RP)}{\kappa \left( S+T-R-P\right)^2}, \frac{(S-P) \left(ST-RP\right)}{\kappa \left( S+T-R-P\right)^2} \right).
\end{equation}
In Sections \ref{sec:StochasticModel} and \ref{sec:PDEModel}, we will consider spatial extensions of this model, where we assume that individuals will play the Hawk-Dove game with all members of the population at their spatial location. We expect that there is an equilibrium state in which the population size of hawks and doves will take the form $(u,v) = (u_0,v_0)$ at each point in space, and we will look to see how spatial movement can lead to instability of the spatially uniform state. Such instability will allow us to identify the formation of spatial patterns and regions at which the population state differs from the coexistence equilibrium point achieved under our ODE model for evolutionary dynamics based on the Hawk-Dove game played in a well-mixed population.

\subsection{Stochastic Spatial Model with Diffusive and Payoff-Driven Motion}
\label{sec:StochasticModel}

We now formulate a stochastic spatial model that combines the demographic events of payoff-dependent and density-dependent birth and death rates with rules for spatial motion due to biased random walks based on payoff gradients. %
Building off of the game-theoretic model formulated in Section \ref{sec:game-background}, we consider game-theoretic interactions following the Hawk-Dove game and use the payoff matrix from Equation \eqref{eq:generalpayoff} to simulate the interaction of individuals at a given site on our spatial lattice. We build our baseline payoff-dependent birth or death rates and constrain population growth using density-dependent regulation as presented in %
Equations \eqref{eq:payoff_equations} and \eqref{eq:densityODEsystem}. %
We will present a conceptual formulation of the structure of our stochastic spatial model in this section, and we also use Section  \ref{sec:StochasticAppendix} of the appendix to present a more detailed description of our model design and the spatial generalization of the Gillespie-type algorithm used to simulate our stochastic model.

We consider a population living on a spatial domain consisting of a one-dimensional lattice containing $N$ patches, and we describe the spatial location of a patch by its index $i \in \{1,\cdots,N\}$. 
All game-theoretic interactions and demographic events occur in one of the $N$ patches, and all birth and death rates can be described by the number of hawks $u_i(t)$ and doves $v_i(t)$ located at a given patch $i$, and all interactions and competition within a patch are assumed to be well-mixed. %
The payoff of hawks and doves are calculated in the same way as in equations \eqref{eq:payoff_equations} above, with each individual playing the game against all other members of the patch and including the possibility of self-interaction. This means that we can define the average payoff achieved by hawks and doves at patch $i$ and time $t$ as the following function of the numbers of hawks and doves $u_i(t)$ and $v_i(t)$:%

\begin{align}\label{eq:PayoffEquationStochastic}
\begin{split}
p_{i, H}(t) &= P \left(\frac{u_i(t) }{u_i(t) + v_i(t) } \right) + T \left(\frac{v_i(t)}{u_i(t) + v_i(t)} \right)  \\
p_{i, D}(t) &= S \left(\frac{u_i(t)}{u_i(t) + v_i(t)} \right) + R \left(\frac{v_i(t)}{u_i(t) + v_i(t)} \right).
\end{split}
\end{align}

We use payoff to characterize the natural growth of the two species. The sign of $p_{i, H}$ and $p_{i, D}$ decides whether the payoff corresponds to the birth or death of an individual, and the absolute value determines the likelihood that such birth or death event occurs.

We further impose constraints $k_{i, u}(t)$ and $k_{i, v}(t)$ on population growth by carrying capacity:
\begin{align}\label{eq:CarryingCapacity}
\begin{split}
k_{i, u}(t) &= \kappa u_i(t) (u_i(t) + v_i(t)) \\
k_{i, v}(t) &= \kappa v_i(t) (u_i(t) + v_i(t))
\end{split}
\end{align}
where $\kappa$ is some constant. At time $t$, a hawk dies at patch $i$ with likelihood $k_{i, U}(t)$, similarly for the dove. 

Interactions among patches are simulated by payoff-driven migration rules. An individual chooses the direction of migration by comparing the payoff of neighboring patches, and migrates with higher possibility to a neighbor with higher payoff. Specifically, the probability that a hawk or dove at patch $i$ migrates to a neighboring patch $i'$ is given by
\begin{align}\label{eq:Migration}
\begin{split}
q_H(i \to i') = \mu_u \cdot \frac{f_u(w_u p_{H,i'})}{\ds\sum_{\tilde{\tilde{i}} \sim \tilde{i}} f_u(w_u p_{H,\tilde{i}})} \\
q_D(i \to i') = \mu_v \cdot \frac{f_v(w_v p_{D,i'})}{\ds\sum_{\tilde{\tilde{i}} \sim \tilde{i}}f_v(w_v p_{D,\tilde{i}})}
\end{split}
\end{align}

where the subscript $\tilde{\tilde{i}} \sim \tilde{i}$ indicates performing a sum over all neighbors $\tilde{\tilde{i}}$ of patch $\tilde{i}$, $\mu_u$ and $\mu_v$ are constants that measure an individual's willingness of migration, $f_u(\cdot)$ and $f_v(\cdot)$ are increasing functions that describes how hawks and doves respectively weight their average payoffs in their choice to move to neighboring patches, and $w_u$ and $w_v$ are parameters describing the sensitivity of payoff differences in determining their choice of patch when moving. Two example classes of movement rules we can consider are movement probabilities based on an affine function of payoff at a given location, given by
\begin{subequations}
\begin{align}
f_u\left( w_u p_{i,H} \right)  &= 1 + w_u p_{i,H} \\
f_v\left( w_v p_{i,D} \right)  &= 1 + w_v p_{i,D},
\end{align}
\end{subequations}
or an exponential weight placed on payoff, given by
\begin{subequations}\label{eq:movement-rule-expo}
\begin{align}
f_u\left( w_u p_{i,H} \right)   &= \exp\left( w_u p_{i,H} \right) \\ 
f_v\left( w_v p_{i,D} \right)  &= \exp\left(w_v p_{i,D} \right). \\ 
\end{align}
\end{subequations}
For our stochastic simulations, we will typically consider the case of the exponential mapping from payoff to movement weight. 
The rules for reproduction and movement defined above in Equations\eqref{eq:PayoffEquationStochastic}, \eqref{eq:CarryingCapacity}, \eqref{eq:Migration} characterize the core features of our stochastic spatial model. We provide additional information on the implementation of this model through Gillespie simulations in %
Section \ref{sec:StochasticAppendix} of the appendix. %


\subsection{PDE Models of Spatial Evolutionary Games}
\label{sec:PDEModel}

Following the introduction of the game-theoretic background in Section \ref{sec:game-background} and the description of payoff-driven stochastic motion in Section \ref{sec:StochasticModel}, we employ discrete-space stochastic models characterized by biased random walks between neighboring patches to derive a corresponding system of partial differential equations (PDEs). This derivation, detailed explicitly in Section \ref{sec:derivation_PDE} of the appendix, results in a system of PDEs the describing the dynamics of the spatial densities of hawks $u(t,x)$ and doves $v(t,x)$ across a one-dimensional interval with $x \in [0,L]$.  %
This system of PDEs is given by
\begin{subequations}
\label{eq:mainPDEsystem}
\begin{align}
        \dsdel{u(t,x)}{t} &=D_u\doubledelsame{u(t,x)}{x} + u\left(p_H(u,v) - k(u+v)\right)\\
        &- 2D_uw_u\left(\dsdel{u}{x}\frac{f_u'( w_u p_H)}{f_u( w_u p_H)}\dsdel{p_H}{x}+u\dsdel{}{x}\left(\frac{f_u'( w_u p_H)}{f_u( w_u p_H)}\dsdel{p_H}{x}\right)\right) \nonumber \\
       \dsdel{v(t,x)}{t} &=D_v\doubledelsame{v(t,x)}{x} + v\left(p_D(u,v) - k(u+v)\right)\\
       &- 2D_vw_v\left(\dsdel{v}{x}\frac{f_v'( w_v p_D)}{f_v( w_v p_D)}\dsdel{p_D}{x}+v\dsdel{}{x}\left(\frac{f_v'( w_v p_D)}{f_v( w_v p_D)}\dsdel{p_D}{x}\right)\right), \nonumber
    \end{align}
\end{subequations}
where $f_u(w_u p_H(\cdot))$ and $f_v(w_v p_D(\cdot))$ again describe the weight that hawks and doves place on payoff to determine movement probabilities when performing payoff-driven directed motion. In this paper, we pair this PDE with zero-flux boundary conditions at the endpoints $x = 0$ and $x = L$.

For the case of the exponential movement rule with $f_u(w_u p_H) = \exp(w_u p_H)$ and $f_v(w_v p_D) = \exp(w p_D)$, we see that the PDE model for payoff-driven motion takes the following simplified form 
\begin{subequations}
    \begin{align}
        \dsdel{u(t,x)}{t} &=D_u\doubledelsame{u(t,x)}{x} + u\left(p_H(u,v) - k(u+v)\right) - 2D_uw_u\left(\dsdel{p_H}{x} \dsdel{u}{x} +u \doubledelsame{p_H}{x}\right) \\
       \dsdel{v(t,x)}{t} &=D_v\doubledelsame{v(t,x)}{x} + v\left(p_D(u,v) - k(u+v)\right) - 2D_vw_v\left(\dsdel{p_D}{x} \dsdel{v}{x} +v\doubledelsame{p_D}{x}\right).
    \end{align}
\end{subequations}
We will typically consider this special case of our PDE model to study the role of payoff-driven motion in numerical simulations. 

To consider the case of purely undirected motion, we can set the payoff sensitivities as $w_u = w_v = 0$ and reduce our model of payoff-driven motion from Equation \eqref{eq:mainPDEsystem} to follow system of reaction diffusion equations %
\begin{subequations} \label{eq:PDEdiffusion-reaction}
    \begin{align}
        \dsdel{u(t,x)}{t} &=D_u\doubledelsame{u(t,x)}{x} + u\left(p_H(u,v) - k(u+v)\right) \\
        \dsdel{v(t,x)}{t}&=D_v\doubledelsame{v(t,x)}{x} + v\left(p_D(u,v) - k(u+v)\right).
    \end{align}
\end{subequations}
We will use this system of PDEs to study the possibility of Turing instability for our spatial hawk-dove game, exploring whether faster hawk diffusion can produce spatial patterns even in the absence of payoff-driven directed motion.

\section{Results from Simulations of the Stochastic Spatial Model}
\label{sec:StochasticResults}

In this section, we present simulation results of our stochastic model. We begin in Section \ref{sec:InitialModelDemo} with a simplified model and basic parameters, illustrating the time-dependent dynamics of our model and the spatial variation promoted by the stochasticity of the model. We then present a more comprehensive set of simulations for the case of purely diffusive motion in Section \ref{sec:PurelyDiffusiveMotion}, and we illustrate the effects of payoff-driven motion in Section \ref{sec:PayoffDrivenMotion}. %

For all of the stochastic simulations in our paper, we consider a common spatial simulation and fix many of the game-theoretic and ecological parameters across all simulations. Our spatial grid consists of 
a one-dimensional space containing a single row with $100$ patches, connected with nearest-neighbor coupling and zero-flux boundary conditions. %

We assume that all game-theoretic interactions within patches follow a hawk-dove game from Equation \eqref{eq:CVpayoff} with total resource $V = 4$ and cost of fighting $C = 6$, and a uniform carrying capacity $\kappa = \frac{1}{1000}$ 
is applied to all patches. Given the above payoff matrix and carrying capacity, we can calculate the expected equilibrium population in the coexistence state by Equation \eqref{eq:uv_equilibium}, finding that $u^{*} = \frac{4000}{9} \approx 444.44$ and $v^{*} = \frac{2000}{9} \approx 222.22$ in each patch. This expected equilibrium state in the large-population limit motivates us to consider initial conditions for the stochastic model consisting of %
this expected equilibrium state with a small noisy perturbation given by %

\begin{subequations}
\begin{align}
  u_i(0) &= 444 + \mathrm{rand(-5, 5)} \\ 
v_i(0) &= 222 + \mathrm{rand(-5, 5)},  
\end{align}
\end{subequations}

where $i \in [1, 100]$ is the patch index and $\mathrm{rand(-5, 5)}$ denotes a random integer in $[-5, 5]$. For our remaining simulations, %
we will consider how changes in the movement rates $\mu_u$ and $\mu_v$ and the payoff sensitivities $w_u$ and $w_v$ for each strategy impact the resulting dynamics in our stochastic spatial model. %

\subsection{Initial Model Demonstration}
\label{sec:InitialModelDemo}

\begin{figure}[H]
    \centering
        \includegraphics[width=0.485\textwidth]{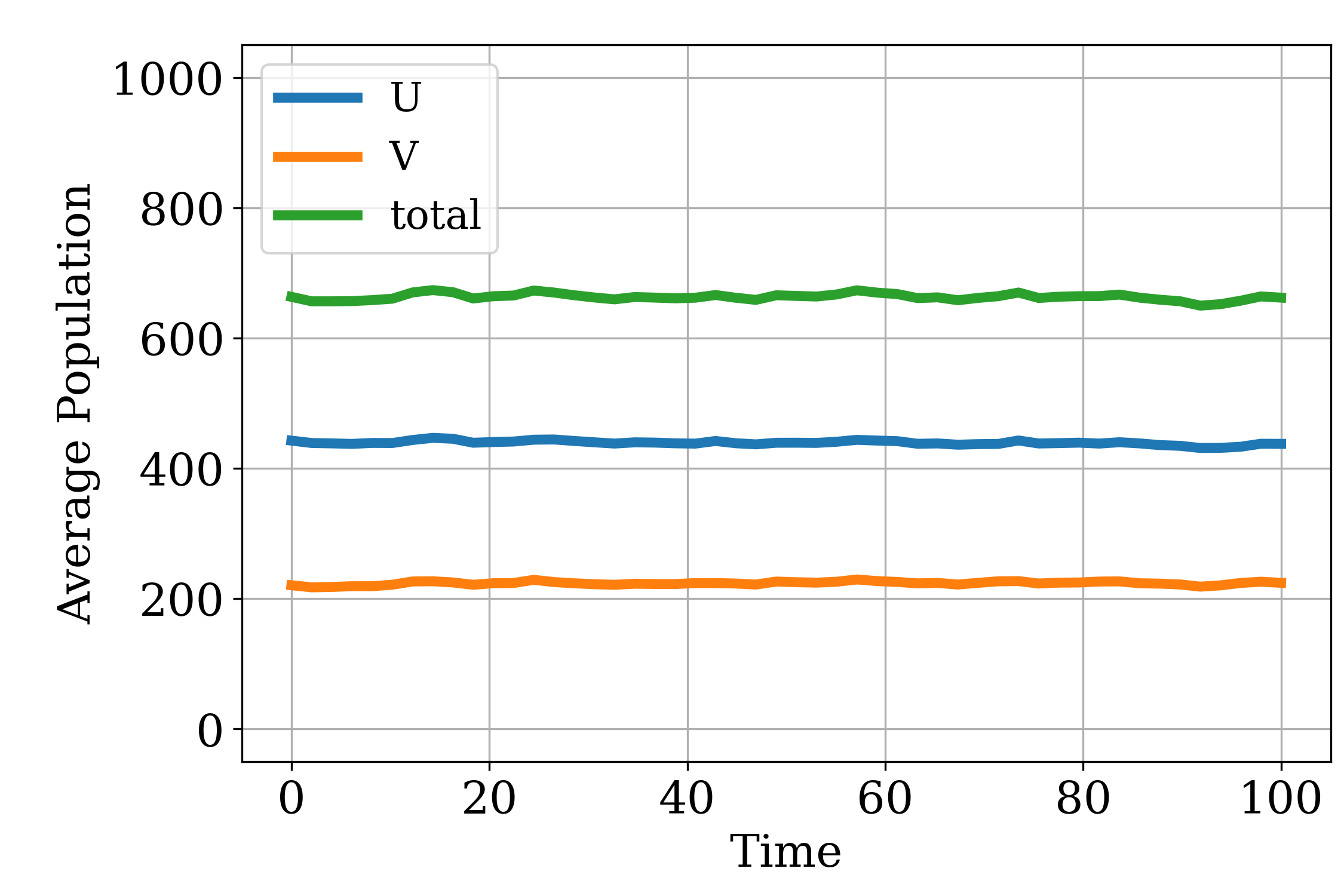}
        \label{fig:demo_popu_dyna}
    \hspace{1mm}
        \includegraphics[width=0.485\textwidth]{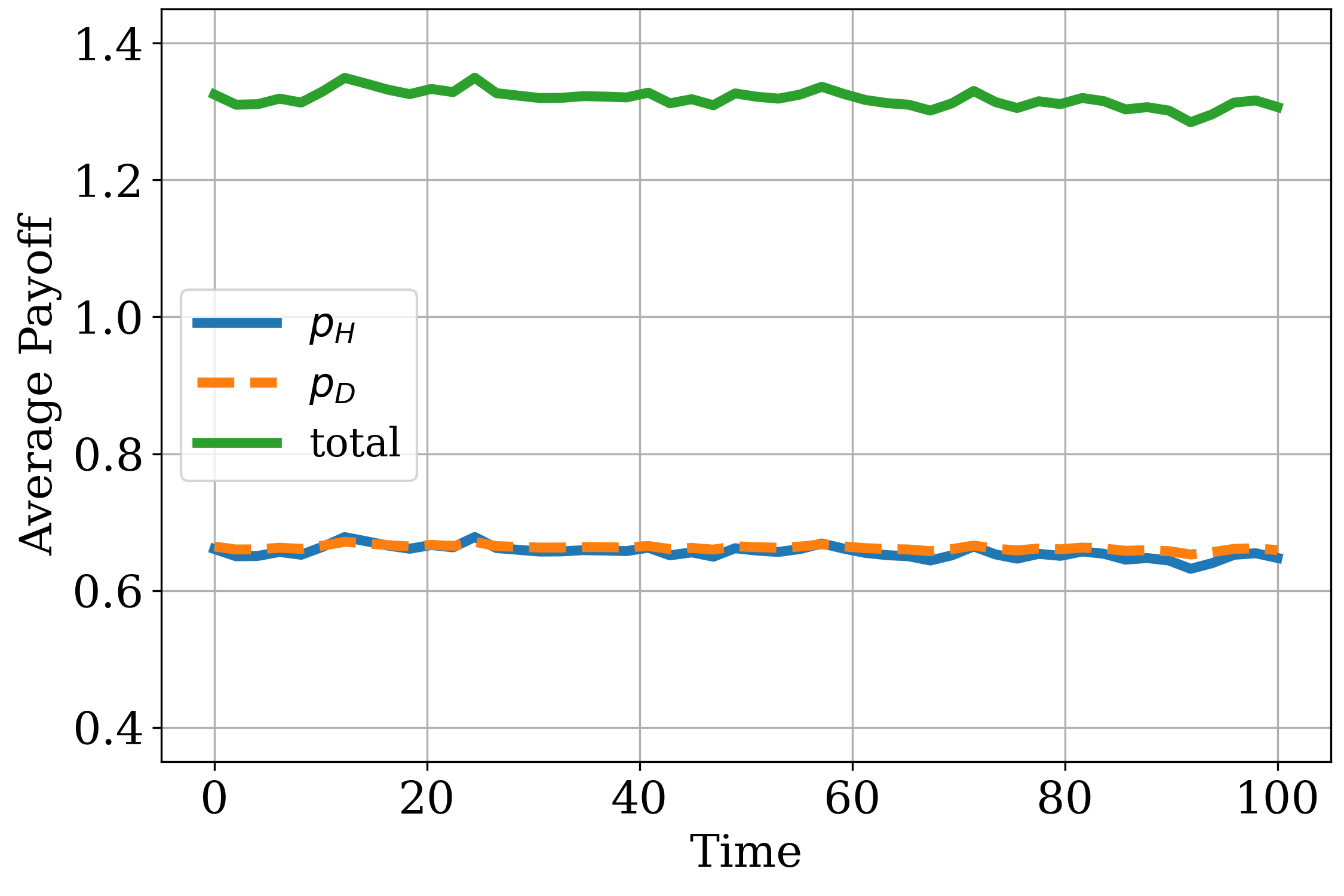}
        \label{fig:demo_payoff_dyna}
    \caption{Dynamics of average population (left) and payoff (right) in our demo model, with a single row of 100 patches and total resource $V=4$ and cost of fighting $C=6$ at each patch. We run a single simulation and track the average population and payoff across the spatial domain over time.}
    \label{fig:demo_dyna}
\end{figure}

To illustrate the temporal and spatial dynamics of our model, we first consider an example simulation of our model with payoff-driven motion. In Figure \ref{fig:demo_dyna}, we show the average population levels and average payoffs of each strategy across our spatial domain for a single simulation featuring equal movement rates and payoff sensitivities for each strategy (specifically, $\mu_u = \mu_v = 2, w_u = w_v = 1$), exploring the temporal dynamics for each time $t \in [0,100]$. The spatially average population sizes and payoffs of both species spend most of the time clustered around the expected equilibrium state, but there is some temporal fluctuation in these averaged quantities due to the stochasticity in our model. The concentration around the equilibrium predicted from the ODE system fits with our intuition that we do not expect deviation from the behavior of a well-mixed system when both hawks and doves have equal mobilities and equal sensitivity to payoff when performing payoff-driven motion.  %

\begin{figure}[H]
    \centering
        \includegraphics[width=1\textwidth]{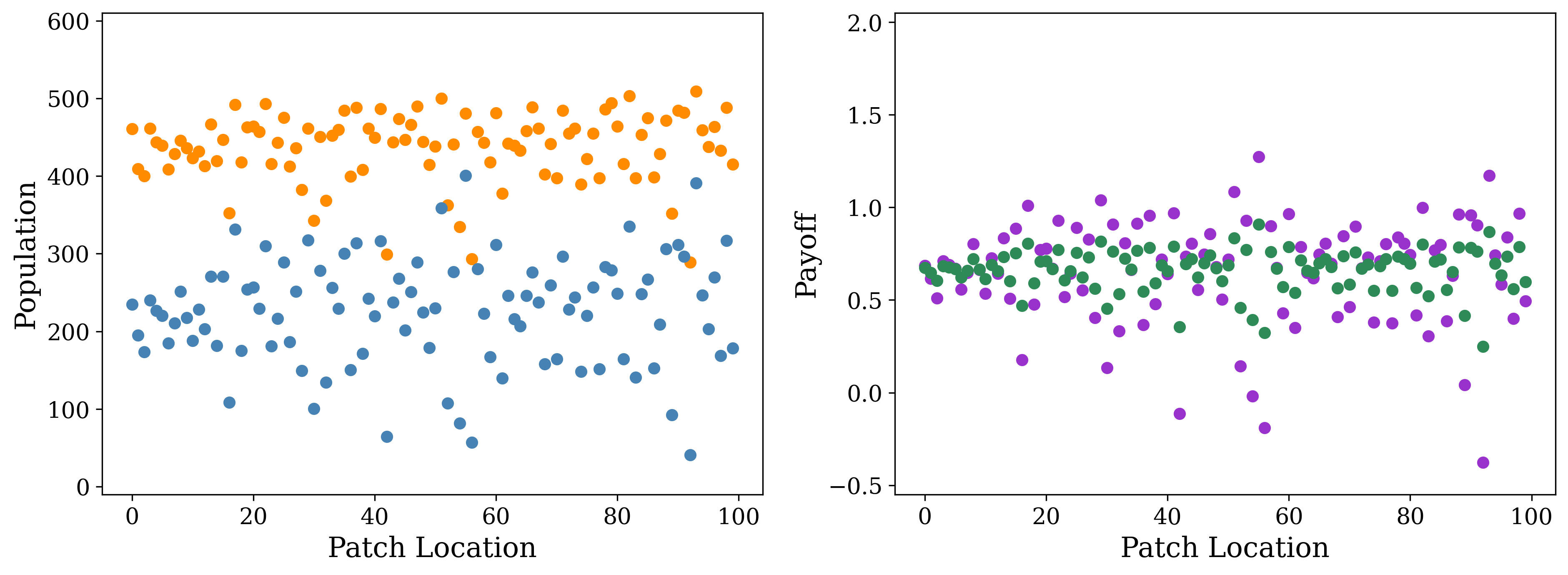}
    \caption{Spatial distribution of population (left) and payoff (right), plotted by averaging over $t \in [95, 100]$ for each patch. The same set of parameters as in Figure~\ref{fig:demo_dyna} is used. The resulting distributions are shown as scatter plots, with each dot representing the population or payoff of one patch. We run the simulation once, which results in the highly stochastic spatial distribution shown in the figures.}
    \label{fig:demo_distribution_end}
\end{figure}

However, we see that the stochastic nature of the model can result in spatial fluctuations even for this case of equal movement rules for each strategy. In particular, we explore in Figure \ref{fig:demo_distribution_end} the spatial distribution of population and payoff averaged over the last five percent of the simulation time (corresponding to times $t \in [95, 100]$), finding that there is substantial spatial variation in the number of hawks and doves at different patches in our domain. To determine whether it is possible to detect coherent spatial structure within our spatial system, 
we will now explore the spatial behavior of our model by averaging the results obtained over 100 simulations for each set of parameter values considered. %
This will allow us to examine the effects of different spatial movement parameters ($\mu_u \ne \mu_v$) or different payoff sensitivities ($w_u \ne w_v$) in promoting spatial patterns and different collective outcomes than achieved in a well-mixed population.

\subsection{Purely Diffusive Motion}
\label{sec:PurelyDiffusiveMotion}

\begin{figure}[!htpb]
    \centering
        \includegraphics[width=0.8\textwidth]{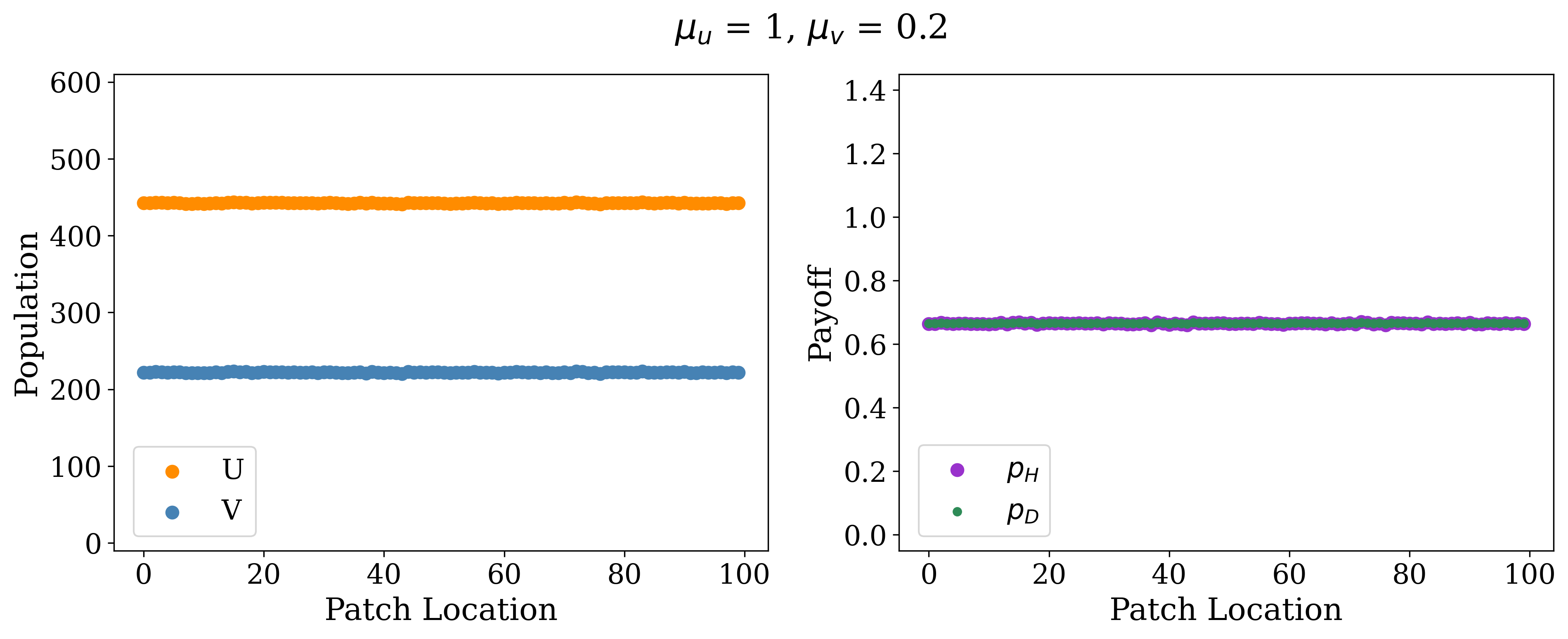}
        \label{fig:w0_1_02}

        \includegraphics[width=0.8\textwidth]{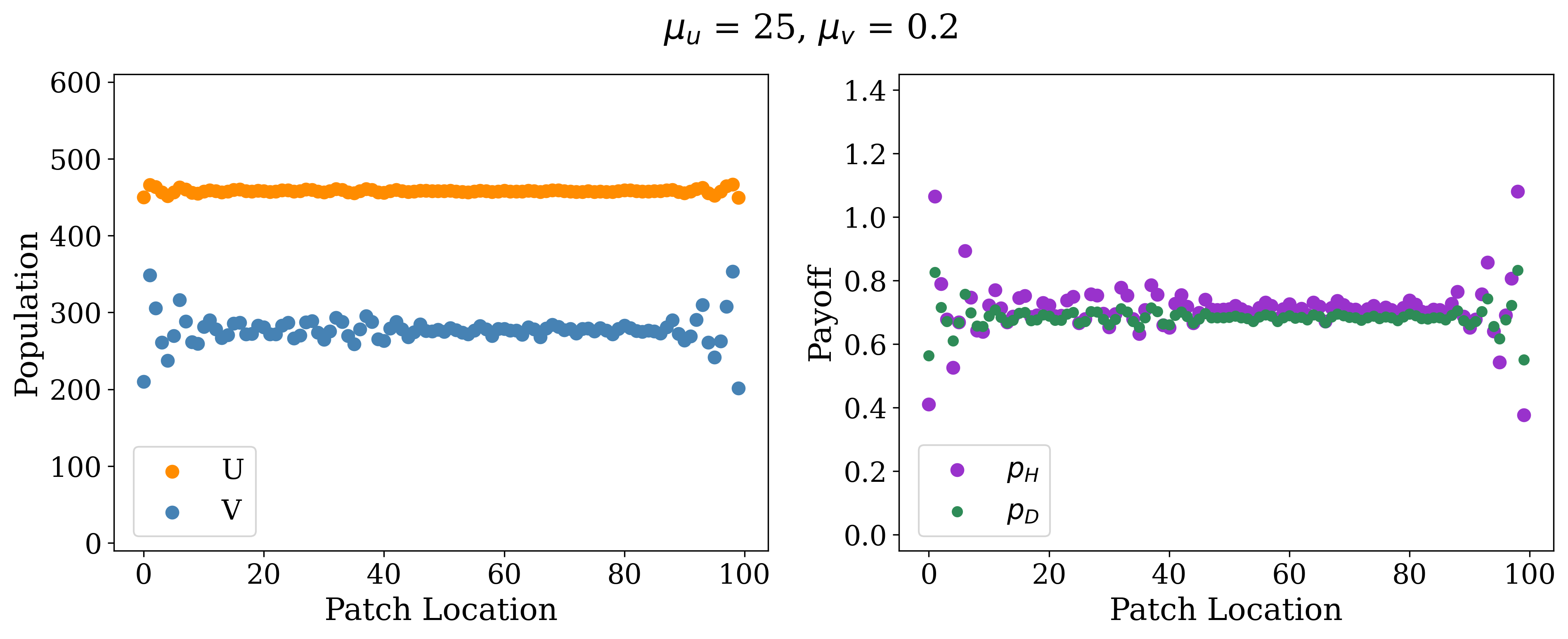}
        \label{fig:w0_10_02}

        \includegraphics[width=0.8\textwidth]{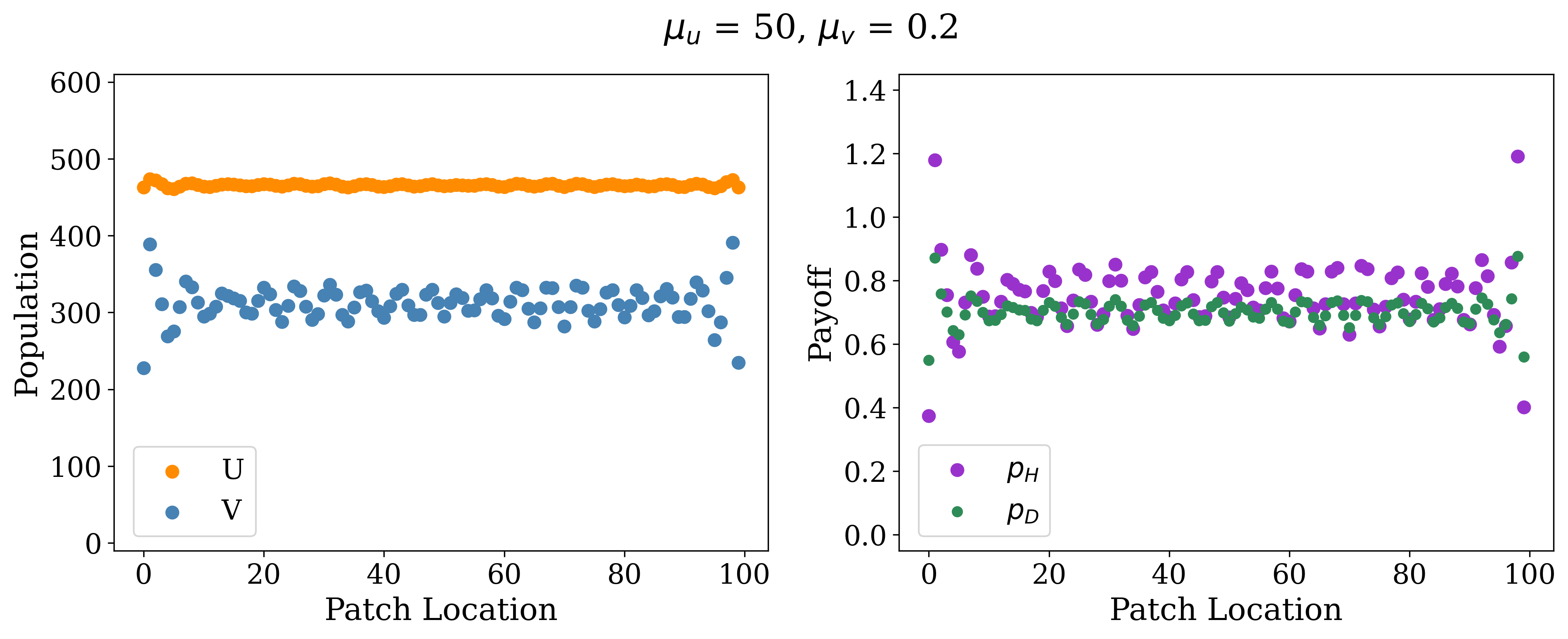}
        \label{fig:w0_50_02}
    \caption{Spatial distribution of population and payoff (left and right in each panel, respectively) observed in models with purely diffusive motion, with $\mu_u = 1, 25, 50$ and $\mu_v$ fixed at 0.2. Considering the stochastic nature of our model, we repeat each simulation 100 times and then take average over the resulting data. We slightly shrink $p_D$ dots for better visibility.}
    \label{fig:w0_spatial_pattern}
\end{figure}

We first consider the case of purely diffusive motion, in which individuals perform simple random walk in the space and payoff has no effect on the direction of migration. In terms of our general model for payoff-driven motion, we can achieve the case of simple random walks by setting $w_u = w_v =0$ in the weight-of-migration functions $f_u$ and $f_v$ in Equation \eqref{eq:Migration}. %
For the simulations of this diffusion model, we will fix the movement rate $\mu_v$ of doves and consider how changing the movement rate $\mu_u$ of hawks impacts emergent spatial patterns in our model. 
We illustrate the spatial profiles of population sizes and payoffs for each strategy achieved across our spatial domain in Figure \ref{fig:w0_spatial_pattern}, displaying these spatial quantities averaged across the last five percent of time-steps in our simulation. For these time-averaged profiles, we see a spatially uniform state for the case of low hawk diffusivity, and see the emergence of a sinusoidal pattern for the simulations with greater values of $\mu_u$. This behavior for large values of $\mu_u$ is reminiscent of the Turing instability in reaction-diffusion equations, with the faster diffusion of the hawks and slow diffusion of doves resembling the Turing mechanism of long-range inhibition and short-range activation \cite{turing1952chemical,dawes2016after,murray2007mathematical}. We find similar sinusoidal patterns in our PDE model for the case of purely diffusive motion in Figure \ref{fig:NP_Du_series} in Section \ref{sec:PDE-purediffusion}.   %

\begin{figure}[H]
\centering
    \subfloat[Population (left) and payoff (right) with $\mu_V = 0.2$] {
        \includegraphics[width=1\textwidth]{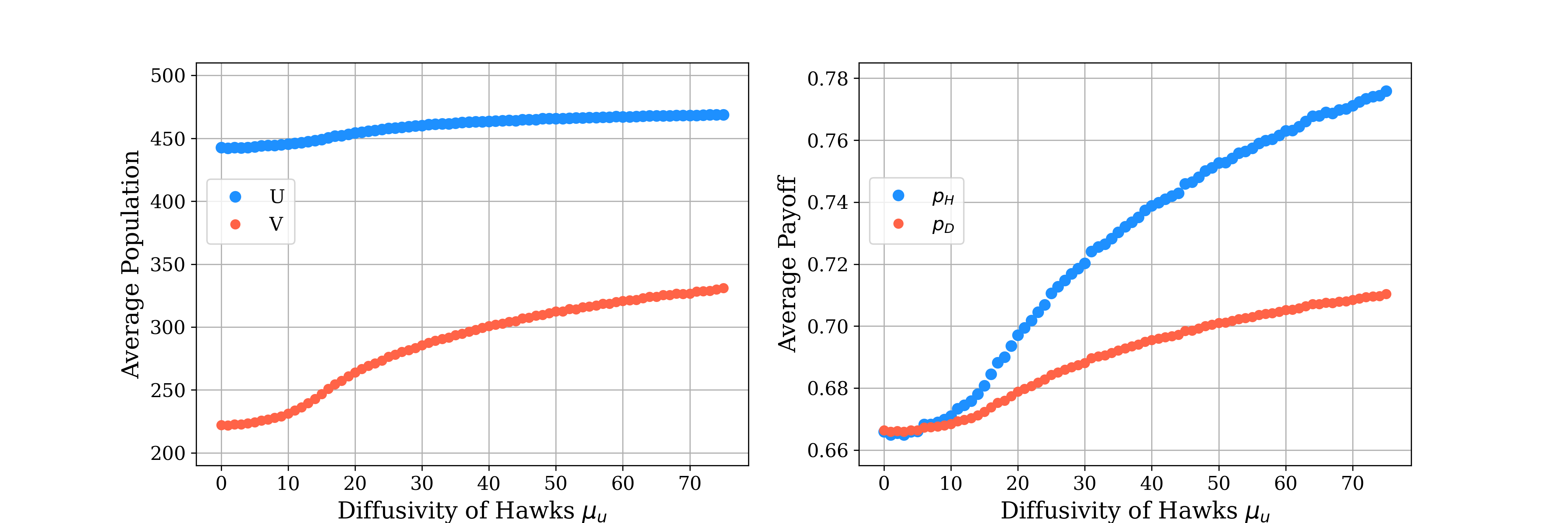}
    }
    \vspace{1mm}
    \subfloat[Population(left) and payoff (right) with $\mu_V = 2.0$] {
        \includegraphics[width=1\textwidth]{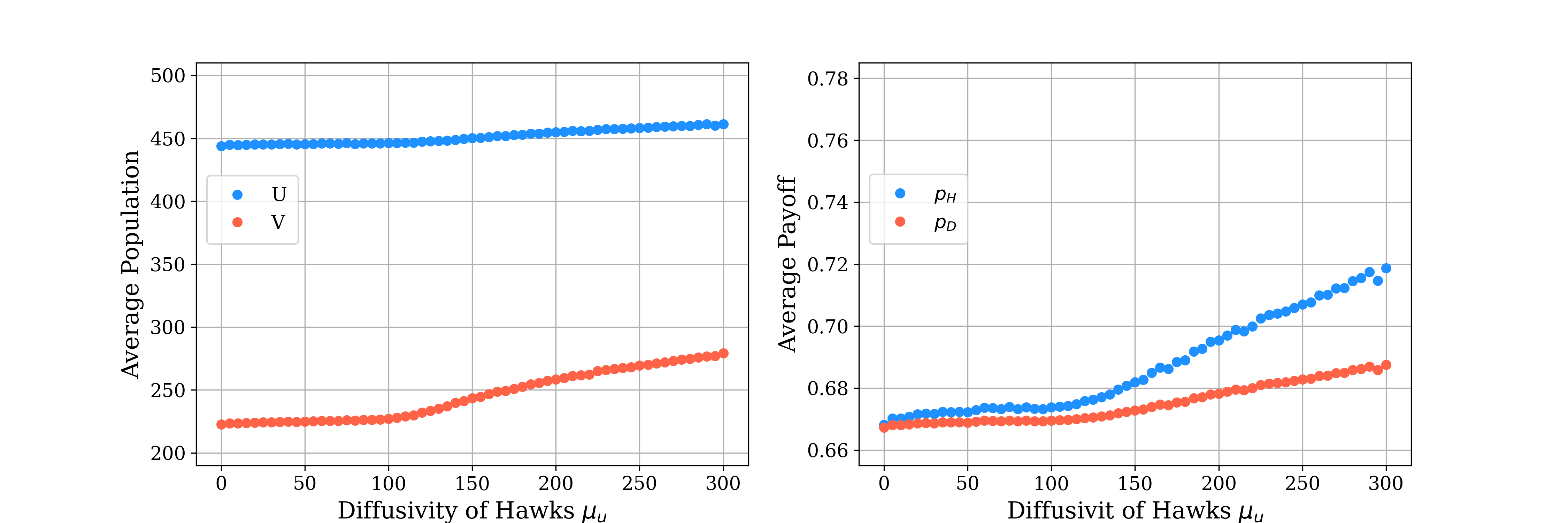}
    }

\caption{Comparison of average population sizes and average payoffs for hawks and doves with purely diffusive motion for the cases of $\mu_v = 0.2$ (top panels) and $\mu_v = 2.0$ (bottom panels). The plots are based on averages computed over 100 simulation runs. On the x-axes are the different values for $\mu_u$: $\mu_u \in [0, 75]$ with step size 1 in the $\mu_v = 0.2$ case, and $\mu_u \in [0, 300]$ with step size 5 in the $\mu_v = 2.0$ case. The y-axes are the average population/payoff across the 100 patches when the simulations reach equilibrium states, specifically the average values over the last five percent of simulation time for each patch. %
}
\label{fig:w_0_muboth}
\end{figure}

In addition, we consider aggregate quantities achieved across our spatial domain, plotting in Figure \ref{fig:w_0_muboth} the average population sizes and payoff achieved for each strategy as a function of the hawk movement rate $\mu_u$ for the cases of two fixed values of the dove movement rate $\mu_u$. In each of these cases, we see that both the average payoff and average population size of each strategy increase with the hawk movement rate $\mu_u$. This indicates that the collective outcome in the spatial population tends to be improved by the faster hawk diffusion, relative to the population and payoff achieved in a well-mixed population.

We also see that the two quantities appear to have a discontinuous change when hawk mobility $\mu_u$ passes approximately the value of $50 \mu_v$, with the quantities seeing roughly constant values when $\mu_u < 50 \mu_v$ and much stronger increases in population and payoff when $\mu_u > 50$. This suggests that the relative ratio $\frac{\mu_u}{\mu_v} \approx 50$ may correspond to a Turing instability in the stochastic model, and we will see that a relatively comparable Turing threshold appears in the PDE reaction-diffusion model in Section \ref{sec:PDE-purediffusion} and analogous game-theoretic and ecological parameters.

\subsection{Payoff-Driven Motion}
\label{sec:PayoffDrivenMotion}

\begin{figure}[!htpb]
    \centering
        \includegraphics[width=0.7\textwidth]{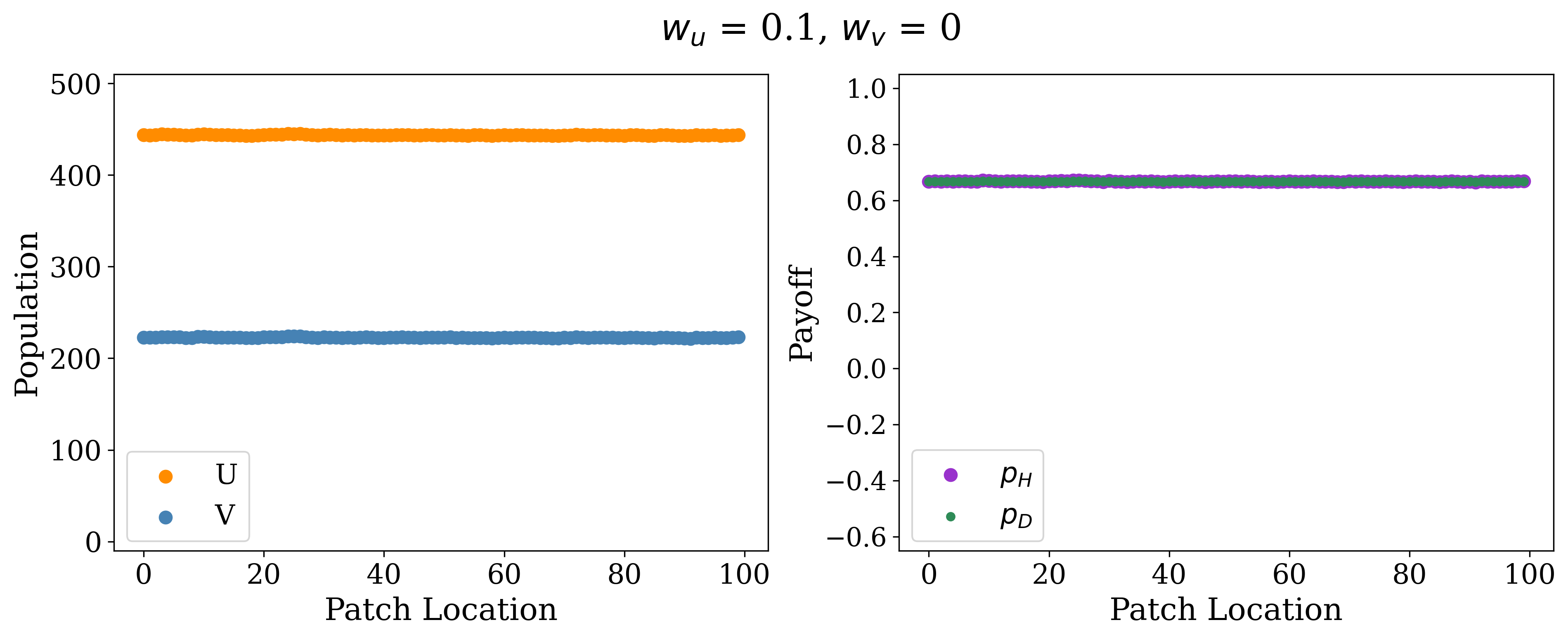}
        
        \includegraphics[width=0.7\textwidth]{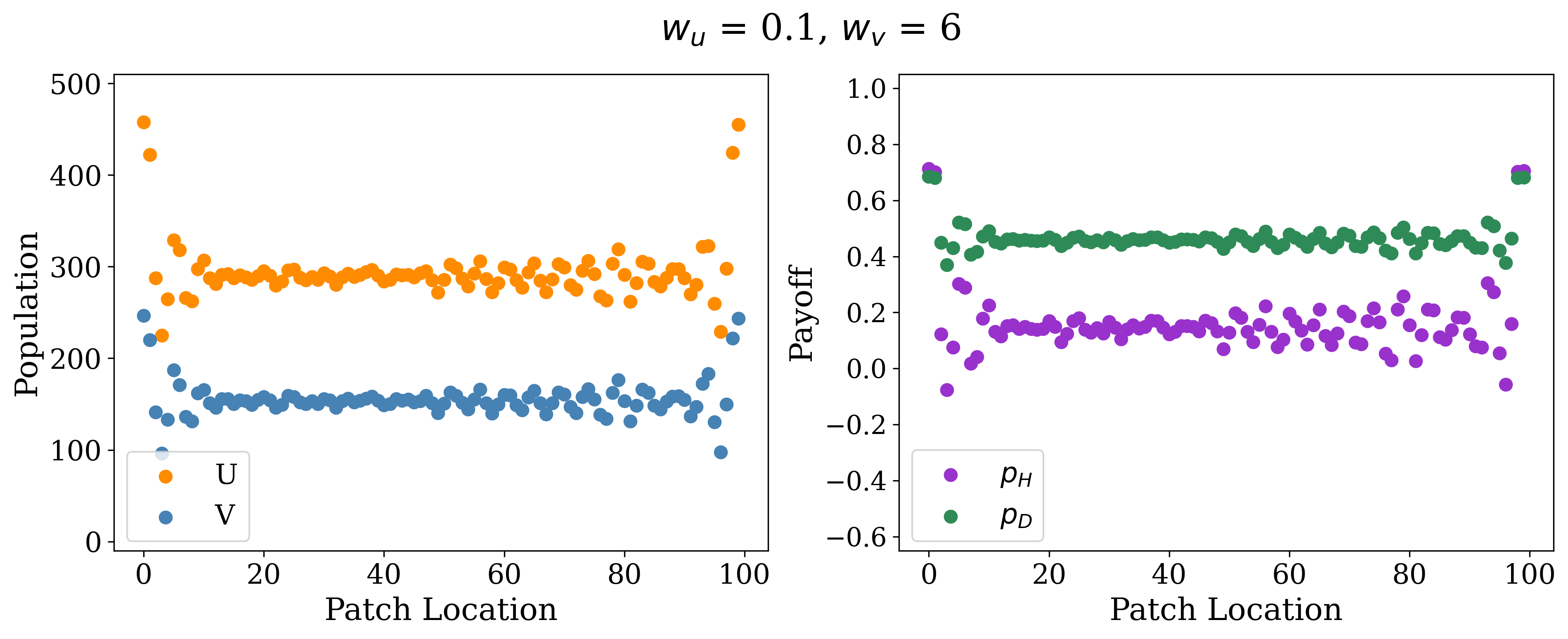}
        
        \includegraphics[width=0.7\textwidth]{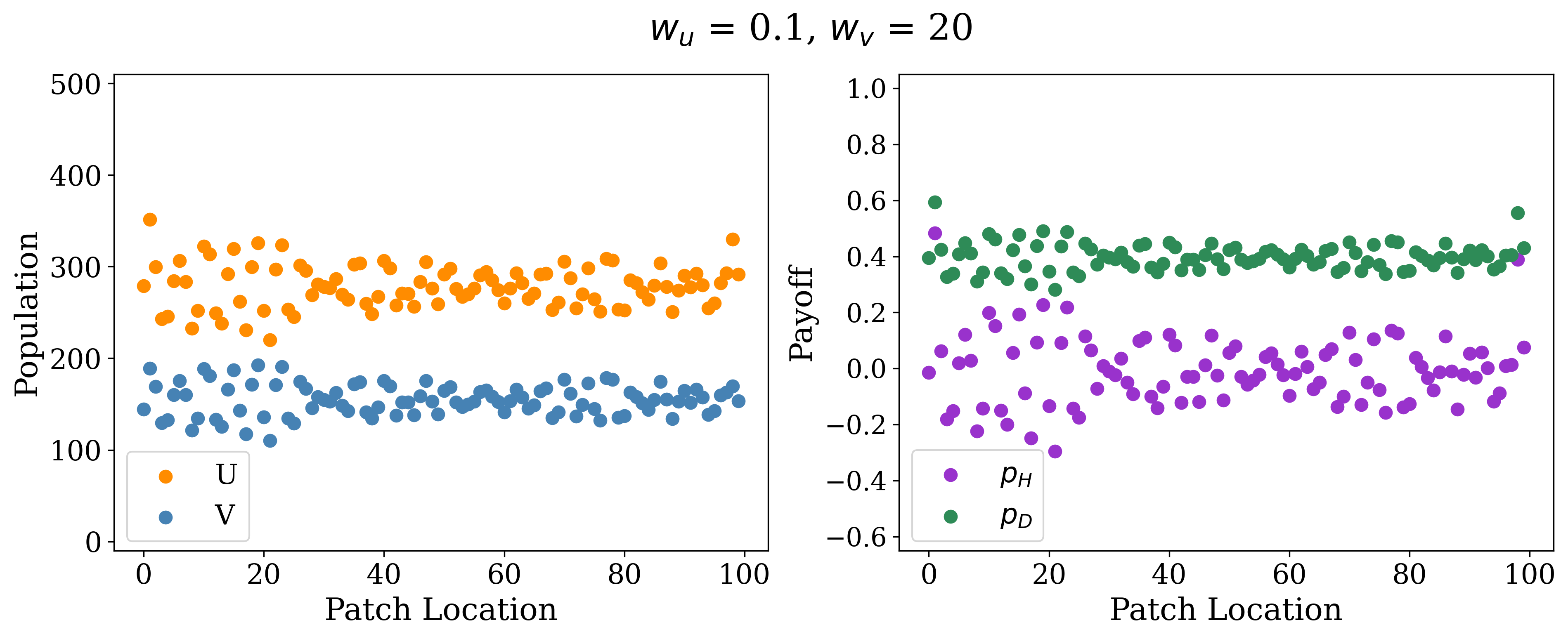}

        \includegraphics[width=0.7\textwidth]{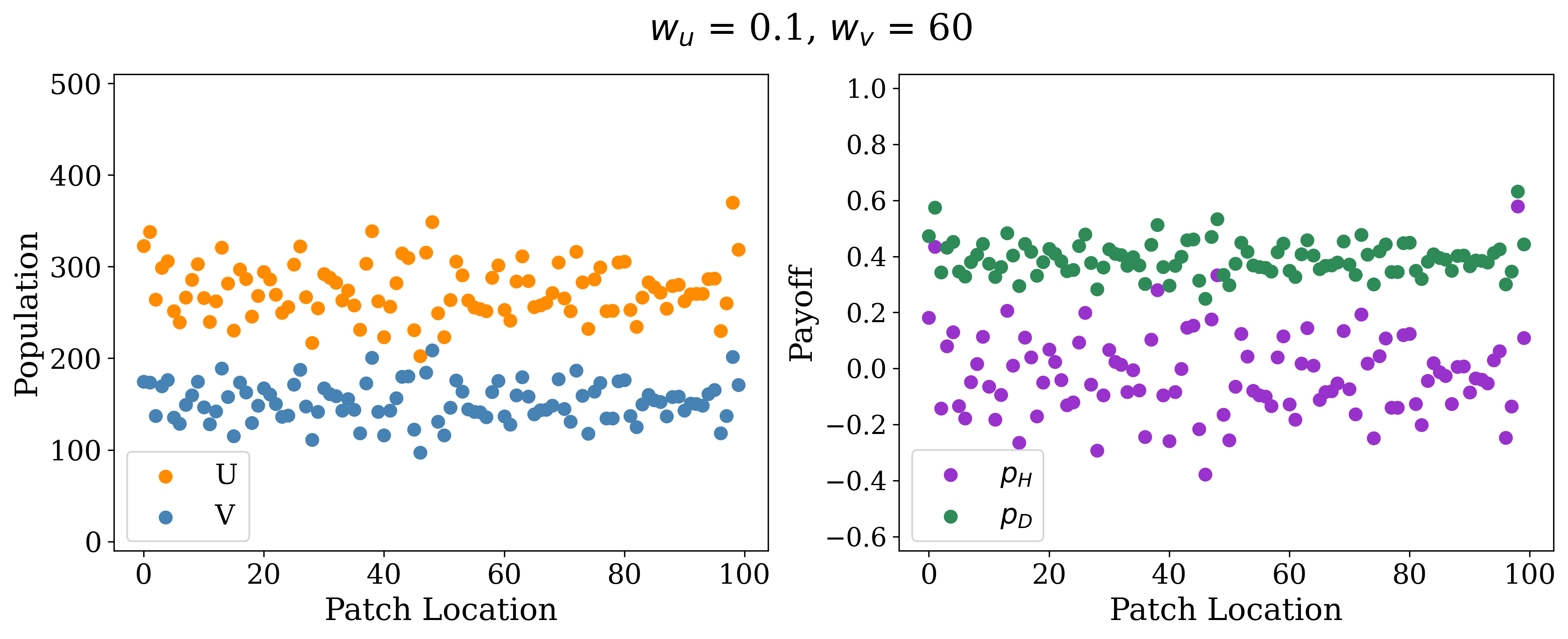}
    \caption{Spatial distribution of population and payoff (left and right in each panel, respectively) observed in models with payoff-driven motion, with $w_u$ fixed at 0.1, $w_v = 0, 6, 20, 60$, and $\mu_u = \mu_v = 1$. Each simulation is repeated 100 times.}
    \label{fig:mueq_spatial_pattern}
\end{figure}

We next consider the case of payoff-driven motion, specifically the case of the exponential movement rules $f_u(w_u p_H) = \exp(w_u p_H)$ and $f_v(w_v p_D) = \exp(w_v p_D)$ with positive payoff sensitivities $w_u$ and $w_v$ for the hawks and doves. In these simulations, we use the same game-theoretic and ecological parameters as in past sections, consider the case equal movement rates $\mu_u = \mu_v = 1$ for the two strategies, fix the hawk payoff sensitivity $w_u = 0.1$, and then explore how varying the dove payoff sensitivity $w_v$ impacts the dynamics of our stochastic spatial models. In Figure \ref{fig:mueq_spatial_pattern}, we present the spatial profiles of the population sizes and average payoffs for each strategy, considering the averaged values achieved at each spatial grid point for the last five percent of time-steps in our simulation. We see that the time-average profiles converge to a spatially uniform state at the equilibrium values expected from the ODE dynamics for our hawk-dove game for the case of low dove payoff sensitivity (for $w_v = 0$), while we see the emergence of patterned spatial profiles as we increase the strength of directed motion for doves. Notably, it appears that the patterns deviate somewhat from the sinusoidal spatial profiles seen in Figure \ref{fig:w_0_muboth} for the case of purely diffusive motion, with relatively rough patterns seen in the case of sufficiently strong payoff-driven motion (for $w_v = 60$). 

\begin{figure}[!htpb]
\vspace*{\fill}
\centering
\makebox[\textwidth][c]{
        \includegraphics[width=1.2\textwidth]{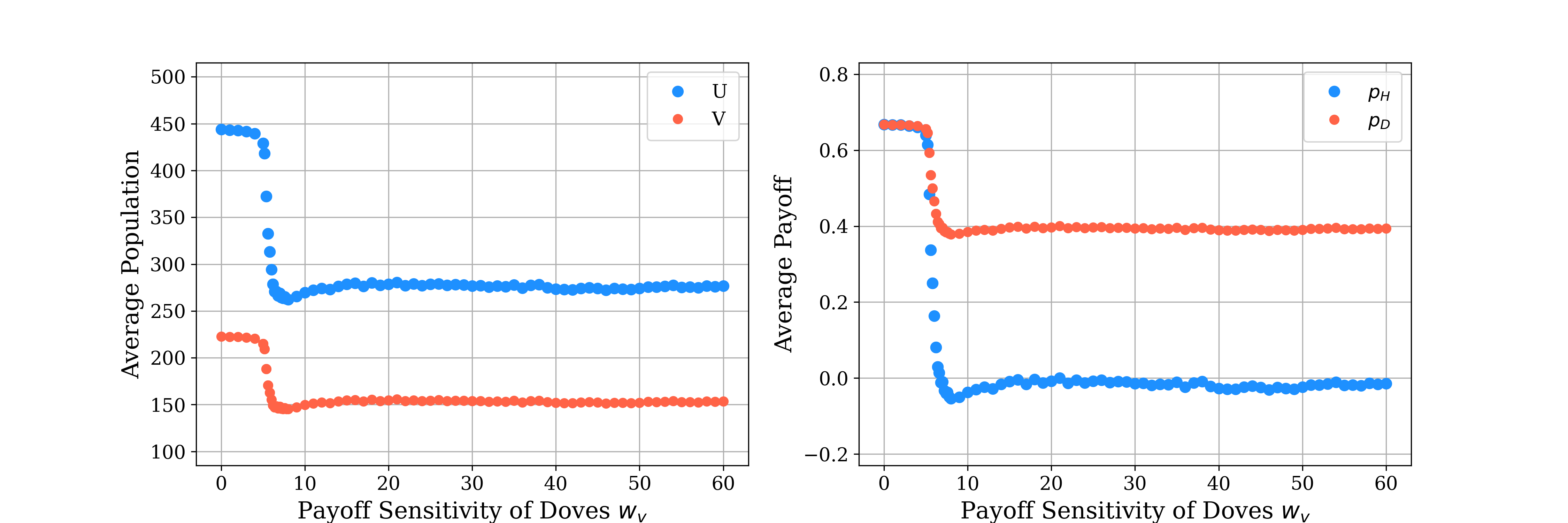}
    }
\caption{Average population (left) and payoff (right) across the spatial domain in payoff-driven case, with $w_u=0.1$, $w_v \in [0, 60]$, and $\mu_u = \mu_v = 1$. In order to capture the sharp decrease within the interval $w_v \in [5, 8]$, we use a finer step size of 0.2 in that region; while a coarser step size of 1 is used in the remainder of the domain. We repeat each simulation 100 times and plot with the average data.}
\label{fig:mu_eq_01}
\end{figure}

We can also consider the average population sizes and payoffs achieved by both strategies under this model of payoff-driven motion, exploring how the dove payoff sensitivity $w_v$ impacts the collective payoff of the population. In Figure \ref{fig:mu_eq_01}, we plot the average values of these quantities achieved over a set of 100 simulations, indicating that both population and size both appear to have a non-monotonic dependence of $w_v$. In particular, we see that the average population size and payoff are both constant for weak payoff-driven motion (for $w_v$ between approximately $0$ and $6$), and then these quantities all rapidly decrease at $w_v \approx 6$. While the payoffs and population sizes later generally appear to experience a slight increase for $w_v > 10$, we still see that the dynamics of the spatial model with sufficiently large $w_v$ produce more lower payoffs and population sizes relative to the equilibrium values predicted in the well-mixed model. Notably, the hawks achieve a much lower average payoff than the doves in the patterned state, so doves may benefit when considering their payoff relative to doves, although both strategies achieve much lower payoffs for large $w_v$ than experienced in the absence of spatial motion.

\section{PDE results for Turing and fully payoff-driven instability}
\label{sec:PDEResults}

We will now consider spatial pattern formation in our PDE model. We first review the conditions required for the stability of the coexistence equilibrium under the hawk-dove dynamics (Section \ref{sec:ODEstabilityreview}), and then we present general results for the linear stability analysis of spatially uniform states under the dynamics of our PDE model (Section \ref{sec:LSAgeneralcriteria}). We then examine the conditions for Turing instability and the resulting spatial patterns for the case of purely diffusive motion (Section \ref{sec:PDE-purediffusion}), and we demonstrate the finite wavenumber patterns are not possible in our model of payoff-driven motion for the case of the $C$-$V$ hawk-dove game and equal diffusivities for hawks and doves (Section \ref{sec:PDE-fullypayoff-driven}).

\subsection{Conditions for Stability of Equilibrium Under Reaction Dynamics}
\label{sec:ODEstabilityreview}

We first consider the equilibrium points of the reaction dynamics from Equation \eqref{eq:densityODEsystem}, which are given by
\begin{align}
\frac{du}{dt} &= a(u,v) := u \left(p_H(u,v) - \kappa(u+v)\right), \\
\frac{dv}{dt} &= b(u,v) := v \left(p_D(u,v) - \kappa(u+v)\right).
\end{align}
Such equilibrium points consist of population states $(u,v) = (u_0,v_0)$ that satisfy the equilibrium condition
\begin{subequations}  \label{eq:uniform_ODE}
\begin{align}
    u_0 \left(p_H (u_0 ,v_0) - \kappa (u_0 +v_0)\right) = 0 \\
    v_0 \left(p_D (u_0, v_0) - \kappa (u_0 +v_0 )\right) = 0,
\end{align}
\end{subequations}
and we recall that a coexistence equilibrium for the two-strategies of a hawk-dove game with resource value $V$ and fighting cost $C$ is given by
\begin{equation}
\left(u_0,v_0 \right) = \left( \frac{\left( C - V \right) V^2}{2 \kappa C^2},  \frac{V \left( C-V\right)^2}{2 \kappa C^2} \right).
\end{equation}
We can assess the stability of an equilibrium by considering the Jacobian matrix of the right-hand side of Equation \eqref{eq:densityODEsystem} evaluated at the point $(u_0,v_0)$, which we can write as
\begin{equation}
J(u_0,v_0) = \bpm \dsdel{a}{u} & \dsdel{a}{v}  \vspace{3mm} \\  \dsdel{b}{u} & \dsdel{b}{v}\epm \bigg|_{(u,v) = (u_0,v_0)} =: \bpm a_1 & a_2 \\ b_1 & b_2 \epm.
\end{equation}

In particular, we see that the linearization of the coexistence equilibrium for the hawk-dove game with the payoff matrix of Equation \eqref{eq:CVpayoff} has entries given by
\begin{subequations}
\label{eq:ODEJacobiansigns}
    \begin{align}
    a_1 &= \frac{\partial}{\partial u} (u (p_H - \kappa (u+v))) \bigg\rvert_{u=u_0, v = v_0}= \frac{V (-C+V)(C+2V)}{2C^2 }< 0\label{eq:coefficient-a1} \\
      a_2 &= \frac{\partial}{\partial v} (u (p_H - \kappa (u+v))) \bigg\rvert_{u=u_0, v = v_0}= \frac{V^3}{C^2} > 0  \label{eq:coefficient-a2} \\
    b_1& =\frac{\partial}{\partial u} (v (p_D - \kappa (u+v))) \bigg\rvert_{u=u_0, v = v_0}=  \frac{-V (V-C)^2}{C^2}< 0 .\label{eq:coefficient-b1}  \\
    b_2 &= \frac{\partial}{\partial v} (v (p_D - \kappa (u+v))) \bigg\rvert_{u=u_0, v = v_0}= \frac{V (V-C)(C-2V)}{2 C^2}. %
    \label{eq:coefficient-b2} 
\end{align}
\end{subequations}
In particular, we see that $b_2 > 0$ if $C < 2V$, while $b_2 < 0$ when $C > 2V$.

\begin{remark}
From the partial derivatives calculated in Equation \eqref{eq:ODEJacobiansigns}, we see that the linearization of our reaction terms will have the activator-inhibitor sign pattern if $C < 2V$, with the dove playing the role of the activator and the hawk playing the role of the inhibitor. This means that it is possible to generate spatial patterns for games with this payoff matrix via a Turing instability. Therefore spatial pattern formation can occur in our spatial model of hawk-dove games in the absence of payoff-driven motion, but that adding payoff-driven motion would be required for other Hawk-Dove games with $C \geq 2V$. 
\end{remark}

\subsection{Linear Stability Analysis of PDE Model with Diffusive and Payoff-Driven Motion}
\label{sec:LSAgeneralcriteria}

We now look to analyze the stability of a spatially uniform state $(u(x),v(x)) = (u_0,v_0)$ with the constant densities corresponding to a stable coexistence equilibrium of hawks and doves under the reaction dynamics. To study the stability of this uniform equilibrium under the PDE model of Equation \eqref{eq:PDE-payoff-driven}, we consider a small perturbation from the uniform state taking the form
\begin{subequations}
\begin{align}
u(t,x)&=u_0+\epsilon u_1(t,x)\\
v(t,x)&=v_0+\epsilon v_1(t,x),
\end{align}
\end{subequations}
and plug these perturbations into both sides of Equation \eqref{eq:PDE-payoff-driven}. We show in Section \ref{sec:linearization_appendix} that, after performing a perturbation expansion of the right-hand side and retaining only terms of order $\epsilon$, the functions $u_1(t,x)$ and $v_1(t,x)$ satisfy the following linear system of PDEs
\begin{subequations}
\begin{align}
\dsdel{u_1}{t} &= D_u \left[ 1 - 2 w_u u_0 c_1 \right] \doubledelsame{u_1}{x} - 2 D_u w_u u_0 c_2 \doubledelsame{v_1}{x} + a_1 u_1 + a_2 v_1 \\
\dsdel{v_1}{t} &= - 2 D_v w_v v_0 c_3 \doubledelsame{u_1}{x}  +  D_v \left[ 1 - 2 w_v v_0 c_4 \right] \doubledelsame{v_1}{x} + b_1 u_1 + b_2 v_1
\end{align}
\end{subequations}
where $a_1$, $a_2$, $b_1$, and $b_2$ are the entries of the linearization for the reaction dynamics and the constants $c_1$, $c_2$, $c_3$, and $c_4$ are given by
\begin{subequations}
\begin{align}
 c_1 &= \frac{f_u'\left( w_u p_H\left( u,v\right) \right)}{f_u\left( w_u p_H\left( u,v\right) \right)}\frac{\partial p_H\left( u,v\right)}{\partial u}\bigg\rvert_{u=u_0, v = v_0} \\
  c_2 &= \frac{f_u'\left( w_u p_H\left( u,v\right) \right)}{f_u\left( w_u p_H\left( u,v\right) \right)}\frac{\partial p_H\left( u,v\right)}{\partial v}\bigg\rvert_{u=u_0, v = v_0} \\
    c_3 &= \frac{f_v'\left( w_v p_D\left( u,v\right) \right)}{f_v\left( w_v p_D\left( u,v\right) \right)}\frac{\partial p_D\left( u,v\right)}{\partial u}\bigg\rvert_{u=u_0, v = v_0} \\
    c_4 &=\frac{f_v'\left( w_v p_D\left( u,v\right) \right)}{f_v\left( w_v p_D\left( u,v\right) \right)}\frac{\partial p_D\left( u,v\right)}{\partial v}\bigg\rvert_{u=u_0, v = v_0} .
\end{align}
\end{subequations}

Due to our assumption of zero-flux boundary conditions, we consider the following ansatz for a solution to this linearized system featuring a sinusoidal spatial profile with wavenumber $m$
\begin{align}
    \begin{split}
        u_1 (t, x) &= k_{u_1}e^{\sigma_m t} \cos\left(\frac{m \pi x}{l}\right),\\
        v_1 (t, x) &= k_{v_1} e^{\sigma_m t} \cos\left(\frac{m \pi x}{l}\right),
    \end{split}
\end{align}
and we note that these expressions for $u_1(t,x)$ and $v_1(t,x)$ will be a solution to linearized system provided that the following linear system is satisfied:
\begin{align} \label{eq:sigma-form PDE}
\begin{split}
    \sigma_m k_{u_1} &= \left[a_1 - D_u \left(\frac{m\pi}{l}\right)^2 + 2D_uw_u c_1u_0  \left(\frac{m\pi}{l}\right)^2 \right]k_{u_1} + \left[a_2 + 2D_uw_u c_2u_0 \left(\frac{m\pi}{l}\right)^2 \right] k_{v_1} \\
    \sigma_m k_{v_1} &= \left[b_1 +  2D_vw_v c_3 v_0  \left(\frac{m\pi}{l}\right)^2 \right] k_{u_1} + \left[b_2 - D_v \left(\frac{m\pi}{l}\right)^2 + 2D_vw_v c_4v_0 \left(\frac{m\pi}{l}\right)^2 \right] k_{v_1}.
\end{split}
\end{align}
Rewriting Equation \eqref{eq:sigma-form PDE} in matrix form
\begin{equation} 
\sigma_m \bpm  k_{u_1}\\
        k_{v_1} \epm  = \underbrace{\begin{pmatrix}
        a_1 - D_u\left[1 -  2 w_u c_1u_0 \right] \left(\frac{m\pi}{l}\right)^2  & a_2 +2D_uw_u c_2u_0\left(\frac{m\pi}{l}\right)^2 \vspace{2mm} \\
        b_1 + 2D_vw_v c_3v_0  \left(\frac{m\pi}{l}\right)^2 & b_2 - D_v \left[1 - 2 w_v c_4v_0  \right] \left(\frac{m\pi}{l}\right)^2 
    \end{pmatrix}}_{:= A(m)} \bpm  k_{u_1}\\
        k_{v_1} \epm,
\end{equation}
so the growth rate $\sigma_m$ is an eigenvalue of the linearization matrix $A(m)$. We therefore expect perturbations with wavenumber $m$ to grow if the matrix $A(m)$ has an eigenvalue with positive real part, so we will to characterize the formation of spatial patterns by evaluating the trace and determinant of $A(m)$ for each wavenumber $m$. 

\subsection{Results for Pure Diffusion Motion}
\label{sec:PDE-purediffusion}

We first consider stability of the uniform state $(u_0,v_0)$ in the absence of payoff-driven motion, which occurs when $w_u = w_v = 0$ and our PDE model reduces to the reaction-diffusion system from Equation \eqref{eq:PDEdiffusion-reaction}.
In the reaction-diffusion case, the linearization $A(m)$ of the PDE system reduces for perturbations of wavenumber $m$ reduces to the following matrix %
\begin{align}
    A_{\mathrm{RD}(m)}= \begin{pmatrix}
        a_1 - D_u \left(\frac{m \pi}{l}\right)^2 & a_2  \\
        b_1   & b_2 - D_v \left(\frac{m \pi}{l}\right)^2 
    \end{pmatrix}.
\end{align}
For the underlying ODE system (without diffusion) to be stable, the standard stability criteria must be satisfied:
\begin{align} \label{eq:ODE-stable-condition}
    a_1 + b_2 &< 0, \quad \text{and} \quad a_1b_2 - a_2b_1 > 0.
\end{align}
The conditions for diffusion-driven (Turing) instability are characterized by the following inequalities:
\begin{subequations}
\begin{align}
    \operatorname{tr}(A_{\mathrm{RD}}(m)) &= a_1 + b_2 - (D_u + D_v)\left(\frac{m\pi}{l}\right)^2 < 0 \label{eq:traceRD} \\
    \det(A_{\mathrm{RD}}(m)) &=  a_1b_2 - a_2b_1 - (a_1D_v + b_2D_u)\left(\frac{m \pi}{l}\right)^2 + D_uD_v\left(\frac{m \pi}{l}\right)^4 > 0.
\end{align}
\end{subequations}
For a homogeneous equilibrium $(u_0,v_0)$ that is stable under the reaction dynamics, we have that $a_1 + b_1 < 0$, so the trace in Equation \eqref{eq:traceRD} will always be negative for any non-negative diffusivities $D_u$ and $D_v$. We therefore seek conditions on the diffusivities $D_u$ and $D_v$ and wavenumber $m$  for which it is possible to achieve instability of the uniform state by examining the condition $\det(A_{\mathrm{RD}}(m)) < 0$. If this inequality holds $\det(A_{\mathrm{RD}}(m))$, we will further have that

\begin{equation} \label{eq:Turinginequality}
    \left(D_v\left(\frac{m \pi}{l}\right)^4 - b_2\left(\frac{m \pi}{l}\right)^2\right)D_u<a_1D_v\left(\frac{m \pi}{l}\right)^2 + b_1a_2 - a_1b_2,
\end{equation}
and we can use this inequality to deduce necessary and sufficient conditions for Turing instablity in our model. 
Because the diffusivities $D_u$ and $D_v$ are non-negative and the term $a_1 b_2 - a_2 b_1 > 0$ due to the stability of the reaction dynamics in the absence of diffusion, this inequality can only hold if the term in parenthesis on the right-hand side is positive. Therefore we see that this inequality can only hold for sufficiently small wavenumbers satisfying
\begin{equation}
m \leq  m_c := \frac{l}{\pi} \sqrt{\frac{b_2}{D_v}}.
\end{equation}
We can then see that for $m \leq m_c$, the uniform state will be unstable to a cosine-wave perturbation with wavenumber $m$ provided that the diffusivity of the hawks $D_u$ satisfies the following inequality
\begin{equation} \label{eq:criticalDu}
D_u>D_u^*(m) := \frac{a_1D_v\left(\frac{m \pi}{l}\right)^2 + b_1a_2 - a_1b_2}{D_v\left(\frac{m \pi}{l}\right)^4 - b_2\left(\frac{m \pi}{l}\right)^2 }.   
\end{equation}

\begin{figure}[!ht]
    \centering
    \subfloat[Critical hawk diffusivity $D_u^*(m)$]{
        \includegraphics[width=0.45\linewidth]{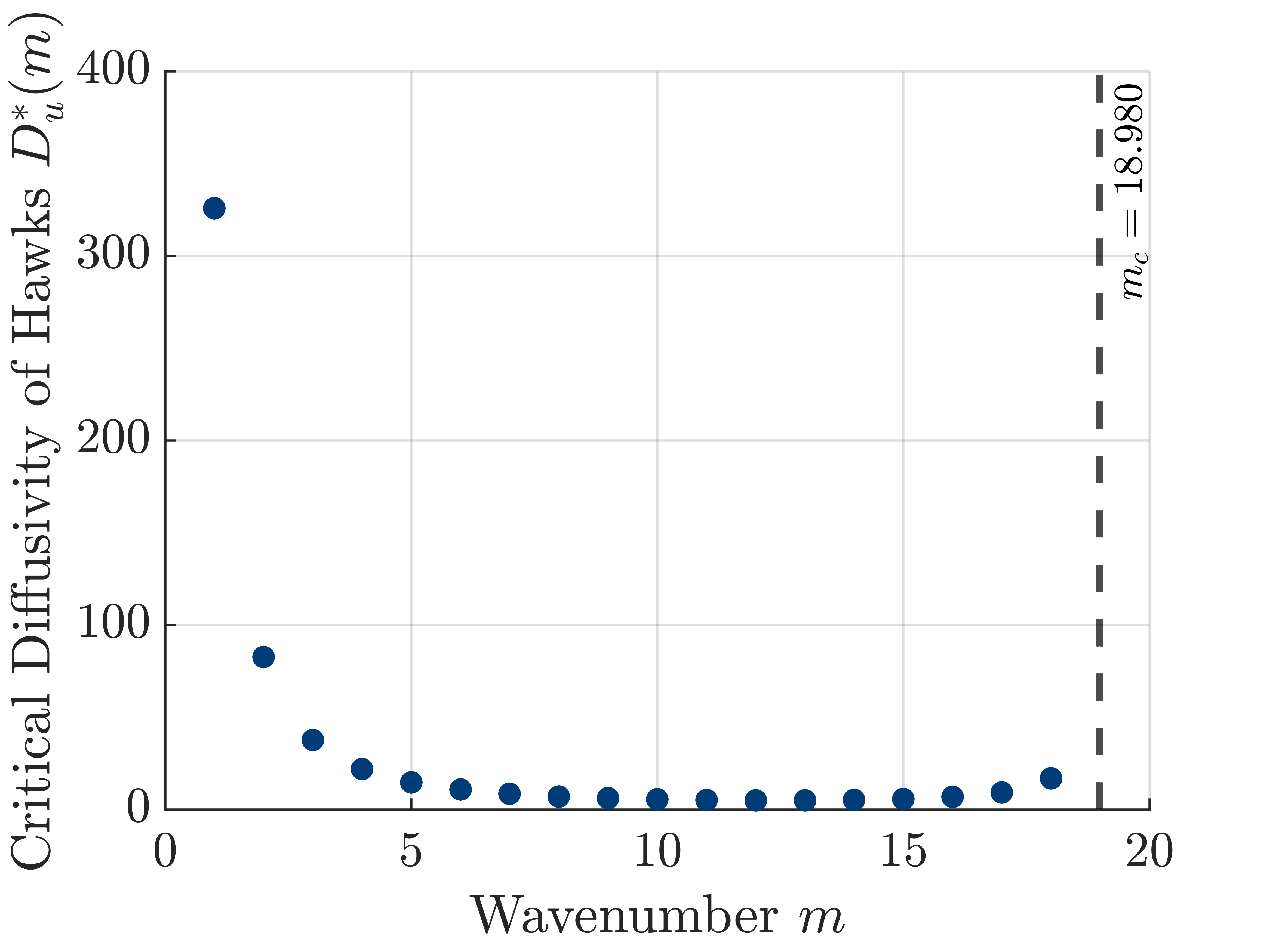}
    }\hspace{5mm}
    \subfloat[Zoomed-in view]{
        \includegraphics[width=0.45\linewidth]{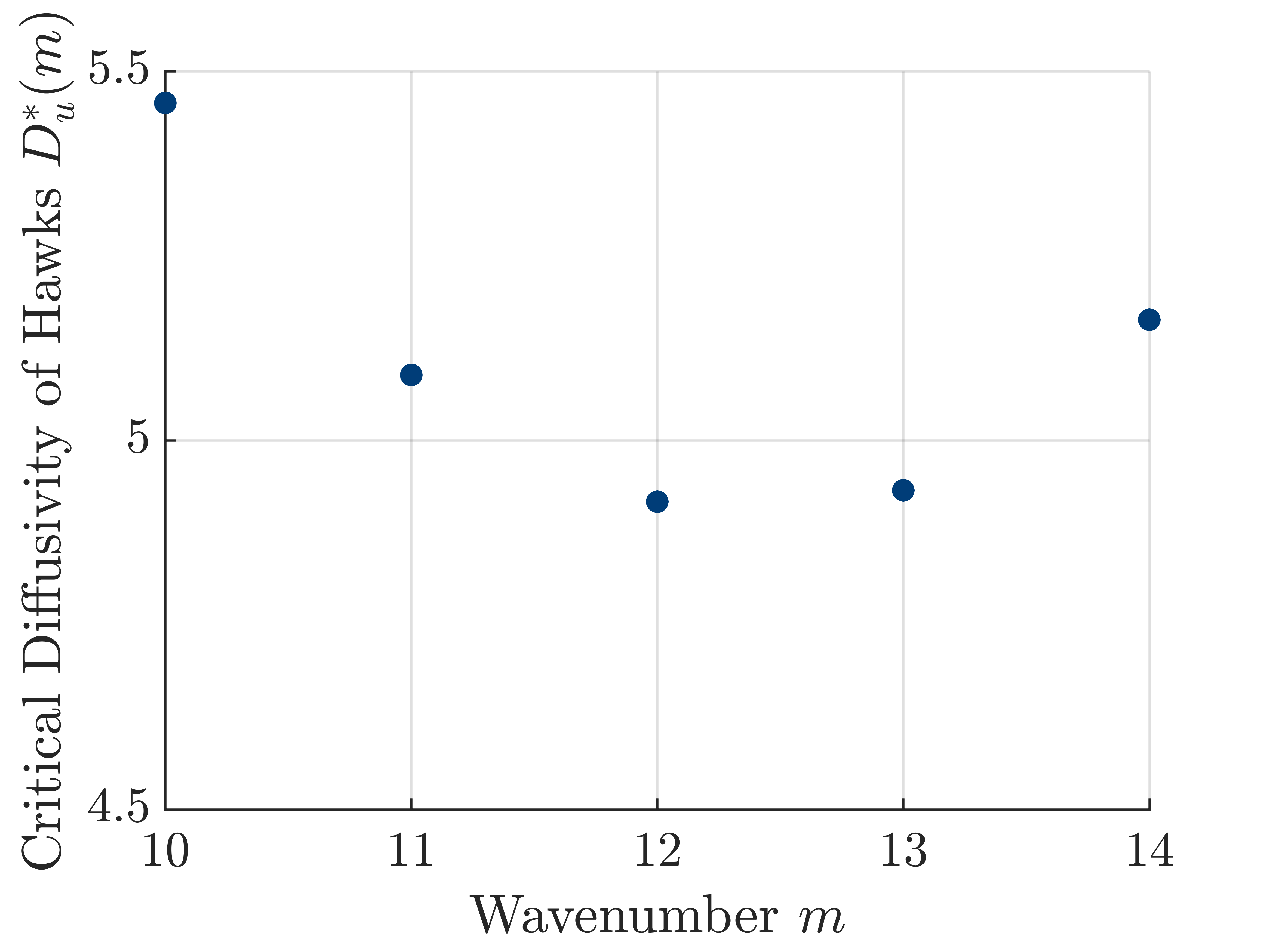}
    }
    \caption{Critical hawk diffusivity $D_u^*$ as a function of wavenumber $m$, characterizing the threshold above (or below) which diffusion-driven pattern formation occurs. (a) Full plot of $D_u^*(m)$ showing that pattern formation is feasible only for $m < m^* \approx 18.98$. The minimum occurs at $m = 12$, yielding $D_u^* \approx 4.917$. (b) Zoomed-in view highlighting the behavior near the critical minimum. The values of $D_u^*(m)$ are calculated using the the hawk-dove payoff matrix with $V = 4$ and $C = 6$,  carrying capacity $\kappa = 0.001$, and domain length $l = 40$.
    }
    \label{fig:nonpayoff_critical_threshold}
\end{figure}

\subsubsection{Illustration of Emergent Patterns Using Numerical Simulations}

\begin{figure}[!ht]
    \centering
    \subfloat[Spatial Distribution of Populations $u$ and $v$]{
        \includegraphics[width=0.45\textwidth]{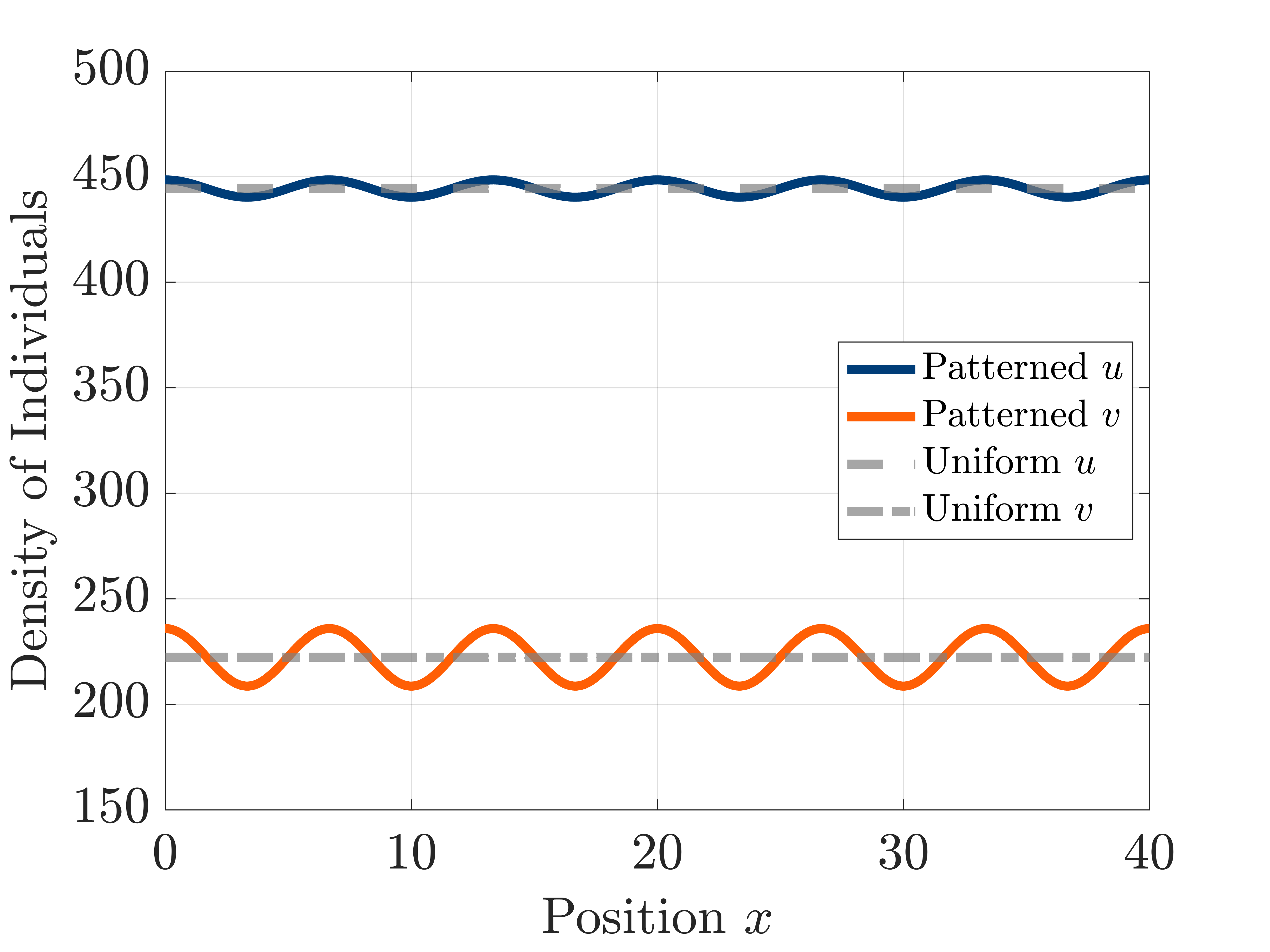}
        \label{fig:nonpayoff_uv_close}
    }\hspace{5mm}
    \subfloat[Payoff Dynamics for $p_H$ and $p_D$]{
        \includegraphics[width=0.45\textwidth]{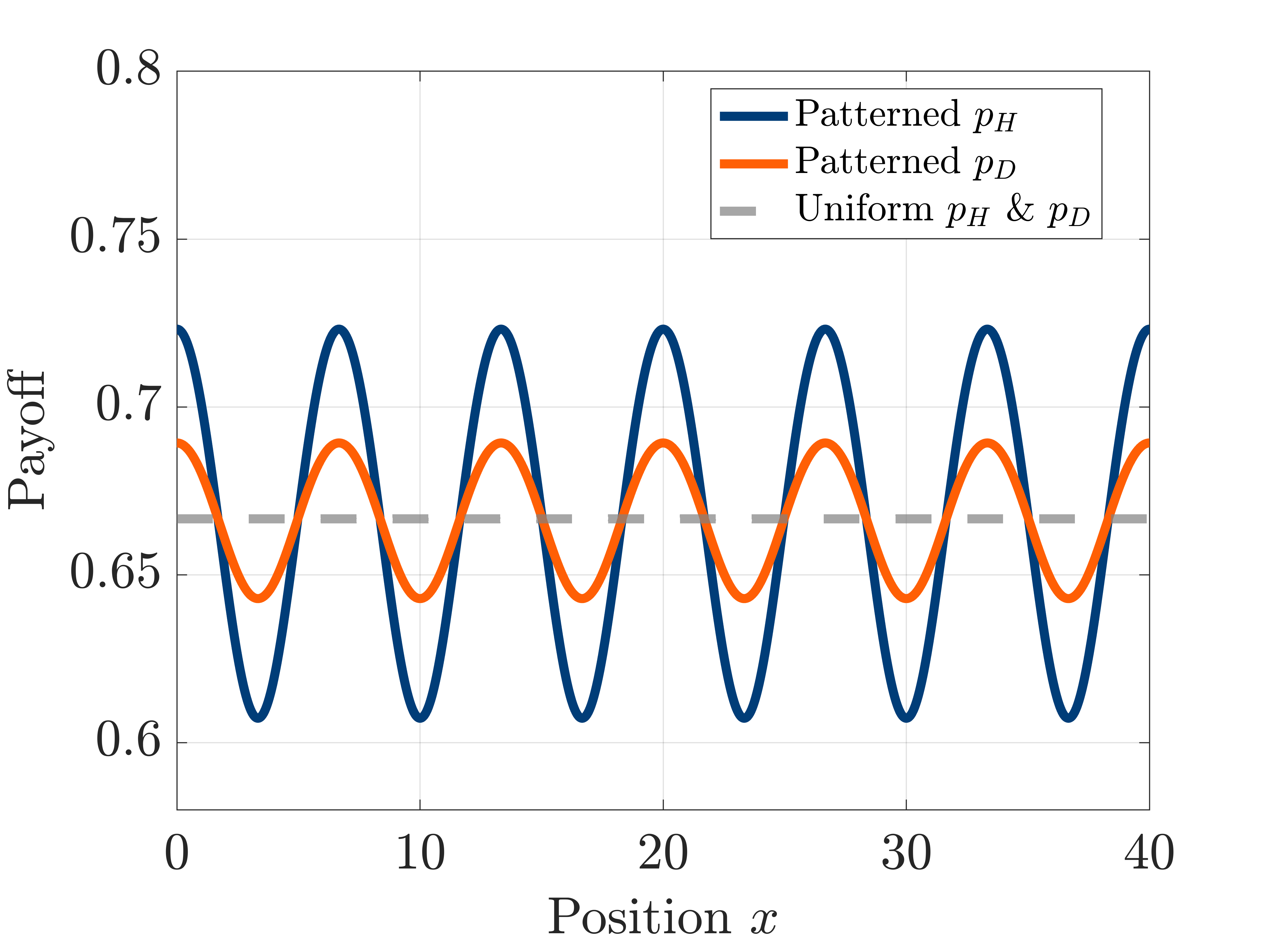}
        \label{fig:nonpayoff_pi_close}
    }
    \caption{Spatial patterns for hawks ($u$) and doves ($v$), and their associated payoffs, under pure diffusion with $D_u = 4.93 > D_u^*$ and $D_v=0.1$ after running numerical simulation until time $t = 10^6$. (a) Cosine-like spatial patterns in population densities. (b) Corresponding periodic patterns in payoffs. All other parameters are identical to those used in Figure~\ref{fig:nonpayoff_critical_threshold}. We note that the patterns in population density and payoff feature six full peaks, corresponding to a wavelength of $\lambda = \frac{l}{6} = \frac{40}{6}$ and a wavenumber $k = \frac{2 \pi}{\lambda} = \frac{3 \pi}{10}$. As we wrote the sinusoidal perturbations proportional to $\cos\left( \frac{\pi m x}{l} \right)$ with corresponding wavenumber $k = \frac{m \pi}{l} = \frac{m \pi}{40}$, we can see that the patterns observed in the numerical simulation correspond to the index $m = m^* = 12$, corresponding to the most unstable wavenumber observed in our linear stability analysis.}
    \label{fig:nonpayoff_pi_uv_close}
\end{figure}

To explore the long-term behavior of spatial patterns in the absence of payoff-driven motion, we perform numerical simulations of the PDE system using a hawk diffusivity value slightly above the critical threshold ($D_u = 4.93$), starting from an initial condition with a small perturbation from the uniform coexistence equilibrium. 
We run the simulation for a large number of time-steps and plot the resulting population densities and spatial profiles of payoff for each strategy are shown in Figure \ref{fig:nonpayoff_pi_uv_close}. We see that the resulting spatial patterns of population densities and payoffs both display sinusoidal shapes and that the distribution of hawks and doves and their respective payoffs are positively correlated, with individuals following each strategy achieving a higher payoff at spatial locations featuring higher population density of the two strategies. The six peaks seen in the spatial patterns also correspond to the most unstable wavenumber of $m^* = 12$ seen from the linear stability analysis in Figure \ref{fig:nonpayoff_critical_threshold}, suggesting that the linear stability analysis provides a helpful prediction of the emergent spatial patterns for diffusivities close to the threshold hawk diffusivity $D^*_u \approx 4.917$.

\begin{figure}[!ht]
    \centering
    \subfloat[$D_u = 10$]{
        \includegraphics[width=0.45\textwidth]{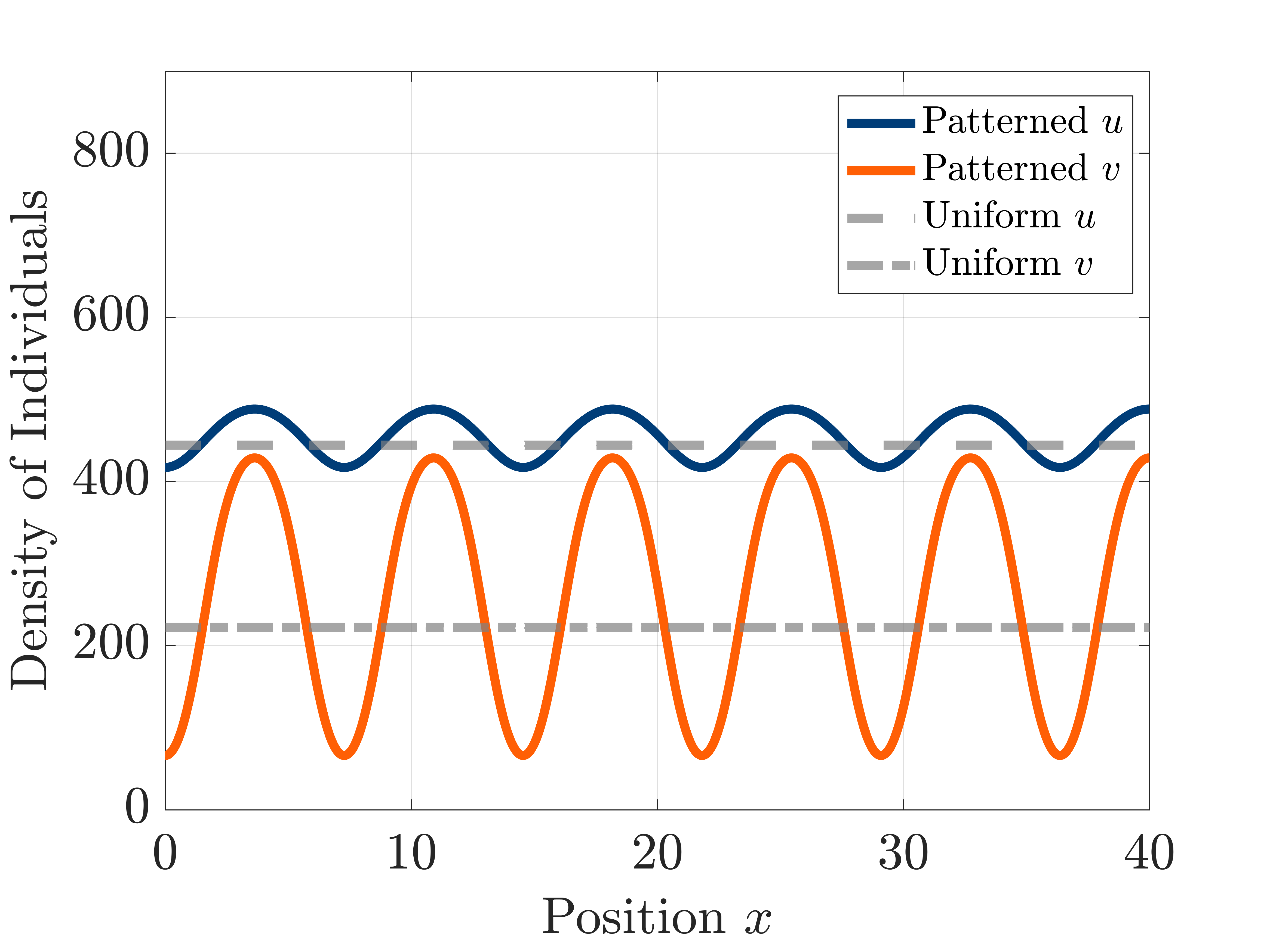}
        \label{fig:NP_uv_10}
    }\hspace{5 mm}
    \subfloat[$D_u = 19.55$]{
        \includegraphics[width=0.45\textwidth]{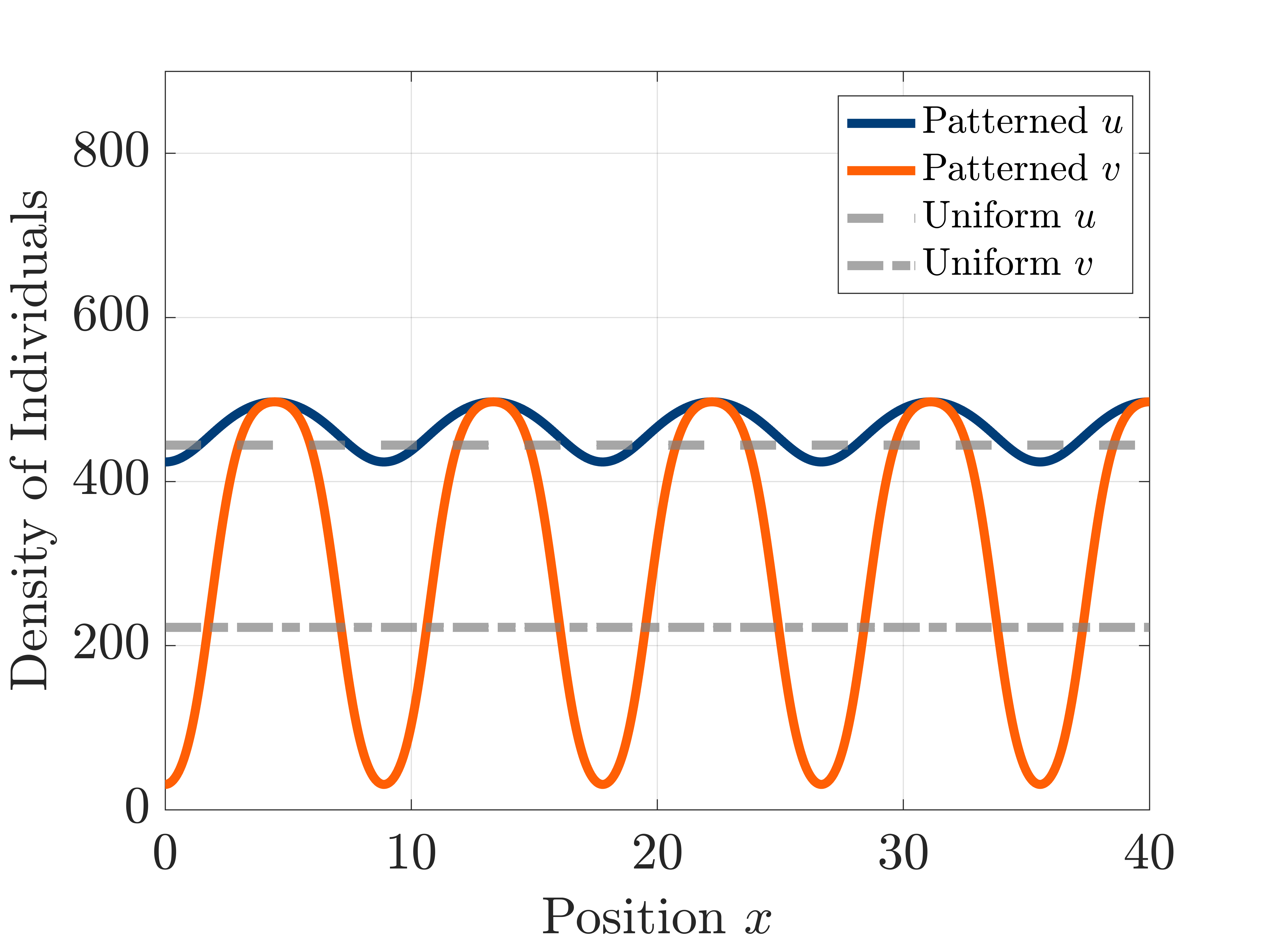}
        \label{fig:NP_uv_19.55}
    }
    \subfloat[$D_u = 40$]{
        \includegraphics[width=0.45\textwidth]{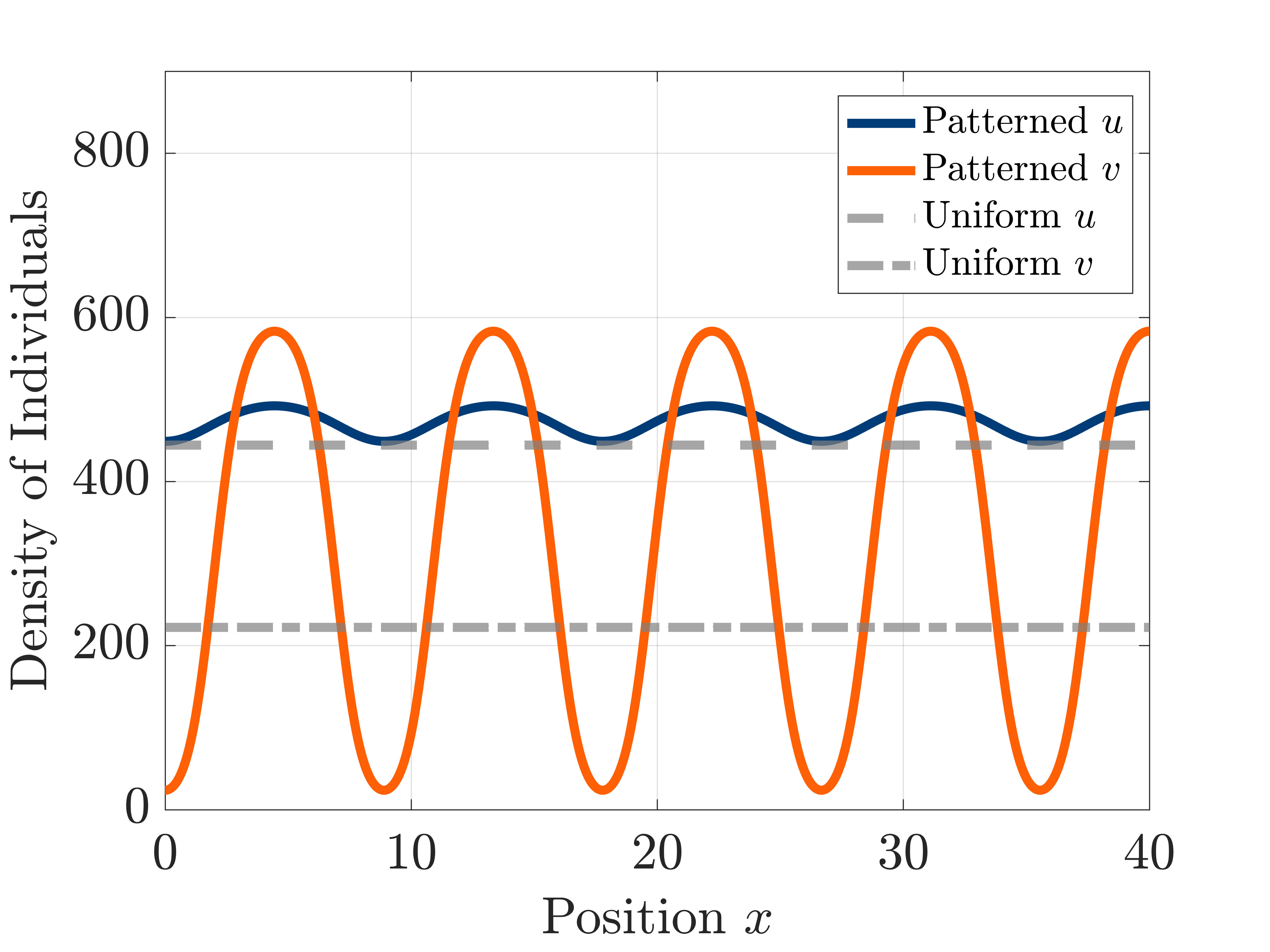}
        \label{fig:NP_uv_40}
    }
   \caption{Steady-state spatial patterns of hawk ($u$, orange curve) and dove ($v$, blue curve) densities under pure diffusion for increasing values of hawk diffusivity $D_u$. We show results for $D_u = 10$, $19.55$, and $40$, with $D_v = 0.1$ fixed. All other parameters are identical to those used in Figure~\ref{fig:nonpayoff_critical_threshold}.}
    \label{fig:NP_Du_series}
\end{figure}

To further examine the influence of hawk diffusivity on the steady-state spatial structure, we conduct additional simulations at progressively larger values of $D_u$ and present the resulting spatial profiles of hawks and doves in Figure~\ref{fig:NP_Du_series}. We see that the number of peaks in the pattern tends to decrease with increased $D_u$, and that the patterns tend to deviate more from the sinusoidal profile seen close to the threshold in Figure \ref{fig:nonpayoff_pi_uv_close}. In particular, we see that the distribution of doves becomes strongly concentrated upon regions of high population density for both strategies, while the hawks maintain a higher baseline population density across the domain and a less pronounced peak at the locations with high concentration of doves.

\begin{figure}[!ht]
    \centering
    \subfloat[Average population densities $\langle u \rangle$ and $\langle v \rangle$]{
        \includegraphics[width=0.45\textwidth]{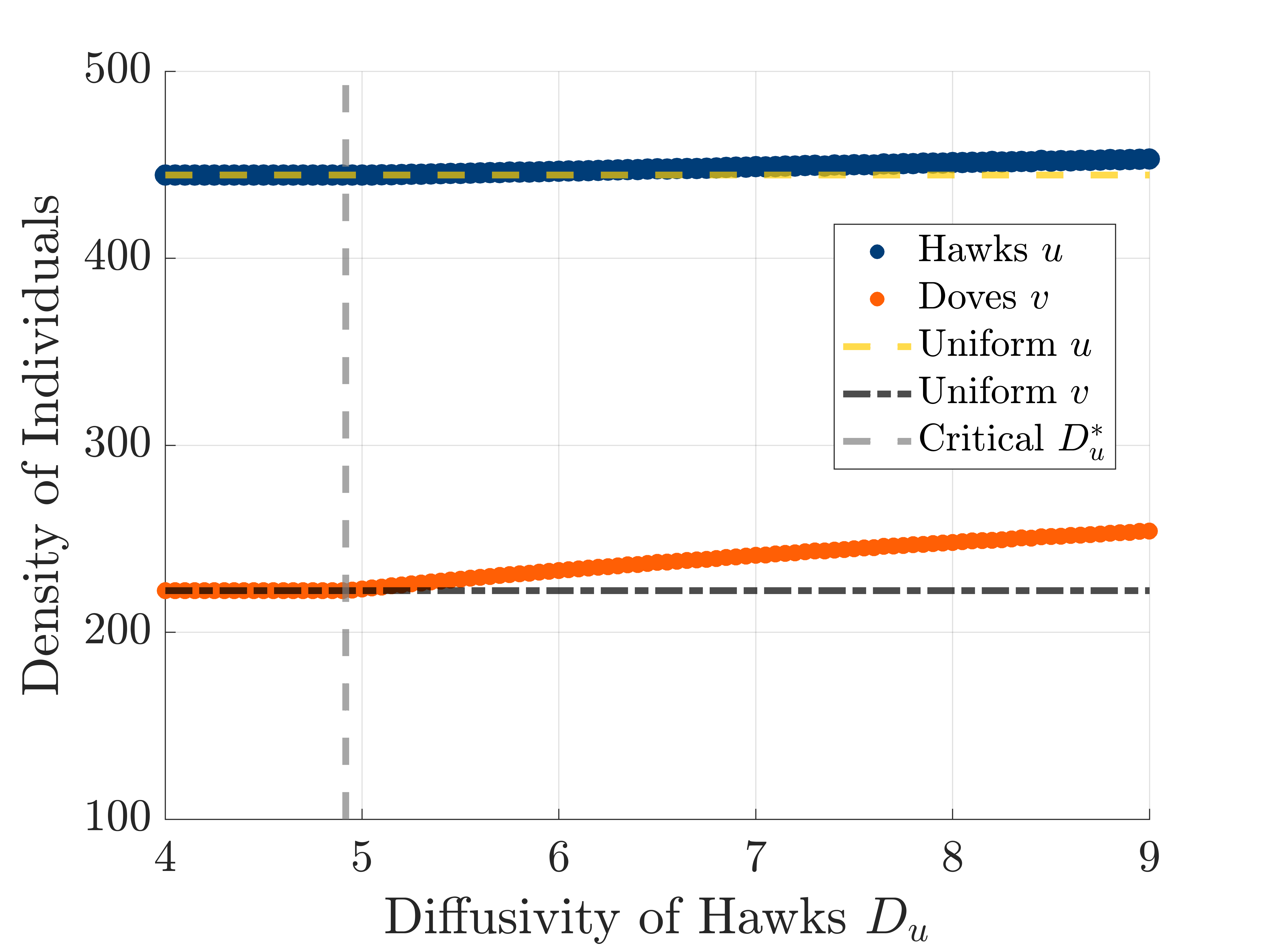}
        \label{fig:NP_uv_Du}
    }\hspace{5mm}
    \subfloat[Average payoffs $\langle p_H \rangle$ and $\langle p_D \rangle$]{
        \includegraphics[width=0.45\textwidth]{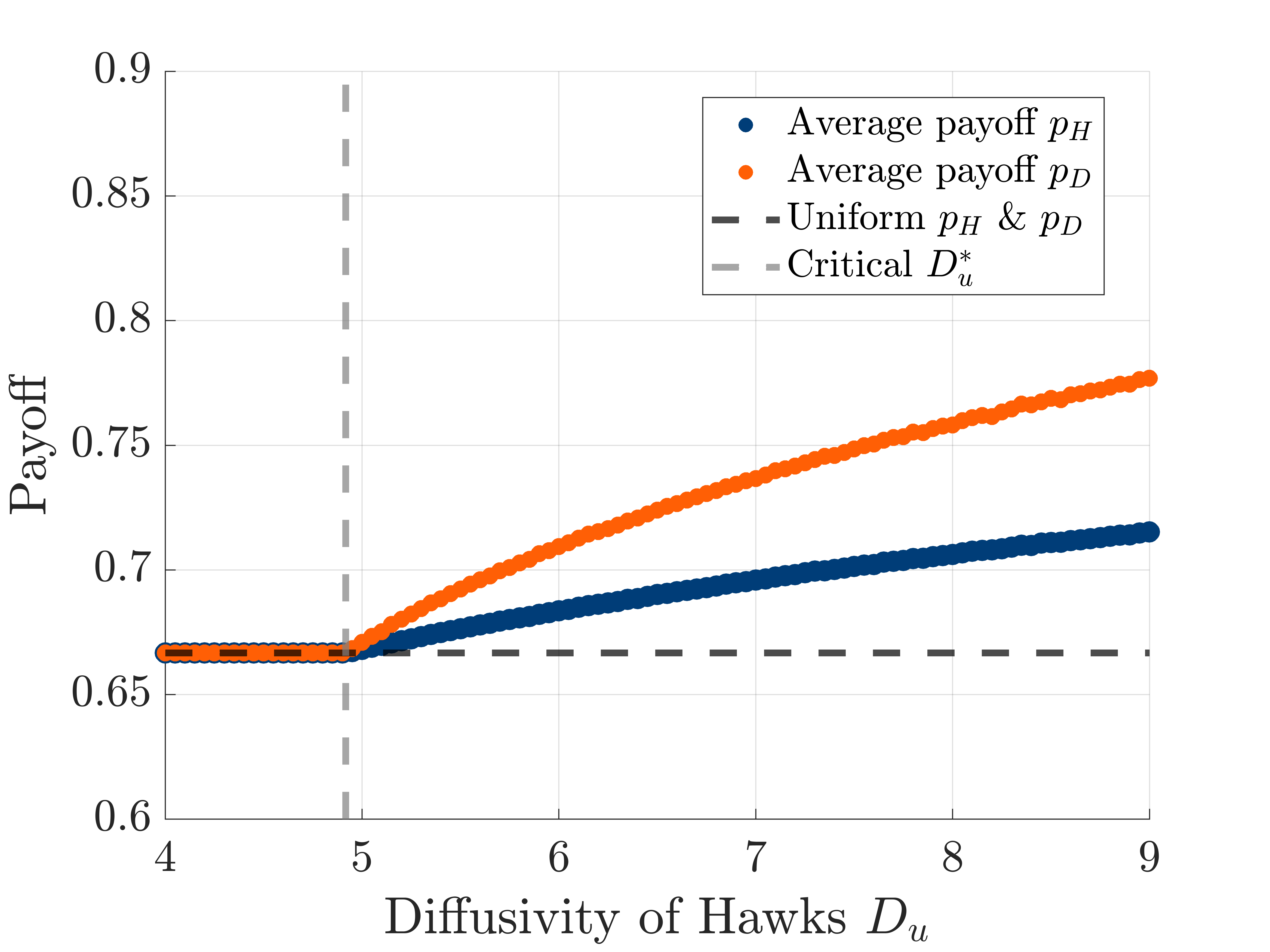}
        \label{fig:NP_pi_Du}
    }
    \caption{Effect of hawk diffusivity $D_u$ on spatially averaged population densities and payoffs. The vertical dashed line indicates the critical threshold $D_u^*$ for pattern formation. Averages are computed over 50 simulations with different initial conditions for each value of $D_u$, where $D_u$ is varied from $4$ to $9$ in increments of $0.05$. All other parameters are identical to those used in Figure~\ref{fig:nonpayoff_critical_threshold}. (a) Densities of hawks ($u$) and doves ($v$). (b) Corresponding average payoffs for hawks ($p_H$) and doves ($p_D$). Uniform equilibrium values are shown as horizontal dashed lines.
 We simulate the system using 50 different initial conditions for each value of $D_u$, generating different perturbabtions of the uniform state with small uniform noise. We calculate the spatial averages for each realization, and then compute the mean over all simulations. }
    \label{fig:NP_Du}
\end{figure}

Given the qualitative changes we have seen in the spatial patterns of hawks and doves as we increase the hawk diffusivity $D_u$, we would also like to quantify how changing hawk diffusivity can impact the collective outcome for each of the two strategies. We provide comparisons of the population and payoff of each strategy in Figure \ref{fig:NP_Du}, and choose to quantify aggregate success of hawks and doves by plotting the average density of each strategy
\begin{subequations}
\begin{align}
\langle u \rangle(t) &= \frac{1}{l} \int_0^l u(t,x) \, dx \\
\langle v \rangle(t) &= \frac{1}{l} \int_0^l v(t,x) \, dx
\end{align}
\end{subequations}
across the spatial domain, as well as the average payoff for each strategy

\begin{subequations}
\begin{align}
\langle p_H \rangle(t) &= \frac{\int_0^{l} p_H\left(u(t,x),v(t,x) \right) u(t,x) \, dx}{\int_0^l u(t,x) \, dx} \\
\langle p_D \rangle(t) &= \frac{\int_0^{l} p_D\left(u(t,x),v(t,x) \right) v(t,x) \, dx}{\int_0^l v(t,x) \, dx}
\end{align}
\end{subequations}
across the spatial domain. We also compare these collective outcomes achieved in the simulations of our reaction-diffusion model to the uniform equilibrium values achieved in the absence of spatial motion, which are given by $\langle p_H \rangle = \langle p_D \rangle = \frac{2}{3}$, $\langle u \rangle = \frac{4000}{9}$, 
and $\langle v \rangle =  \frac{2000}{9}$ 
 for our given hawk-dove game and strength of density-dependent regulation.

We find that as $D_u$ increases beyond the critical threshold $D_u^*$, the average population densities and payoffs for both hawks and doves increase. In particular, this behavior agrees with the outcome seen in the stochastic spatial model shown in Figure \ref{fig:w_0_muboth}, suggesting that the Turing patterning mechanism in the diffusive model helps to promote population and payoff for hawks and doves in both the individual-based and mean-field descriptions of spatial hawk-dove games with diffusive motion. Another point of agreement we see between the PDE and stochastic model is that the threshold diffusivity $D^*_u \approx 4.917$ for pattern formation is slightly less than fifty times the diffusivity $D_v = 0.1$, similar to the ratio of movement rates required to generate patterns in the stochastic spatial model. %

\subsection{Results for Payoff-Driven Motion with Equal Mobilities for Each Strategy}
\label{sec:PDE-fullypayoff-driven}

We now consider the possibility of spatial pattern formation in the presence of payoff-driven motion. We focus on the case of $D_u = D_v$ with equal diffusivities for the two strategies, looking to explore how the relative sensitivities to payoff $w_u$ and $w_v$ impact the possibility of forming spatial aggregations. 
In particular, we look to find conditions on the sensitivity $w_v$ of payoff-driven motion for doves to allow the formation of spatial patterns, and whether such patterns can result in a finite dominant wavenumber that characterizes the pattern near onset of instability.

In the presence of payoff-driven motion, the two possible conditions for the growth of perturbations with wavenumber $m$ are
\begin{equation} \label{eq:payofftracecondition}
\begin{aligned}
      \operatorname{tr}(A(m))&= \left[ - \left(D_u +  D_v \right) + 2 w_u c_1 u_0 + 2 w_v c_4 v_0 \right] \left( \frac{m \pi}{l} \right)^2 + a_1 + b_2 %
     \end{aligned}
\end{equation}
or
\begin{equation}
\begin{aligned}
    \det (A(m)) &= \left(a_1 - D_u \left[ 1 - 2 w_u c_1 u_0 \right] \left(\frac{m\pi}{l}\right)^2 \right)\left(b_2 - D_v \left[1 - 2 w_v c_4 v_0 \right] \left(\frac{m\pi}{l}\right)^2 \right) \\ &- \left(a_2 + 2D_uw_u c_2u_0  \left(\frac{m\pi}{l}\right)^2\right)\left(b_1 +  2D_vw_v c_3v_0 \left(\frac{m\pi}{l}\right)^2\right) < 0.
\end{aligned}
\end{equation}

For simplicity, we will consider the conditions for instability in the case of the exponential movement rules given by $f_u(w_u p_H) = e^{w_u p_H}$ and $f_v(w_v p_D) = e^{w_v p_D}$. For this movement rule and the hawk-dove game with resource $V$ and fighting cost $C$,  we see that the constants $c_1$, $c_2$, $c_3$, and $c_4$ take the form 
\begin{equation} \label{eq:parameters-c1c2c3c4}
    \begin{aligned}
        c_1 &= -\frac{(V+C)v_0}{2\left(u_0+v_0\right)^2} < 0 \\
        c_2 &= \frac{(V+C)u_0}{2\left(u_0+v_0\right)^2}  > 0 \\
        c_3 &= -\frac{Vv_0}{2\left(u_0+v_0\right)^2} < 0 \\
        c_4 &= \frac{Vu_0}{2\left(u_0+v_0\right)^2}  > 0,
    \end{aligned}
\end{equation}

Because we require that $a_1 + b_2 < 0$ to ensure stability of the Hawk-Dove coexistence equilibrium in the absence of spatial motion, we see from Equation \eqref{eq:payofftracecondition} that it will only be possible to obtain a positive trace provided that the coefficient of the quadratic term in Equation \eqref{eq:payofftracecondition} is positive. Using this observation and the expressions from Equation \eqref{eq:parameters-c1c2c3c4}, we see that a positive trace will only be possible if
\begin{equation}\label{eq:tracecondition}
 -D_u w_u u_0 \frac{(V+C)v_0}{\left(u_0+v_0\right)^2} + D_v w_v v_0 \frac{V u_0}{\left(u_0+v_0\right)^2} - D_u - D_v + a_1 + b_2 > 0.
\end{equation}
We can rearrange this inequality to see that a necessary condition for $\operatorname{tr}(A(m)) > 0$ is that the payoff sensitivity $w_v$ for doves satisfies
\begin{equation}\label{eq:inf-wavenum-1}
    w_v > w_v^{\textnormal{I}} := \frac{\left(u_0 + v_0\right)^2}{D_v u_0 v_0 V} \left[ D_u + D_v  + \frac{D_u w_u u_0 v_0 (V+C)}{\left(u_0 + v_0\right)^2} \right].
\end{equation}
In particular, we see from Equation \eqref{eq:payofftracecondition} that the $\mathrm{tr}(A(m))$ is a quadratic function of the wavenumber $m$, and therefore we see that the condition $w_v > w_v^{\textnormal{I}}$ would imply that the $\mathrm{tr}(A(m))$ would be positive for infinitely many wavenumbers $m$.

We can now look to analyze the possibility of pattern formation through the determinant becoming negative for some wavenumber $m$. We note that the determinant is given by the following quartic polynomial of the wavenumber $m$
\begin{equation}
\label{eq:detAmpayoff}
\det(A(m)) = \alpha \left(\frac{m\pi}{l}\right)^4 + \beta \left(\frac{m\pi}{l}\right)^2 + \gamma,
\end{equation}
where the coefficients $\alpha$, $\beta$, and $\gamma$ are given by
\begin{subequations}
\label{eq:Detcondition}
\begin{align} 
    \alpha &= \left(-D_u - \frac{D_u w_u u_0 v_0 (V+C)}{(u_0 + v_0)^2}\right)\left(-D_v + \frac{D_v w_v u_0 v_0 V}{(u_0 + v_0)^2}\right) \label{eq:Detconditionalpha} \\
    &\quad + \frac{D_u w_u u_0^2 (V+C)}{(u_0 + v_0)^2} \cdot \frac{D_v w_v v_0^2 V}{(u_0 + v_0)^2}  \nonumber \\
    \beta &= a_1 \left(-D_v + \frac{D_v w_v u_0 v_0 V}{(u_0 + v_0)^2}\right) + b_2 \left(-D_u - \frac{D_u w_u u_0 v_0 (V+C)}{(u_0 + v_0)^2}\right)  \label{eq:betaexpression}\\
    &\quad + \frac{a_2 D_v w_v v_0^2 V}{(u_0 + v_0)^2} - \frac{b_1 D_u w_u u_0^2 (V+C)}{(u_0 + v_0)^2}
   \nonumber \\
    \gamma &= a_1 b_2 - b_1 a_2.
\end{align}
\end{subequations}
In addition, we know that $\gamma > 0$ due to our assumption on the stability of the coexistence equilibrium under the reaction dynamics. 

From the form of this determinant, we see that there are two possible ways to achieve $\det(A(m)) < 0$ for a given wavenumber $m$: either we have that $\alpha < 0$ or we have that $\alpha > 0$ and $\beta$ is sufficiently negative to allow a band of unstable wavenumbers.    %
From Equation \eqref{eq:Detconditionalpha}, we see that $\alpha < 0$ provided that the payoff sensitivity of doves satisfies
\begin{align}
    w_v > w_v^{\textnormal{II}} := \frac{(u_0 + v_0)^2}{D_v u_0 v_0 V} \left( D_u + \frac{D_u w_u u_0 v_0 (V+C)}{(u_0 + v_0)^2} \right).
\end{align}
In this case, the quartic term of Equation \eqref{eq:detAmpayoff} has a negative coefficient, so we expect that $\det(A(m)) < 0$ for sufficiently large wavenumbers $m$ when $\alpha < 0$. This again produces infinitely many unstable wavenumbers, so we must turn to the case of $\alpha > 0$ to explore conditions in which biologically feasible spatial patterns can emerge due to the effects of payoff-driven motion. 
To explore the possibility of pattern formation with a finite dominant wavenumber, we look to find a positive root to the expression of $\det(A(m))$ from Equation \eqref{eq:detAmpayoff} when interpreted as a quadratic polynomial in the variable $q := \left(\frac{m \pi}{l}\right)^2$. For this polynomial, the roots are given by
\begin{equation}
\label{eq:qpmroots}
q_{\pm} := \frac{- \beta \pm \sqrt{\beta^2 - 4 \alpha \gamma}}{2 \alpha}.
\end{equation}
For the case in which $\alpha > 0$, we therefore see that there are two necessary conditions to have a positive root in the form of Equation \eqref{eq:qpmroots}:
\begin{enumerate}[(i)]
    \item $\beta < 0$
    \item $\beta^2 - 4 \alpha \gamma > 0$.
\end{enumerate}

In particular, the emergence of a finite wavenumber pattern can only occur if $w_v$ is sufficiently large for both of these conditions to hold. Using the expression for $\beta$ from Equation \eqref{eq:betaexpression}, we can see that $\beta < 0$ when $w_v$ satisfies the following condition
\begin{equation}
w_v > w_v^{\textnormal{III}a} := \frac{\left(u_0 + v_0 \right)^2}{\left(a_1 u_0 + a_2 v_0\right) D_v v_0 V} \left( a_1 D_v + b_2 D_u + \frac{\left(b_1 u_0 + b_2 v_0 \right) D_u u_0 \left( V+C\right) w_u}{\left(u_0 + v_0 \right)^2} \right).
\end{equation}
We can derive a similar lower bound $w_v^{\textnormal{III}b}$ to ensure that the condition $\beta^2 - 4 \alpha \gamma > 0$ holds, but we postpone the derivation of this threshold quantity to Section \ref{sec:secondcondtion} of the appendix. By combining these two threshold quantities, we see that a necessary condition for the emergence of a finite range of unstable wavenumbers provided that 
\begin{equation}
  w_v > w_v^{\textnormal{III}}:= \max\left(w_v^{\textnormal{IIIa}} , w_v^{\textnormal{IIIb}}\right).  
\end{equation}

\begin{figure}[!ht]
    \centering
    \includegraphics[width=0.65\linewidth]{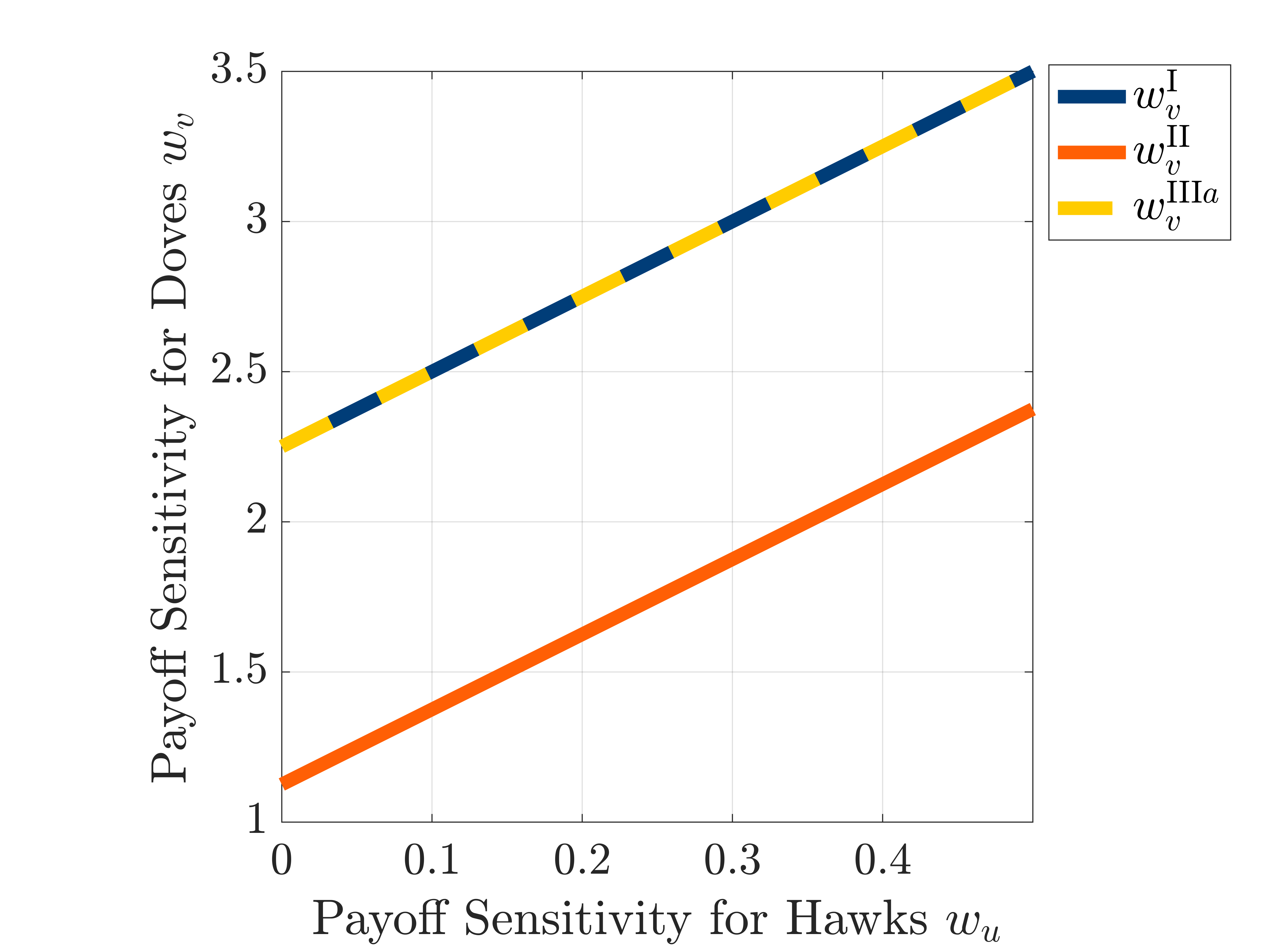}
   \caption{Critical thresholds $w_v^{\textnormal{I}}$ (blue solid line), $w_v^{\textnormal{II}}$ (orange solid line), and $w_v^{\textnormal{III}}$ (maize dashed line) for the emergence of spatial patterns, plotted as functions of the payoff-driven weight for hawks $w_u$. The diffusivities are set to $D_u = D_v = 0.1$, and $w_u$ is varied from $0$ to $0.5$. All other parameters are identical to those used in Figure~\ref{fig:nonpayoff_critical_threshold}. The complete overlap of $w_v^{\textnormal{II}}$ and $w_v^{\textnormal{III}}$ suggests that the pattern-forming instability emerges simultaneously for infinitely many wavenumbers.}
    \label{fig:P_wv_bounds_purePD}
\end{figure}

At this point, we can compare three of the thresholds on $w_v$ that we have found to see whether finite wavenumber patterns are possible by payoff-driven motion for the case of equal diffusivities for the hawk-dove game with the $C$-$V$ payoff matrix. For this class of games, we can apply the expressions from Equation \eqref{eq:parameters-c1c2c3c4} to the threshold quantities $  w_v^{\textnormal{I}}$, $  w_v^{\textnormal{II}}$, and $  w_v^{\textnormal{IIIa}}$ to see that

\begin{subequations}
\begin{align}
    w_v^{\textnormal{I}} &= 
\frac{C^2 \left(D_u + D_v\right) + D_u V w_u \left(C - V\right)\left(C + V\right)}
{D_v V^2 \left(C - V\right)} \\ 
    w_v^{\textnormal{II}} &= 
\frac{D_u \left[ C^2 + V w_u \left(C - V\right)\left(C + V\right) \right]}
{D_v V^2 \left(C - V\right)} \\
    w_v^{\textnormal{III}a} &= 
\frac{C \left( D_u \left(C - 2V\right) + D_v \left(C + 2V\right) \right)
+ D_u V w_u \left(C - V\right)\left(C + V\right)}
{D_v V^2 \left(C - V\right)}.
\end{align}
\end{subequations}
For the case of equal diffusivities $D_u = D_v$, we see that the threshold quantities satisfy $w_v^{\textnormal{IIIa}} =  w_v^{\textnormal{I}}$ and $w_v^{\textnormal{IIIa}} > w_v^{\textnormal{II}}$, and we illustrate the behavior of these three threshold quantities as functions of hawk payoff sensitivity $w_u$ in Figure \ref{fig:P_wv_bounds_purePD}. Because $w_v > w_v^{\textnormal{IIIa}}$ is a necessary condition to obtain finitely many unstable wavenumbers and $w > w_v^{\textnormal{II}}$ is a sufficient condition for having infinitely many unstable wavenumbers, we see that it is not possible to achieve a finite wavenumber pattern for the $C$-$V$ Hawk-Dove game and equal diffusivity of the two strategies.

However, we can also use the threshold quantity $w_v^{\textnormal{IIIb}}$ derived in Section \ref{sec:secondcondtion} to explore whether it is possible to achieve finite wavenumber patterns if $D_u \ne D_v$. %
To do this, we must find a senstitivity of payoff-driven motion satisfying
\begin{equation}
w_v^{\textnormal{III}} < w_v < \min\left( w_v^{\textnormal{I}},  w_v^{\textnormal{II}} \right).
\end{equation}
We show in Section \ref{sec:mixedeffectsresults} that it is possible to find such a value of $w_v$ provided that $D_u > D_v$, allowing us to combine the effects of differential diffusivity and different sensitivites to payoff for hawks and doves.

\section{Comparing the Stochastic and PDE Models with Mixed Effect of Diffusion and Directed Motion}
\label{sec:mixedeffectsresults}

As we saw in Section \ref{sec:PDEResults}  that the PDE model produces an infinite wavenumber instability in the case of payoff-driven motion with equal diffusivity of hawks and doves, we have not yet been able to get much of a quantitative or qualitative understanding of the effects of payoff-driven motion on the collective behavior of the population in the large population limit. To remedy this, we look to explore how a mix of different movement rates ($\mu_u \ne \mu_v$ or $D_u \ne D_v$) and different payoff sensitivities ($w_u \ne w_v$) can help to produce spatial patterns and impact the collective outcomes of each strategy. In particular, we consider how a mix of increased diffusivity of the hawks (with $\mu_u > \mu_v$ or $D_u > D_v$) and increased payoff sensitivity of doves ($w_v > w_u$) can contribute to the formation of spatial patterns in both the PDE model. We first illustrate how the combination of these two mechanisms can promote spatial pattern formation with a finite wavenumber in Section \ref{sec:MixDiscussions}, and then we provide a comparison between the collective outcomes achieved under these mixed effects in the PDE and stochastic spatial models in Section \ref{sec:MixResults}. 
\subsection{PDE Results for Pattern Formation under Mixed Diffusion and Payoff Effects}
\label{sec:MixDiscussions}

\begin{figure}[!ht]
    \centering
    \subfloat[Critical thresholds of $w_v$]{
        \includegraphics[width=0.45\textwidth]{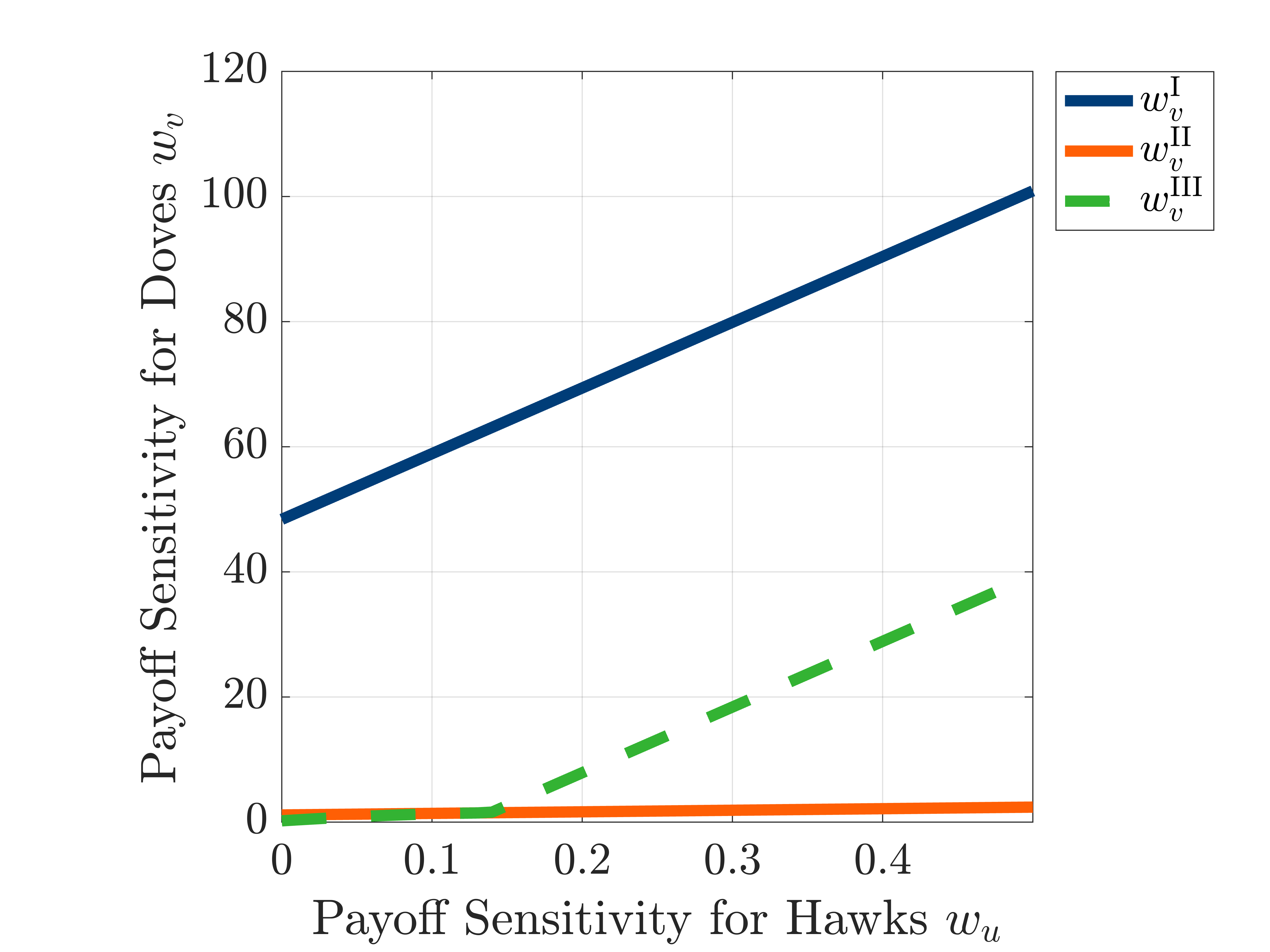}
        \label{fig:P_wv_bounds_4.2}
    }\hspace{5mm}
    \subfloat[Zoomed-in view]{
        \includegraphics[width=0.45\textwidth]{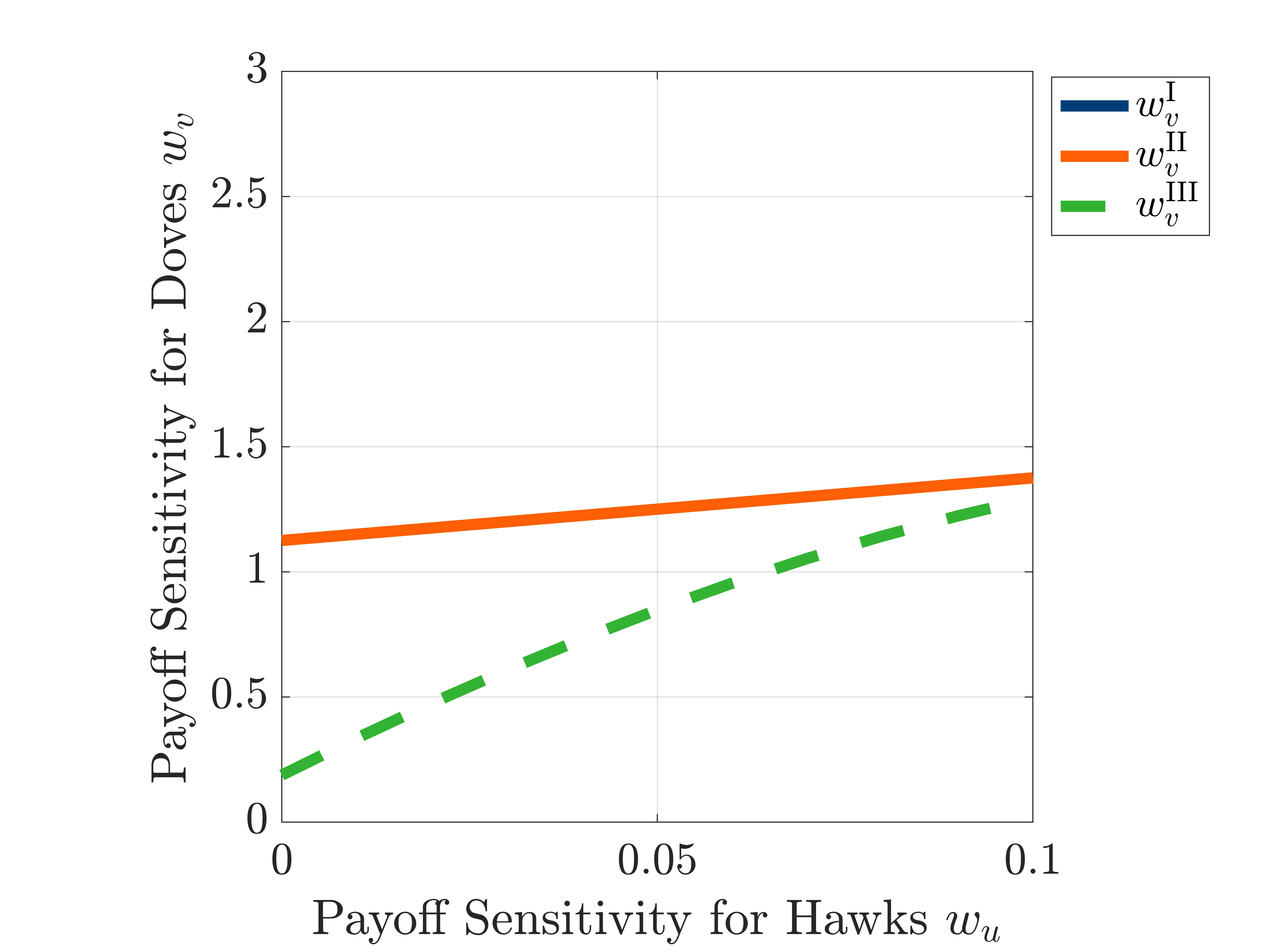}
        \label{fig:P_wv_bounds_4.2_zoomin}
    }
    \caption{Critical thresholds $w_v^{\textnormal{I}}$ (blue solid line), $w_v^{\textnormal{II}}$ (orange solid line), and $w_v^{\textnormal{III}}$ (green dashed line) as functions of the payoff-driven weight for hawks $w_u$, plotted for the case where $D_u = 4.2 < D_u^*$ and $D_v = 0.1$. All other parameters are identical to those in Figure~\ref{fig:nonpayoff_critical_threshold}. Panel (a) shows the full range of thresholds, while panel (b) provides a zoomed-in view of the region where $w_v^{\textnormal{II}}$ and $w_v^{\textnormal{III}}$ slightly diverge, indicating the presence of a finite-wavenumber instability.}
    \label{fig:P_wv_bounds_4.2_combined}
\end{figure}

We now look to explore the possibility of spatial pattern formation in the PDE model for parameter regimes for which it is not possible to destabilize the uniform state due to the effects of either differences in diffusivity or differences in payoff sensitivity on their own. In particular, we recall that we found the critical threshold for dove diffusivity $D_u^* = 4.917$ using Equation~\ref{eq:criticalDu} for the reaction-diffusion model and our choice of game-theoretic and ecology parameters. If we fix the dove diffusivity at $D_v = 0.1$ and $D_u = 4.2$, we cannot expect the emergence of spatial patterns by the Turing mechanism alone, 
so we look to numerically compute the three critical thresholds $w_v^{\textnormal{I}}$, $w_v^{\textnormal{II}}$, and $w_v^{\textnormal{III}}$ as functions of $w_u$ in Figure~\ref{fig:P_wv_bounds_4.2_combined} required to achieve a spatial pattern for sufficiently strong payoff sensitivity $w_v$ of doves. As shown in Figure~\ref{fig:P_wv_bounds_4.2}, we observe a small gap between $w_v^{\textnormal{II}}$ (blue solid line) and $w_v^{\textnormal{III}}$ (yellow dashed line), indicating the possibility achieving spatial patterns with a finite dominant wavenumber for these values of $w_u$ and $w_v$. Notably, the range of dove payoff sensitivities $w_v$ that allow finite wavenumber patterns decreases with the payoff sensitivity of hawks $w_u$, so the possibility of achieving physically meaningful spatial patterns due to payoff-driven motion is most feasible when hawks move faster than doves (with $D_u > D_v$) and doves are more capable of moving towards locations with greater payoff (with $w_v > w_u$).  %

\subsection{Exploration of PDE and Stochastic Models for Different Mobilities and Payoff Sensitivities}
\label{sec:MixResults}

\begin{figure}[!ht]
    \centering
    \subfloat[Average population densities $\langle u \rangle$ and $\langle v \rangle$]{
        \includegraphics[width=0.45\textwidth]{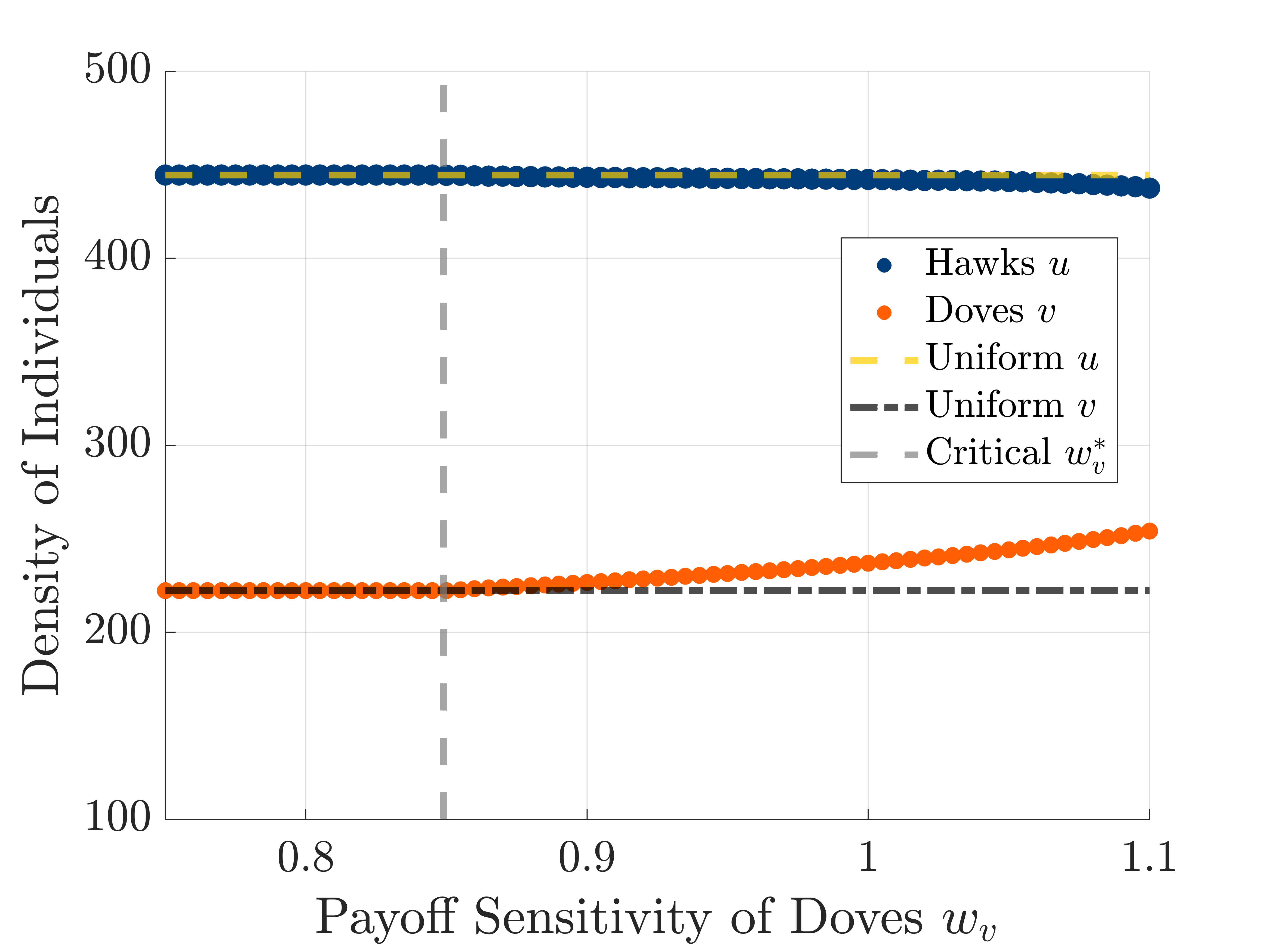}
        \label{fig:P_uv_wv}
    }\hspace{5mm}
    \subfloat[Average payoffs $\langle p_H \rangle$ and $\langle p_D \rangle$]{
        \includegraphics[width=0.45\textwidth]{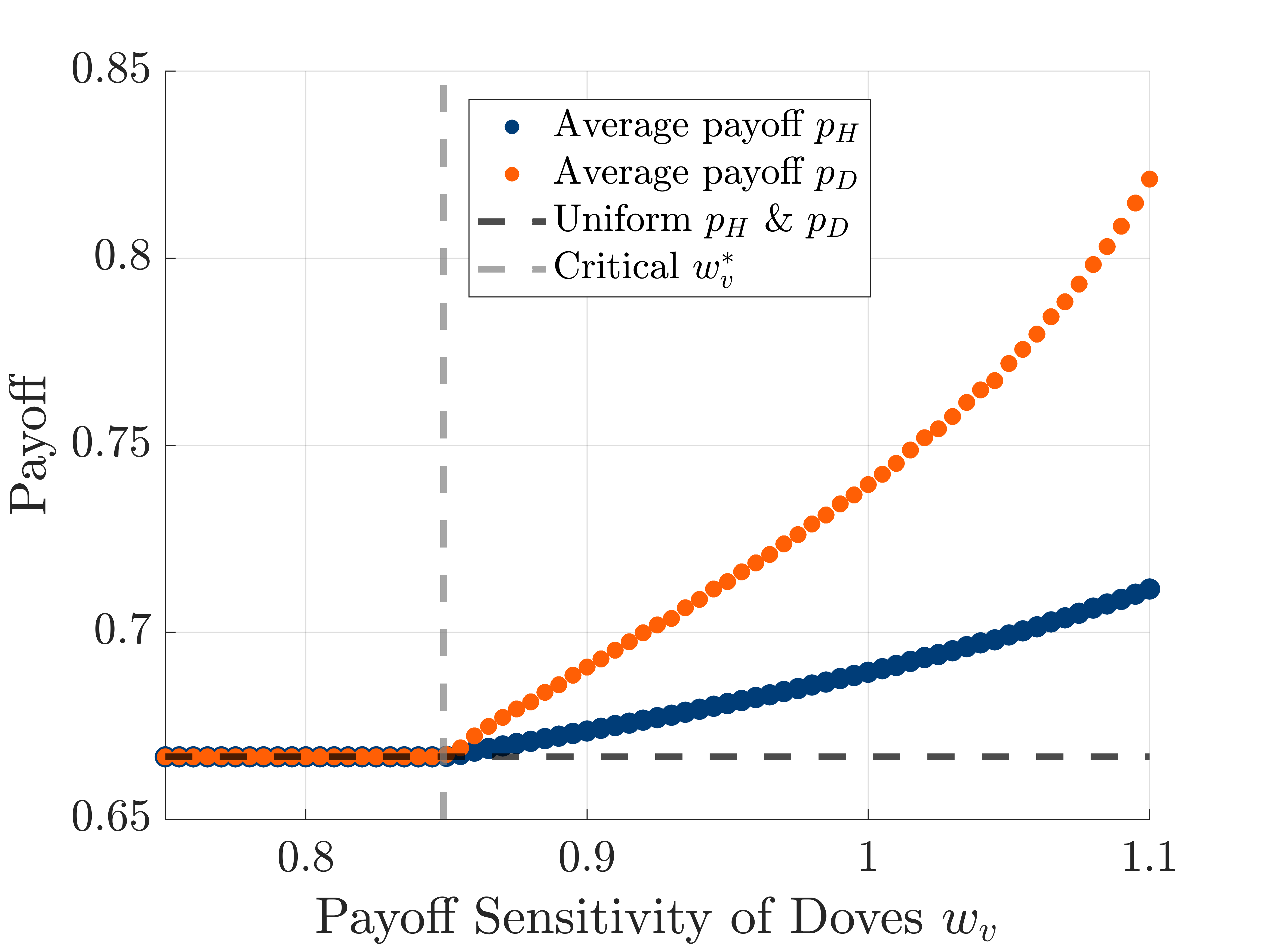}
        \label{fig:P_pi_wv}
    }
    \caption{Effect of the payoff sensitivity of doves $w_v$ on spatially averaged population densities and payoffs under mixed diffusion and payoff-driven motion. We fix $w_u = 0.05$ and vary $w_v$ from $0.75$ to $1.1$ in increments of $0.005$. The vertical dashed line marks the critical threshold $w_v^* = 0.848$ for pattern formation. (a) Average densities of hawks ($u$) and doves ($v$). (b) Corresponding average payoffs of hawks ($p_H$) and doves ($p_D$). All other parameters are identical to those used in Figure~\ref{fig:nonpayoff_critical_threshold}.}
    \label{fig:NP_wv}
\end{figure}

Now that we have seen that the interaction of differing diffusivities and differing payoff sensitivities can promote finite wavenumber patterns, we can look to explore how increasing the dove payoff sensitivity can impact spatial patterns in both the PDE and stochastic model. We start by presenting in Figure \ref{fig:NP_wv} the average population densities and average payoffs achieved by each strategy as we increase the payoff sensitivity $w_v$ of doves. We find that the number of hawks decreases as we increase $w_v$ above the pattern formation threshold, while the number of doves increases across the spatial domain. The average payoff increases with dove payoff sensitivity for each strategy for $w_v > w_v^*$, but we see that dove achieves a higher average payoff than hawks over the spatial domain in the pattern-forming regime.

\begin{figure}[H]
\vspace*{\fill}
\centering
\makebox[\textwidth][c]{
        \includegraphics[width=1.0\textwidth]{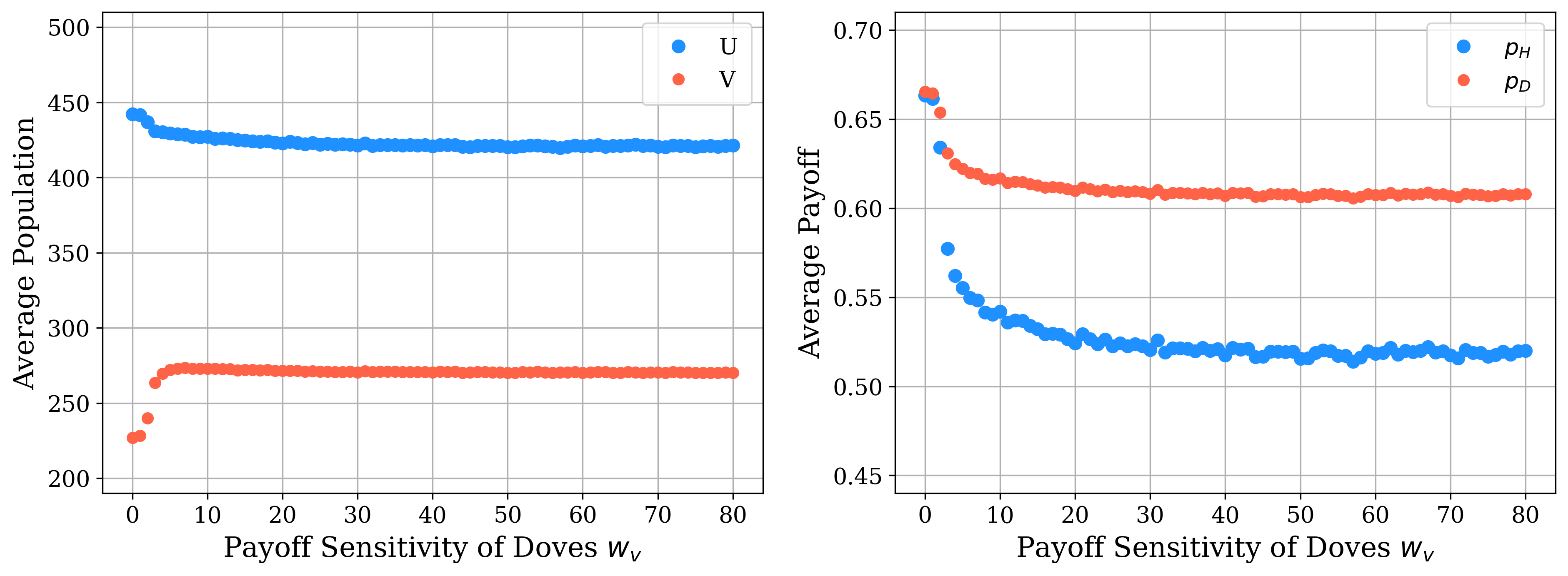}
    }
\caption{Average population (left) and payoff (right) versus $w_v$ in the stochastic model, using $\mu_u=4.8, \mu_v=0.1, w_u=0.1$, and $w_v \in [0, 80]$ with step size 1. We repeat each simulation 100 times and plot with the average data.}
\label{fig:mix_stochastic}
\end{figure}

We also perform analogous simulations for the stochastic spatial model in Figure \ref{fig:mix_stochastic}. We find that the average population sizes for the hawks and doves follow a similar trend to the PDE model in this case, with the number of doves increasing and the number of hawks decrease as the payoff sensitivity $w_v$ of doves is increased. However, we see a contrast with the results from the PDE simulation for the average payoffs achieved by each strategy, as both the hawks and doves experience decreasing average payoff across the spatial domain with increasing payoff sensitivity for doves. As with the case of the PDE model, we see that the average payoffs of the doves exceeds the payoff achieved by hawks in this parameter regime. In particular, we see that the simulations of our PDE and stochastic model agree overall on the relative collective success of hawks and doves across the spatial domain, but the two modeling frameworks differ regarding whether payoff-driven motion can help to promote the average outcome relative to the payoff that would be achieved in the absence of spatial motion.

\section{Extension of the PDE Model Incorporating nonlocal Payoff-Driven Motion}
\label{sec:nonlocalPDEmodels}

As we have seen that the PDE model of payoff-driven motion displays a short-wave instability in the case of equal diffusivities for the two strategies, we now look to consider modifications of our model of payoff-driven motion to allow the possibility of pattern formation with a finite characteristic wavenumber. To do this, we we consider an extension of our model for payoff-driven motion in which individuals consider the average payoff that their strategy achieves over a small interval of width $2 \rho$ around their current location. We assume that individuals consider the  gradient of their strategy's payoff relative to this spatially averaged payoff, and we can model the spatial distribution of hawks $u(t,x)$ and doves $v(t,x)$ using the following system of nonlocal PDEs:%
\begin{subequations}
\begin{align}\label{eq:PDE_nonlocal_integral}
    \frac{\partial u(t,x)}{\partial t} &= D_u \frac{\partial^2 u(t,x)}{\partial x^2}  + u \left(p_H(u(t,x),v(t,x)) - \kappa \left(u(t,x)+v(t,x)\right)\right) \\
    &- 2D_u \dsdel{}{x} \left( u(t,x)  \dsdel{}{x}  \int_{x - \rho}^{x + \rho} \frac{1}{2 \rho} \log\left[ f_u \left(w_u p_H \left(u(t,y),v(t,y) \right) \right) \right] dy\right) \nonumber \\
    \frac{\partial v(t,x)}{\partial t} &= D_v \frac{\partial^2 v(t,x)}{\partial x^2}  + v \left(p_D(u(t,x),v(t,x)) - \kappa \left(u(t,x)+v(t,x)\right)\right)    \\
    &- - 2D_v \dsdel{}{x} \left( v(t,x)  \dsdel{}{x}  \int_{x - \rho}^{x + \rho} \frac{1}{2 \rho} \log\left[ f_v \left(w_v p_D \left(u(t,y),v(t,y) \right) \right) \right] dy\right)  . \nonumber
\end{align}
\end{subequations}
We can also evaluate the inner derivative in this system of PDEs to remove the integral term, resulting in the follow system of PDEs for the nonlocal evaluation of the payoff gradient: 
\begin{subequations}
\begin{align}\label{eq:PDE_nonlocal_before}
    \frac{\partial u(t,x)}{\partial t} &= D_u \frac{\partial^2 u(t,x)}{\partial x^2}  + u \left(p_H(u(t,x),v(t,x)) - \kappa \left(u(t,x)+v(t,x)\right)\right) \\
    &- 2D_u \frac{\partial}{\partial x} \left( u(t,x) \left[ \frac{1}{2 \rho} \left( \log(f_u(u(t,x+\rho), v(t,x+\rho))) - \log(f_u(u(t,x-\rho), v(t,x-\rho))) \right) \right] \right)  \nonumber \\
    \frac{\partial v(t,x)}{\partial t} &= D_v \frac{\partial^2 v(t,x)}{\partial x^2}  + v \left(p_D(u(t,x),v(t,x)) - \kappa \left(u(t,x)+v(t,x)\right)\right)    \\
    &- 2D_v \frac{\partial}{\partial x} \left( v(t,x) \left[ \frac{1}{2 \rho} \left( \log(f_v(u(t,x+\rho), v(t,x+\rho))) - \log(f_v(u(t,x-\rho), v(t,x-\rho))) \right) \right] \right). \nonumber
\end{align}
\end{subequations}
Due to the dependence of the change in the hawk and dove densities at location $x$ depending on the payoffs of individuals at locations $x - \rho$ and $x + \rho$, it is natural consider our system of PDEs on a one-dimensional interval with $x \in [0,L]$ with periodic boundary conditions. 

This choice of nonlocal sensing is motivated by nonlocal models for chemotaxis, which biologically allow for the possibility of finite precision of sampling the environmental and mathematically helps to smooth out the feedback between payoff-driven motion and climbing the gradient of a strategy's own population density \cite{hillen2007global,hillen2009user,buttenschon2018space}. We will show in Section \ref{sec:LSAnonlocal} that this version of nonlocal payoff evaluation can allow for spatial instability of the uniform state, and that such patterns can only be achieved by finitely many wavenumbers. 

\subsection{Linear stability analysis}
\label{sec:LSAnonlocal}
As with our local PDE model, we will consider the stability of the uniform coexistence equilibrium $(u_0,v_0)$ in the Hawk-Dove game by linearizating our nonlocal PDE model around $(u_0,v_0)$. To do this, we consider solutions $u(t,x)$ and $v(t,x)$ taking the form
\begin{subequations}
\begin{align}
    u(t,x) &= u_0 + \epsilon u_1(t,x), \\
    v(t,x) &= v_0 + \epsilon v_1(t,x).
\end{align}
\end{subequations}
for a small parameter $\epsilon$.Substituting  these expressions for $u(t,x)$ and $v(t,x)$ into the nonlinear system of PDEs for Equation \eqref{eq:PDE_nonlocal_before}, we can neglect the nonlinear terms of order $\epsilon^2$ and use the fact that $(u_0,v_0)$ is an equilibrium of the ODE system to obtain the following linearization of our PDE model:
\begin{subequations}
\begin{align}
      \frac{\partial u_1(t,x)}{\partial t} &= D_u \frac{\partial^2 u_1(t,x)}{\partial x^2}  +a_1 u_1(t,x) + a_2 v_1(t,x) \\
    &- \frac{D_uw_u u_0}{ \rho} \frac{\partial}{\partial x} \left( c_1(u_1(t,x+\rho)-u_1(t,x-\rho)) + c_2(v_1(t,x+\rho)-v_1(t,x-\rho))\right), \\
    \frac{\partial v_1(t,x)}{\partial t} &= D_v \frac{\partial^2 v_1(t,x)}{\partial x^2}  + b_1 u_1(t,x) + b_2 v_1(t,x) \\
    &- \frac{D_vw_v v_0}{ \rho} \frac{\partial}{\partial x} \left( c_3(u_1(t,x+\rho)-u_1(t,x-\rho)) + c_4(v_1(t,x+\rho)-v_1(t,x-\rho))\right),
\end{align}
\end{subequations}
where $a_1$, $a_2$, $b_1$, $b_2$ are the entries of the Jacobian matrix for the reaction terms and $c_1$, $c_2$, $c_3$, and $c_4$ are the constants from Equation \eqref{eq:parameters-c1c2c3c4} describing the partial derivatives for the weight functions $f_u(\cdot)$ and $f_v(\cdot)$ for payoff-driven motion evaluated at the uniform coexistence state. We present the full details for the derivation of  Section \ref{sec:nonlocal-linearization} of the appendix.

To assess the stability of the equilibrium, we consider the following ansatz for solutions to the linearized system with wavenumber $2m$:
\begin{align}
    \begin{split}
        u_1 (t, x) &= \hat{u}_1e^{\sigma_m t} \cos\left(\frac{2 m \pi x}{l}\right),\\
        v_1 (t, x) &= \hat{v}_1 e^{\sigma_m t} \cos\left(\frac{2 m \pi x}{l}\right).
    \end{split}
\end{align}
where $\hat{u}_1$ and $\hat{v}_1$ are constants and and $\sigma$ is the growth rate of the pattern with wavenumber $2m$, and we choose this form of the sinusoidal perturbation to satisfy the periodic boundary conditions for our nonlocal PDE model. To write our solutions in terms of this ansatz, we first substitute these expressions into the term payoff-driven motion term and apply the sum and difference identities for sine and cosine to see that
\begin{align}
\begin{split}
     &-\frac{D_uw_u u_0}{ \rho} \frac{\partial}{\partial x} \left( c_1(u_1(t,x+\rho)-u_1(t,x-\rho)) + c_2(v_1(t,x+\rho)-v_1(t,x-\rho))\right) \\
     &=-\frac{D_uw_u u_0}{\rho} \frac{\partial}{\partial x}\left(\left(c_1 \hat{u_1}e^{\sigma_m t}+c_2 \hat{v_1}e^{\sigma_m t}\right)\left[\cos\left(\frac{2 m\pi\left(x+\rho\right)}{l}\right)-\cos\left(\frac{2 m\pi\left(x-\rho\right)}{l}\right)\right]\right) \\
     &=2D_uw_u u_0\left(\rho^{-1}\sin\left(\frac{2 m \pi\rho}{l}\right)\right)\left(\left(c_1 \hat{u_1}e^{\sigma_m t}+c_2 \hat{v_1}e^{\sigma_m t}\right)\frac{\partial }{\partial x}\sin\left(\frac{2 m \pi x}{l}\right)\right)\\
     &=2D_uw_u u_0 \left(\frac{2 m \pi }{l}\right)\left(\rho^{-1}\sin\left(\frac{2 m \pi\rho}{l}\right)\right)\left(c_1u_1(t,x)+c_2v_1(t,x)\right).
\end{split}
\end{align}
Similarly, for the payoff-driven motion term for $v_1$, we see that
\begin{align}
\begin{split}
     &- \frac{D_vw_v v_0}{ \rho} \frac{\partial}{\partial x} \left( c_3(u_1(t,x+\rho)-u_1(t,x-\rho)) + c_4(v_1(t,x+\rho)-v_1(t,x-\rho))\right)\\
     &=2D_vw_v v_0\frac{m \pi }{l}\left(\rho^{-1}\sin\left(\frac{2 m \pi\rho}{l}\right)\right)\left(c_3u_1(t,x)+c_4v_1(t,x)\right).
\end{split}
\end{align}
Using these expressions, we can see rewrite our system of linear PDEs in the following form%
\begin{align}
\begin{split}
    \frac{\partial u_1(t,x)}{\partial t} &= D_u \frac{\partial^2 u_1(t,x)}{\partial x^2}  + a_1u_1(t,x)+ a_2v_1(t,x) \\
    &+ 2D_uw_u u_0\frac{m \pi }{l}\left(\rho^{-1}\sin\left(\frac{2 m \pi\rho}{l}\right)\right)\left(c_1u_1(t,x)+c_2v_1(t,x)\right), \\
    \frac{\partial v_1(t,x)}{\partial t} &= D_v \frac{\partial^2 v_1(t,x)}{\partial x^2}  +b_1u_1(t,x)+ b_2v_1(t,x) \\
    &+2D_vw_v v_0\frac{m \pi }{l}\left(\rho^{-1}\sin\left(\frac{2 m \pi\rho}{l}\right)\right)\left(c_3u_1(t,x)+c_4v_1(t,x)\right).
\end{split}
\end{align}
Using our ansatz for the expressions for $u_1(t,x)$ and $v_1(t,x)$ to see that our solutions will take this form provided that the the growth rate $\sigma_m$ is a solution to the following eigenvalue problem
\begin{align}
\sigma_m \begin{pmatrix}
    \hat{u_1} \\
    \hat{v_1}
\end{pmatrix}
=B(m)\begin{pmatrix}
   \hat{u_1} \\
   \hat{v_1}
\end{pmatrix},
\end{align}
where
\begin{align}
\resizebox{\textwidth}{!}{$
    B(m)=\begin{pmatrix}
    -\left(\frac{2m \pi }{l}\right)^2D_u +a_1+ \frac{2m \pi }{l}2D_uw_u  u_0 c_1 \rho^{-1}\sin\left(\frac{2m \pi\rho}{l}\right) & a_2 + \frac{2m \pi }{l}2D_uw_u  u_0 c_2  \rho^{-1}\sin\left(\frac{2m \pi\rho}{l}\right)\\
    b_1+ \frac{2m \pi }{l}2D_uw_u  u_0 c_3 \rho^{-1}\sin\left(\frac{2m \pi\rho}{l}\right)& -\left(\frac{2m \pi }{l}\right)^2D_v+b_2 + \frac{2m \pi }{l}2D_vw_v u_0 c_4  \rho^{-1}\sin\left(\frac{2m \pi\rho}{l}\right)
\end{pmatrix}
$}.
\end{align}

\begin{figure}[!htbp]
    \centering
    \subfloat[$\operatorname{tr}(B)(m)$]{
        \includegraphics[width=0.45\textwidth]{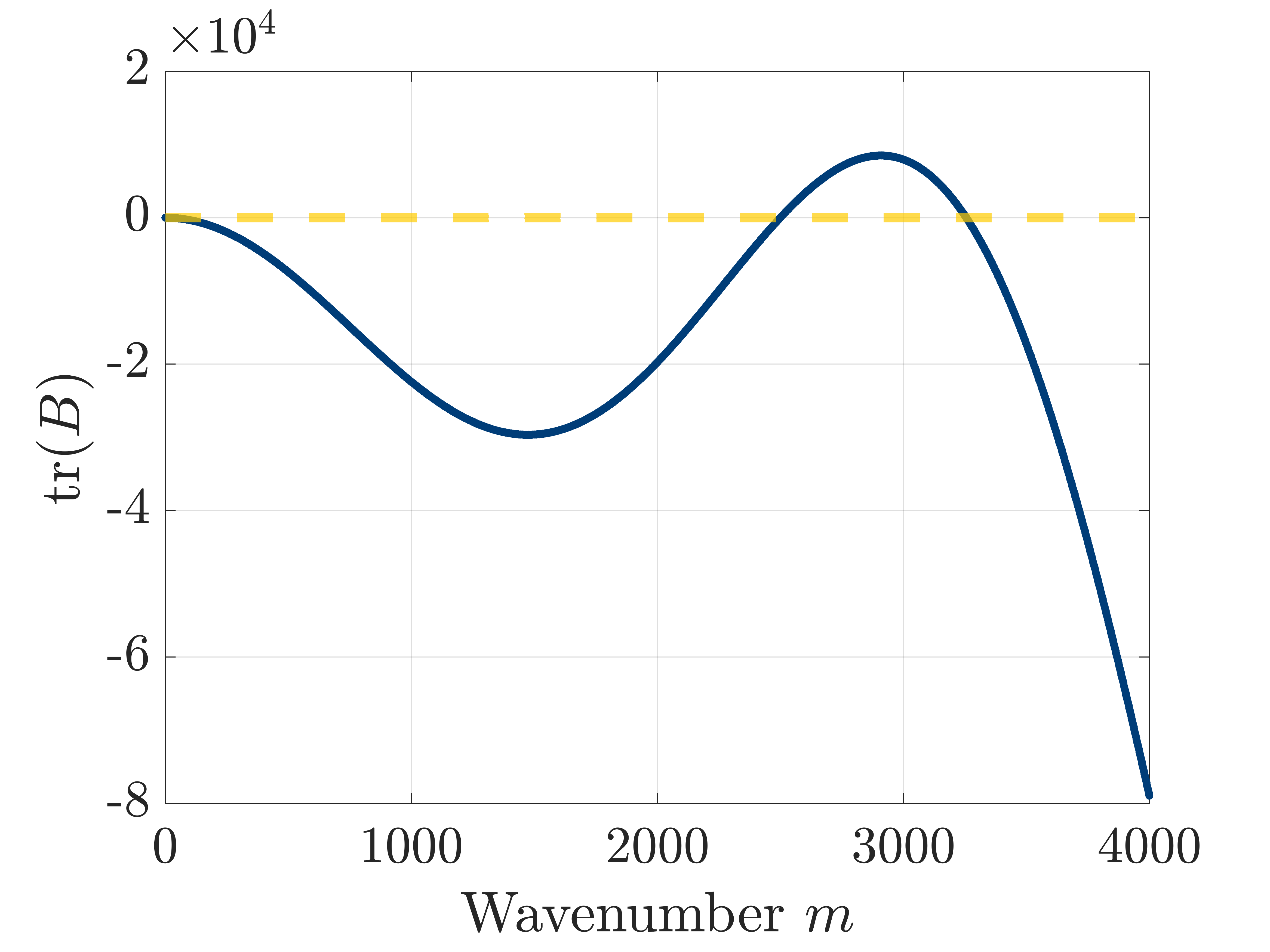}
        \label{fig:Nonlocal-tr}
    }\hspace{5mm}
    \subfloat[$\det(B)(m)$]{
        \includegraphics[width=0.45\textwidth]{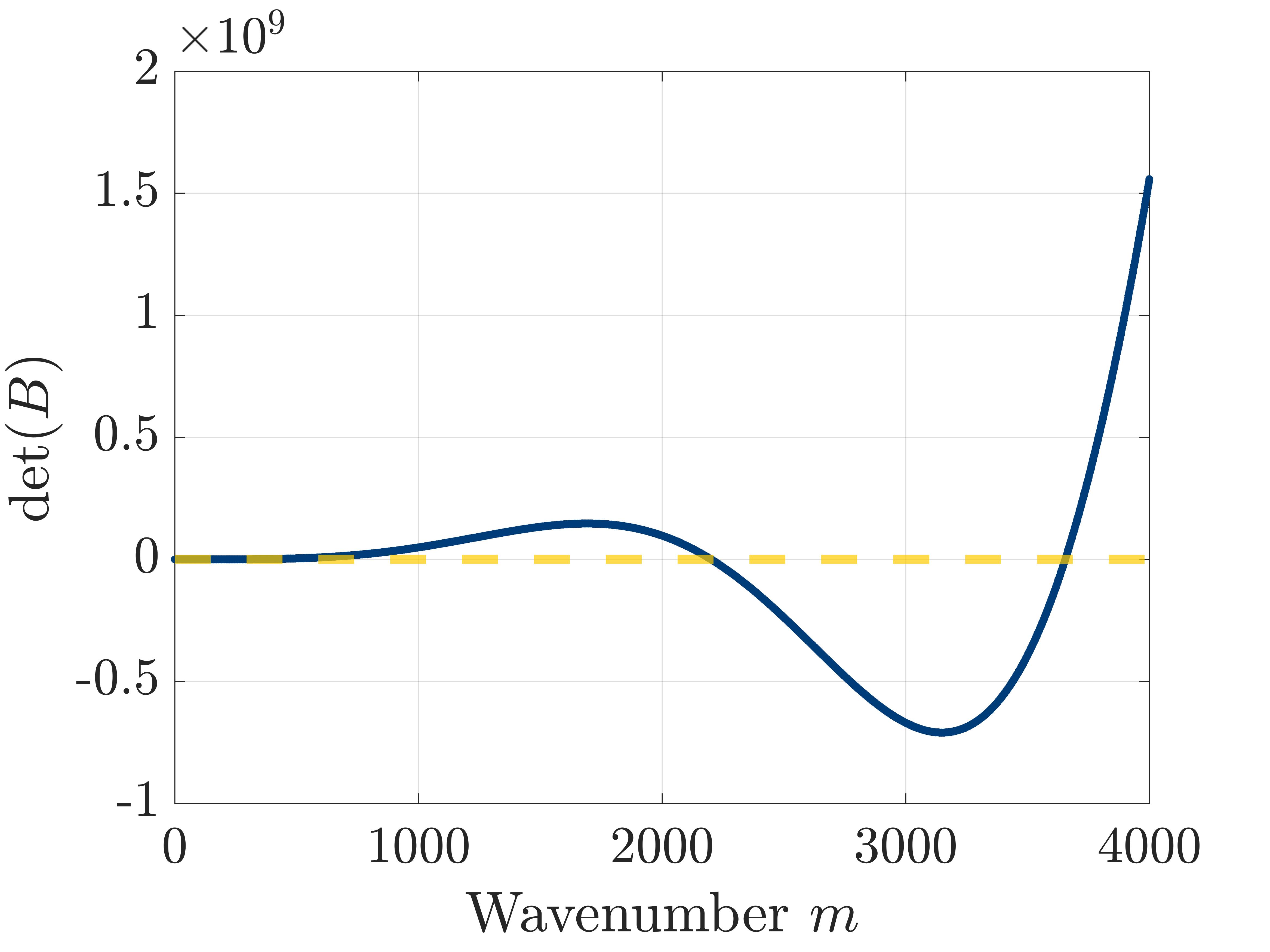}
        \label{fig:Nonlocal-det}
    }
    \caption{Plots of the trace $\operatorname{tr}(B)$ (left panel) and determinant $\det(B)$ (right panel) against the wavenumber $m$. Positive values of $\operatorname{tr}(B)$ and negative values of $\det(B)$, required for the formation of spatial patterns, only occur at finite values of $m$. We set $D_u=D_v=0.1$,  $w_u=1$,  $w_v=10$, and $\rho=0.01$. All other parameters are identical to those used in Figure~\ref{fig:nonpayoff_critical_threshold}.}
    \label{fig:Nonlocal-det-tr}
\end{figure}

In particular, the uniform equilibrium will be unstable provided that the matrix $B(m)$ has an eigenvalue with positive real part, so we turn to evaluating $\operatorname{tr}(B(m))$ and $\det(B(m))$ to determine the possibility of finding unstable wavenumbers for our nonlocal system. We note that the trace of $B(m)$ is given by
\begin{align}
    \operatorname{tr}(B(m)) &= a_1 + b_2 - \left(\frac{2 m \pi }{l}\right)^2(D_u + D_v) \\ &+ \frac{m \pi }{l} \left( 2D_uw_u u_0 c_1 \rho^{-1}\sin\left(\frac{2 m \pi\rho}{l}\right) + 2D_vw_v u_0 c_4 \rho^{-1}\sin\left(\frac{2 m \pi\rho}{l}\right) \right). 
\end{align}
For spatial patterns to form, we require $\operatorname{tr}(B) > 0$. %
As the wavenumber index $m$ becomes very large, the term $m^2$ dominates the expression for $\operatorname{tr}(B(m))$, since $\sin\left(\frac{2 m \pi \rho}{l}\right)$ is a bounded function of $m$. This means that $\operatorname{tr}(B)$ will be negative for sufficiently large values of $m$, and therefore it is not possible to have infinitely many unstable wavenumbers due to the trace of $B(m)$ becoming negative. We also numerically explore the behavior of $\operatorname{tr}(B(m))$ as a function of $m$ for the case of equal diffusivities $D_u = D_v$, showing in %
Figure \ref{fig:Nonlocal-tr} that it is possible to achieve $\operatorname{tr}(B(m)) > 0$ for a range of wavenumber indices $m$ in the presence of sufficiently strong payoff sensitivity $w_v$ for doves. %

We can also examine the conditions under which patterns can emergence due to the Jacobian matrix $B(m)$ having negative determinant for some wavenumbers. The determinant can be written in the following form
\begin{align}
    \det(B(m)) &= \alpha_4 \left(\frac{m \pi}{l}\right)^4 + \alpha_3(m) \left(\frac{m \pi}{l}\right)^3 + \alpha_2(m)\left(\frac{m \pi}{l}\right)^2 + \alpha_1(m)\left(\frac{m \pi}{l}\right) + \alpha_{0},
\end{align}
where $\alpha_{0} = a_1 b_2 - b_1 a_2 > 0$, $\alpha_4 > 0$, and $\alpha_1(m)$, $\alpha_2(m)$, and $\alpha_3(m)$ are bounded functions of $m$. The leading coefficient $\alpha_4 = D_u D_v$ is positive and the other terms grow at most at order $\mc{O}(m^3)$ as $m \to \infty$, so we can deduce that the determinant will be positive for sufficiently large values of $m$, and that the condition $\det(B(m)) < 0$ can be achieved for only finitely many wavenumbers %
We can further see by plotting $\det(B)$ as a function $m$ in Figure \ref{fig:Nonlocal-det}, showing that there exist a finite range of unstable wavenumbers for our chosen hawk-dove game, equal diffusivities $D_u = D_v$, and sufficiently large payoff sensitivity for doves. Together, our analytical observations and the numerical results from Figure \ref{fig:Nonlocal-det-tr} shows that the nonlocal version of our PDE model allows for the formation of feasible spatial patterns due to payoff-driven motion, even for the case in which the hawk and doves are equally mobile. %

\section{Discussion}
\label{sec:discussion}

In this paper, we have considered spatial pattern formation in evolutionary games in both discrete and continuous space, exploring the role of purely diffusive motion or payoff-driven directed motion in establishing spatially heterogeneous profiles of cooperators and defectors. We found that both the stochastic spatial model and the PDE model can display Turing instability for hawk-dove games when the diffusivity of defectors exceeds that of cooperators, with the unbiased random walks for each strategy resulting in higher average payoffs achieved by hawks and doves across the spatial domain. Incorporating directed motion towards locations with higher payoff resulted in more subtle pattern-forming behavior, with the stochastic spatial model suggesting the formation of patterns with decreased average payoff relative to the payoff achieved in the uniform state and with the PDE model displaying an short-wave / infinite-wavenumber instability indicating the formation of arbitrarily fine patterns and possible blowup of the PDE solution. By combining increased diffusivity of hawks and more sensitive payoff-driven motion for doves, we were able to characterize different pattern-forming effects provided by payoff-driven motion in the case of the stochastic and PDE models.  

One interesting feature of our models was that the case of undirected motion appeared to improve the population size and collective payoff of both strategies, despite the fact that undirected motion places no emphasis on improving payoff for either strategy. In contrast, the stochastic model with payoff-driven motion suggested that individuals moving to improve their personal payoff could result in a collective loss of payoff across the population. This decrease in the collective outcome when doves are better at searching for locations with increase payoff bears some resemblance to results seen on the evolution of dispersal strategies has shown that movement strategies that ignore fitness can be evolutionarily stable when individuals sample fitness of neighboring patches \cite{galanthay2012generalized}. As we derived a system of ODEs for metapopulation patch dynamics in Section \ref{sec:ODEderivation} as an intermediate step in the derivation of our PDE from the agent-based model, it may be possible to apply the approach applied by Galanthay and Flaxman to the patch obtained in this paper. In addition, generalizing of our agent-based model and the resulting patch ODE system to incorporate connectivity beyond nearest neighbors on a lattice may be helpful for exploring payoff-driven motion on metapopulation networks, allowing to build on existing approaches for exploring spatial patterns using the structure of metapopulation graphs \cite{othmer1969interactions,othmer1971instability,levin1974dispersion,segel1976application,fahimipour2022sharp,coclite2025replicator}. Such analysis may also be interesting to explore in the context of stochastic pattern formation on networks, as the role of stochasticity has often been shown to alter the parameters thresholds for pattern-forming instability related to PDE models obtained in the PDE limit in the case of reaction-diffusion models with nearest-neighbor connectivity \cite{butler2009robust,butler2011fluctuation,woolley2011stochastic,biancalani2010stochastic,asslani2012stochastic,cao2014stochastic,karig2018stochastic,kim2020stochastic}.

Another interesting feature of the linear stability analysis of our main PDE model was that the pattern-forming instability often featured infinitely many unstable wavenumbers. This behavior has been seen before in the work of payoff-driven motion by Helbing and coauthors \cite{helbing2008migration,helbing2009pattern}, who introduced a modified version of their model with a fourth-order diffusion term to stabilize high-frequency modes and allow for stable numerical simulations. In this paper, we took alternate approaches to explore modified models of payoff-driven motion featuring finitely many unstable wavenumbers, considering both nonlocal gradient sensing with an integral term and a fourth-order PDE obtained by Taylor expansion of the integral kernel for gradient sensing of payoff. These two modified models share qualitative features with nonlocal models for collective cell motion with cell-to-cell adhesion \cite{armstrong2006continuum,hillen2020nonlocal} and to associated higher-order local models \cite{falco2023local}. A recent model of evolutionary games with environmental feedback also considered the case of environmental-driven motion with a chemotaxis-like term, showing that patterns with finite wavenumber can emerge when individuals climb the gradient of a quantity indicating higher environmental quality (and corresponding higher payoff), rather than directly climbing the payoff gradient itself 
\cite{yao2025spatial}. This distinction between the emergence of spatial patterns featuring characteristic pattern scale and infinite wavenumber instabilities suggests that loosening the feedback between local payoff and directed motion towards regions of increasing success may be helpful for achieved well-behaved spatial patterns in PDE models for spatial evolutionary games.  

While the focus in this paper was on game-theoretic interactions following a Hawk-Dove game, the analysis of spatial patterns arising from payoff-driven motion can be studied in a range of games featuring a coexistence state of multiple strategies in the absence of spatial. A lot of existing work has focused on the presence of Turing instabilities or instabilities due to directed motion towards or away from various strategies in the context of ecological public goods games \cite{funk2019directed,hauert2008ecological,wakano2009spatial,wakano2011pattern,park2019ecological,gokhale2020eco}, so it may be useful to apply the techniques explored in this paper to such models of ecological public goods as well. In particular, it could be interesting to explore spatial patterns generated by payoff-driven motion when game-theoretic interactions follow an ecological public goods game, determining whether the behavior we saw of pattern formation decreasing the average payoff of the population generalizes from what was seen in this paper for the Hawk-Dove game. In addition, by considering other classes of matrix games like the Prisoners' Dilemma or coordination games, it may be possible to study how behaviors like diffusion-driven coexistence \cite{wakano2009spatial,wakano2011pattern} or phase separation \cite{hwang2013deterministic,li2025phase} can be altered or amplified due to mechanisms like payoff-driven motion. 

Although we have primarily focused on the role played by spatial movement rules in the establishment of spatial patterns and resulting impacts on payoff and population size, it can also be important to understand the role that the local reaction dynamics play in the qualitative behavior of emergent spatial patterns. In our models, we followed the approach used by both Brown and Hansell and Durrett and Levin \cite{brown1987convergence,durrett1994importance,fahimipour2022sharp} to model local population dynamics, considering a baseline per-capita birth or death rate based on the value and sign of the average payoff obtained by hawks and doves, as well as a density-dependent logistic regulation imposed by the total number of hawks and doves at a spatial location. Other models of spatial evolutionary games have considered a range of different possible assumptions of reaction dynamics with a range of assumptions of frequency-dependent and frequency-dependent factors, with Helbing and other authors considering imitation dynamics with a total constant population size across space \cite{helbing2009pattern,bratus2014replicator}, Belmonte and coauthors considering only payoff-based birth rates and population regulation through negative payoffs in same-strategy interactions, and the assumption of a maximum population density imposed in spatial models of ecological public goods games \cite{wakano2009spatial,wakano2011pattern,funk2019directed,park2019ecological,gokhale2020eco}. Overall, these different assumptions on the roles of strategic frequencies and population density may result in different possibilities in pattern-forming instability or the impact of spatial patterns on collective outcomes across the different models considered, and opens the possibility of a systematic exploration of these effects in future work. In particular, this motivates further exploration in how spatial patterns will interact with various proposed modifications of replicator equations and other models of evolutionary game dynamics to incorporate a range of density-dependent effects \cite{durrett1994importance,cressman1997spatial,novak2013density,kvrivan2018beyond}, which can play an especially important role in patterned states in which hawks and doves tend to cluster in the same population aggregates. 

Other work on spatial pattern formation and spatial phenomena in evolutionary game theory has explored the possibility of nonlocal game-theoretic interactions in shaping the spatial distribution of strategies \cite{aydogmus2018discovering,hwang2013deterministic}. While our nonlocal model only addresses the evaluation of spatial movement rules through the term governing spatial movement, it may be interesting to explore how spatial feedback with nonlocal interactions would also impact the establishment and stability of spatial patterns. One model of nonlocal interactions could arise by turning to models of evolutionary games with environmental feedback and environmental-driven motion \cite{weitz2016oscillating,lin2019spatial,antonioni2019individual,tilman2020evolutionary,yao2025spatial}, with a nonlocal model arising under the assumption that the spatiotemporal dynamics of an environmental resource occur on a faster timescale than strategic dynamics. In addition, the use of nonlocal interactions and nonlocal payoff evaluations may suggest an alternative Lagrangian individual-based description of spatial game-theoretic dynamics featuring a finite population of individuals performing payoff-driven motion in continuous space via nonlocal interaction and sensing kernels. Such models have been explored in the context of evolutionary games between players living in an ambient fluid flow \cite{uppal2018shearing,herrerias2018stirring,herrerias2019motion,krieger2020turbulent}, and similar off-lattice agent-based models have been used to study pattern formation and collective behavior in settings ranging from developmental biology \cite{volkening2015modelling,volkening2018iridophores,martinson2024linking} to collective motion in groups of humans or other animals \cite{cucker2007emergent,ha2008particle,helbing1995social,benson2023classroom,hein2015evolution}.

Another question for future work is to consider the coevolution of spatial movement rules with strategic behavior in game-theoretic interactions. There is a rich literature in the evolution of spatial movement rules, where diffusion or advection rates can be studied as quantitative traits using adaptive dynamics and pairwise invasibility analysis \cite{hastings1983can,chen2012dynamics,hutson2003evolution}. As we have seen that spatial movement appears to decrease the average payoff of the population but allow for certain individuals to obtain greater payoffs than is achieved in the well-mixed population, it would be useful to explore whether such spatial movement rules would evolve. In particular, one direction for future work would be to explore is whether the individual advantage conferred by biased spatial movement could result in an evolutionary social dilemma resulting in a lower total long-time population than if individuals did not have the ability to perform payoff-driven motion. These evolutionary questions related to migration strategies may be particularly relevant in the case of costly migration \cite{galanthay2012generalized,lee2022costly}, in which the tradeoffs of benefits of potentially reaching a location with an individual advantage in game-theoretic payoff could be partly diminished due to an additional cost incurred to moving to such a location. Beyond the evolution of dispersal strategies, it may also be interesting to explore how payoff-driven motion and other mechanisms for spatial motion interact with the evolution of a continuous strategy for game-theoretic actions, such as the evolution of a level of effort in contributing to a public good \cite{hermsen2022emergent,doekes2024multiscale}.

In addition to prior work on spatial motion on evolutionary game theory, the models explored in our paper provide connection with work on the PDE models achieved as mean-field limits of agent-based models with individual-based descriptions of biased random walks \cite{short2008statistical,alsenafi2018convection}, as well as models in the mathematical biological literature that derive mean-field PDE models from individual-level birth-death dynamics \cite{champagnat2006unifying,luo2014unifying,luo2017scaling}. The derivation of our PDE model from a continuous-time model of random walkers on a grid built upon the approach from Wang and coauthors involving characterizing the infinitesimal mean and variance of the continuous-time model \cite{wang2021stochastic}, incorporating the approach used in models of evolutionary biology in which the PDE limit is derived as length scale of spatial grid and the influence of each individual jointly tend to zero \cite{champagnat2006unifying,luo2014unifying}. In addition, our use of the game-theoretic payoff achieved by individuals in the patterned states of our models provides a new perspective on the costs and benefits of spatial pattern formation, allowing us to measure whether individual cooperators and defectors improve their payoff by following a given rule for spatial movement.  

\newpage
\renewcommand{\abstractname}{Acknowledgments}
\begin{abstract}
This paper started as an undergraduate research project as part of the Spring 2024 Illinois Mathematics Lab (IML), and we greatefully acknowledge support from this program. The authors would like to thank Jirui Liu, Seokhwan Moon, Ahmad Rahman, Edwin Shen, and Muhammad Raza for helpful discussions in the early stages of this project. We would particularly like to thank Seokhwan Moon for many helpful discussions regarding the derivation and stability analysis of PDE models.

TY was supported by the University of Illinois Undergraduate Research Office via a Research Support Grant, and DBC's research is partially supported by the Simons Foundation through the Travel Support for Mathematicians Program (grant MPS-TSM-00007872). 
\end{abstract}

\renewcommand{\abstractname}{Statement on Code Availability}
\begin{abstract}
All code used to generate figures is archived on GitHub (see \href{https://github.com/YuzuruTesla/Stochastic-PDE-Payoff-Driven}{GitHub Repository})) and licensed
for reuse, with appropriate attribution/citation, under a BSD 3-Clause Revised License. This
repository contains the Python code to simulate the stochastic spatial model and the Matlab code used to run numerical simulations of the PDE model, as well as all scripts that were used to generate the figures in the paper. The simulations of the stochastic spatial model used the Python package \href{https://pypi.org/project/piegy/}{piegy}, which was developed by author Chenning Xu. 
\end{abstract}

\bibliographystyle{unsrt}
\bibliography{References}

\appendix
\appendixpage
\addtocontents{toc}{\protect\setcounter{tocdepth}{1}}

In this appendix, we provide additional description of our stochastic model and present additional derivations and analysis of our PDE models. In Section \ref{sec:StochasticAppendix}, we present full specification of the rates for our stochastic spatial model, and we present the Gillespie algorithm we use to simulation birth, death, and movement events. We then present the derivation of our PDE model as a continuum limit of the stochastic spatial model in Section \ref{sec:derivation_of_diff_eqs}, and we then provide additional calculations for the linear stability analysis of local PDE model and our model with nonlocal payoff sensing in Sections \ref{sec:linearization_appendix} and \ref{sec:nonlocal-linearization}, respectively.

\section{Details on Simulation of Stochastic Model with Spatial Gillespie Algorithm}
\label{sec:StochasticAppendix}

The approach we will use to simulate our stochastic spatial model is based on the Gillespie algorithm for simulating continuous-time Markov chains, and we use a formulation of the rates of demographic and movement events following the approach used by Seri and Shnerb to study a similar spatial game model by with undirected motion \cite{seri2012sustainability}. We first characterize the all the possible events that are described in Section \ref{sec:StochasticModel}, and we formulate the reaction rates of each event to determine the probability that each event occurs next in our stochastic simulation. %

In our stochastic model, we consider a spatial domain consisting of a line of $n$ patches in one-dimension, with each patch having two neighbors expect for the two patches at the ends of the line. We assume that all game-theoretic interactions and ecological competition within patches are well-mixed. We simulate the growth and decay of hawks and doves and measure their payoff by considering each patch individually. Migration behaviors are modeled by considering pairs of neighboring patches. Within each patch, four types of discrete events are defined:

\begin{itemize}
    \item Birth/death of an individual due to payoff. 
    \item Death of an individual due to carrying capacity / density-dependent regulation
    \item Migration of a hawk
    \item Migration of a dove
\end{itemize}

Each of these events is governed by one of the three sets of equations in Section \ref{sec:StochasticModel}: the payoff functions in Equation\eqref{eq:PayoffEquationStochastic}, the %
the density-dependent regulation described in Equation \eqref{eq:CarryingCapacity}, and the migration rates provided by Equation \eqref{eq:Migration}. At some patch $i$ and time $t$, the rate of a payoff-triggered birth/death event is proportional to the produce of the number of individuals of a given strategy at patch $i$ and the average payoff achieved for that strategy at patch $i$, taking the form%

\begin{align*}
\begin{split}
r_{i, H-p}(t) = \lvert p_{i, H}(t) \cdot u_i(t) \lvert \\
r_{i, D-p}(t) = \lvert p_{i, D}(t) \cdot v_i(t) \lvert,
\end{split}
\end{align*}

where $p_{i, H}(t)$ and $p_{i, D}(t)$ are payoffs given by the payoff functions in Equation \eqref{eq:PayoffEquationStochastic}. The signs of the payoffs $p_{i, H}(t)$ and $p_{i, D}(t)$ determine whether the event corresponds to the birth or death of an individual, with a positive payoff corresponding to a birth event and a negative payoff corresponding to a death event. %

The rates of death events due to density-dependent regulation  for the hawks and doves are described by $r_{H-\kappa}$ and $r_{D-\kappa}$, which are given by %
\begin{align*}
\begin{split}
r_{i, H-\kappa}(t) = k_{i, H}(t) \\
r_{i, D-\kappa}(t) = k_{i, D}(t),
\end{split}
\end{align*}

where $k_{i, H}(t)$ and $k_{i, D}(t)$ are given by the regulation terms in Equation\eqref{eq:CarryingCapacity}. Since $k_{i, H}(t)$ and $k_{i, D}(t)$ already include a population term, we do not multiply by the population size again when calculating the likelihoods.

The migration behaviors, described in Section \ref{sec:StochasticModel} in the main paper, correspond to a biased random walk based on differences in payoffs, with the per-capita rate that an individual migrates to some neighboring patch $i'$ given by the Equation \eqref{eq:Migration}. We can map this per-capita migration rate to a total rate of a migration event from patch $i$ to a neighboring patch $i'$ by multiplying by the total number of individuals at patch $i$, %
resulting in a movement rate of the form
\begin{align*}
\begin{split}
r_{i, H-i'}(t) = u_i(t)q_H(i \to i')(t) \\
r_{i, D-i'}(t) = v_i(t)q_D(i \to i')(t).
\end{split}
\end{align*}

Additionally, to address concerns regarding boundary conditions, we scale $\mu_u$ and $\mu_v$ in Equations \ref{eq:Migration} according to the number of neighbors each patch has. For a patch with $a$ neighbors, we use $\frac{a}{2} \mu_u$ and $\frac{a}{2} \mu_v$ as the effective diffusivity parameters for individuals at that patch.
To implement a zero-flux boundary condition, we assume that individuals in the boundary patches $i=1$ and $i = N$ move inward during any successful migration event, never leaving the spatial domain. 

So far, we have presented all the discrete events considered for a single patch and discussed how their rates are determined. To restate briefly, there are 8 events per patch: 2 birth/death events triggered by payoff (one for each species), 2 death events due to carrying capacity, 2 migration events for hawks (one in each direction) and 2 migration events for doves. This gives a total of $8n$ rates that we need to simulate our $n$-patch one-dimensional model, and we can now use these rates to determine how events are selected to simulate our stochastic process through a Gillespie algorithm. %

Let $R_i(t)$ denote the tuple of the reaction rates that correspond to the discrete events of patch $i$ and time $t$. We can then describe the state of the entire spatial domain by taking the union over all patches: $R(t) = \ds\bigcup_i R_i(t)$, which incorporates all events in the space and characterizes the state of our stochastic system at time $t$. We index the rates in $R(t)$ as $r_1(t), r_2(t), \cdots, r_{8n-1}, r_{8n}(t)$ for notational convenience. The Gillespie algorithm is then applied to simulate event selection and time advancement. Specifically, the operations at some time $t=t_1$ are as follows.

We first choose some random number $a_1 \in (0, 1)$ and find the smallest index $b$ such that

\begin{equation*}
    \sum_{j=1}^{b} r_j(t) > a_1 \sum_{j} r_j(t),
\end{equation*}

where $r_b$ denotes the index of the event selected at the current time $t$. 

A second random number $a_2 \in (0, 1)$ is drawn, which allows us to determine the following time-step p $\Delta t$ that is taken after the chosen event:

\begin{equation*}
    \Delta t = \frac{1}{\sum_{j} r_j(t)}\log\left(\frac{1}{a_2}\right)
\end{equation*}

We then update time $t \to t_1 + \Delta t$ and store data if the time $t$ passes a check point. We repeat the above process until time $t$ reaches the specified time limit $t_{\max}$. Furthermore, when we perform an ensemble of simulations for the same parameter values, we loop the simulation $s_{\max}$ times and average the resulting population and payoff quantities to make aggregate statements about the stochastic model. %

We now summarize the key steps for our stochastic simulation using the pseudocode presented in Algorithm \ref{alg:StochasticAlgorithm}.

\begin{algorithm}[!htpb]
\caption{\textbf{Stochastic model for evolutionary dynamics}}
\label{alg:StochasticAlgorithm}
\begin{algorithmic}
\State \textbf{Inputs:} Number of patches $n$, time limit $t_{\max}$, number of simulation rounds $s_{\max}$, \\ initial populations $I$, payoff matrices $M$, and other constants (diffusivity $\mu$, payoff sensitivity $w$, and carrying capacity $\kappa$ as defined in Section \ref{sec:StochasticModel})

\State
\For{$s = 1$ to $s_{\max}$}
    \State Initialize $t = 0$
    \State Initialize each patch with parameters $I$, $M$, $\mu$, $w$, and $\kappa$.
    \State At each patch, initialize payoff, rates of payoff-triggered birth/death events, rates of carrying
    \State capacity-triggered death events. 
    
    \State Initialize migration rates at each patch.
    
    \While{$t < t_{\max}$}
        \State Pick two random numbers $a_1 , a_2 \in (0, 1)$
        \State Compute $\Delta t = \frac{1}{\sum_{j} r_j(t)}\log(\frac{1}{a_1})$
        \State Compute $b$ : the smallest integer s.t. $\sum_{j=1}^{b} r_j(t) > a_2 \sum_{j} r_j(t)$
        \State Update population, payoff, and rates based on event index $b$.
        \State Update $t \rightarrow t + \Delta t$
        \State Record population and other data if $t$ passes a check point
    \EndWhile
\EndFor
\State Compute the mean and other statistical info over the $s$ simulations
\end{algorithmic}
\end{algorithm}

\section{Derivation of PDE Model with Payoff-Driven Motion from Stochastic Individual-Based with Payoff-Driven Biased Random Walk}
\label{sec:derivation_of_diff_eqs}

In this section, we provide derivations connecting our individual-based stochastic model and our PDE model for evolutionary games with payoff-driven motion. We complete this derivation in two steps, first using techniques from stochastic processes to derive a system of ODEs for the densities of hawks and doves at each discrete spatial patch in the limit of infinitessimal mass of individuals, which allows us to describe deterministic population dynamics for a population living on $n$ patches. We then use techniques inspired by numerical analysis to derive of partial differential equations for the spatial profile of hawks and doves across a one-dimensional domain, which we obtain by taking the limit of infinity many patches and infinitessimal distance between patches. We present our derivation of the ODE patch model in Section \ref{sec:ODEderivation}, and we present the derivation of the PDE model in Section \ref{sec:derivation_PDE}.

\subsection{Derivation of System of ODEs on Discrete Spatial Lattice}
\label{sec:ODEderivation}

In this section, we present a derivation of our PDE model for evolutionary game dynamics with payoff-driven motion from our stochastic model with a biased random walk between neighboring patches. %
For simplicity, we will describe the derivation for a one-dimensional spatial domain, but this approach can be readily generalized to study higher-dimensional spatial domains and slight differences in neighborhood configurations (such as considering the von Neumann and Moore neighborhoods).%

We consider a family of stochastic models a spatial domain of length $L$ divided into $n$ evenly divided patches, and we assume that each individual has a mass $\frac{1}{m}$ that contributes to the total population size of their patch. We denote the number of hawks and doves at patch $i$ at time $t$ by $u_i^{m,n}(t)$ and $v_i^{m,n}(t)$, respectively, and we look to consider how these numbers of hawks and doves changes in time. In particular, we will use the payoffs $p_{i,H}(t)$ and $p_{i,D}(t)$ to denote the payoffs of hawks and doves at a given patch $i$, and we look to show how these payoffs impact the rate of demographic and migration events at patch $i$. 
The number $u_i^{m,n}(t)$ of hawks increases by $\frac{1}{m}$ if one of the following three events occurs:
\begin{itemize}
    \item A hawk reproduces within patch $i$, which occurs with rate
    \[ m u_i^{m,n}(t) \left[ p_{i,H}(t)\right]_{+},\]
where $[h(x)]_+$ denotes the positive part of the function $h(x)$. 
    
    \item A hawk moves from patch $i-1$ to patch $i$, which occurs with rate
\[
 m u_{i-1}^{m,n}(t) \left( \frac{f_u\left(w_u p_{i,H}(t) \right)}{ f_u\left(w_u p_{i,H}(t)  \right) + f_u\left(w_u p_{i-2,H}(t) \right) } \right) 
\]

    \item A hawk moves from patch $i+1$ to patch $i$, which occurs with rate
   \[
 m u_{i+1}^{m,n}(t) 
  \left( \frac{f_u\left(w_u p_{i,H}(t) \right)}{ f_u\left(w_u p_{i,H}(t)  \right) + f_u\left(w_u p_{i+2,H}(t) \right) } \right) 
  \]
\end{itemize}
Similarly, we can see that the number of hawks decreases by $\frac{1}{m}$ when one of the three events occurs
\begin{itemize}
    \item A hawk reproduces within patch $i$, which occurs with rate
    \[ m u_i^{m,n} \left[ \kappa \left( u_i^{m,n}(t) + v_i^{m,n}(t) \right) \right] + m u_i^{m,n}(t) \left[ p_{i,H}(t)\right]_{-} 
    \]
where $[h(x)]_-$ denotes the negative part of the function $h(x)$. 
    
    \item A hawk moves from patch $i$ to patch $i+1$, which occurs with rate
   \[
 m u_{i}^{m,n}(t) \left( \frac{f_u\left(w_u p_{i+1,H}(t) \right)}{ f_u\left(p_{i+1,H}(t) \right) + f_u\left(w_u p_{i-1,H}(t) \right)} \right) 
  \]

    \item A hawk moves from patch $i$ to patch $i-1$, which occurs with rate
\[
 m u_{i}^{m,n}(t) \left( \frac{f_u\left(w_u p_{i-1,H}(t) \right)}{ f_u\left(p_{i+1,H}(t) \right) + f_u\left(w_u p_{i-1,H}(t) \right)} \right) 
\]
 
\end{itemize}
We can then calculate that, conditioned on the value $u_i^{m,n}(t)$, the expected value of the number of hawks in patch $i$ at time $t + \Delta t$ is given by
\begin{equation}
\begin{aligned}
& E\left[ u_i^{m,n}\left( t + \Delta t \right) - u_i^{m,n}\left(  t \right)   \bigg|  u_i^{m,n}(t) \right] \\ &= \frac{1}{m} P\left( u_i^{m,n}\left( t + \Delta t \right) - u_i^{m,n}\left(  t \right) = \frac{1}{m} \right) -  \frac{1}{m} P\left(  u_i^{m,n}\left( t + \Delta t \right) - u_i^{m,n}\left(  t \right)  = - \frac{1}{m} \right) + o\left(\Delta t \right) \\
&= \frac{1}{m} \left[m u_i^{m,n}(t)  \left[  p_{i,H}(t) \right]_+ \Delta t -  m u_i^{m,n}(t)  \left[ p_{i,H}(t)  \right]_- \Delta t \right]  - \frac{1}{m} \left[ m u_i^{m,n}  \kappa \left( u_i^{m,n}(t) + v_i^{m,n}(t) \right) \right] \Delta t
\\ &+ \frac{1}{m} \left[m u_{i-1}^{m,n}(t) \mu_u \left( \frac{f_u\left(w_u p_{i,H}(t) \right)}{ f_u\left(w_u p_{i,H}(t) \right) + f_u\left(w_u p_{i-2,H}(t)\right)} \right) \right] \Delta t  \\ &+ \frac{1}{m}  \left[ m u_{i+1}^{m,n}(t) \mu_u \left( \frac{f_u\left(w_u p_{i,H}(t)\right)}{ f_u\left(w_u p_{i,H}(t)\right) + f_u\left( w_u p_{i+2,H}(t) \right)} \right) \Delta t    \right] \\
&- \frac{1}{m} \left[m u_{i}^{m,n}(t) \mu_u \left( \frac{f_u\left(w_u p_{i-1,H}(t)\right)}{ f_u\left(w_u p_{i-1,H}(t) \right) + f_u\left(w_u p_{i+1,H}(t) \right)} \right) \right] \Delta t \\ &- \frac{1}{m} \left[ m u_{i+1}^{m,n}(t) \mu_u \left( \frac{f_u\left(w_u p_{i+1,H}(t) \right)}{ f_u\left(w_u p_{i-1,H}(t) \right) + f_u\left(w_u p_{i+1,H}(t) \right)} \right)  \right] \Delta t +  o\left( \Delta t \right),
\end{aligned}
\end{equation}
which we can rearrange to see that
\begin{equation}
\begin{aligned}
&\frac{E\left[ u_i^{m,n}\left( t + \Delta t \right) - u_i^{m,n}\left(  t \right)   \bigg|  u_i^{m,n}(t) \right]}{\Delta t}  \\
&= u_i^{m,n}(t)  \left[  p_{i,H}(t) - \kappa \left(u_i^{m,n}(t) + v_i^{m,n}(t) \right) \right] \\
&+ \mu_u u_{i-1}^{m,n}(t) \left( \frac{f_u\left(w_u p_{i,H}(t) \right)}{ f_u\left(w_u p_{i,H}(t) \right) + f_u\left(w_u p_{i-2,H}(t) \right)} \right) + \mu_u  u_{i+1}^{m,n}(t) \left( \frac{f_u\left(w_u p_{i,H}(t) \right)}{ f_u\left(w_u p_{i,H}(t) \right) + f_u\left(w_u p_{i+2,H}(t) \right)} \right)   \\
&- \mu_u u_{i}^{m,n}(t) \left( \frac{f_u\left(w_u p_{i-1,H}(t)\right)}{ f_u\left(w_u p_{i-1,H}(t) \right) + f_u\left(w_u p_{i+1,H}(t)\right)} \right) - \mu_u u_{i}^{m,n}(t) \left( \frac{f_u\left(w_u p_{i-1,H}(t) \right)}{ f_u\left(w_u p_{i-1,H}(t)\right) + f_u\left(w_u p_{i+1,H}(t) \right)} \right) \\ &+ \frac{o\left(\Delta t\right)}{\Delta t}.
\end{aligned}
\end{equation}
Because the rates at which each of the events occur are $O(m)$ as $m \to \infty$, we see that it is possible to show that the infinitessimal variance of the process takes the form
\begin{equation}
\begin{aligned}
&\lim_{\Delta t \to 0} \frac{V\left[u_i^{m,n}\left( t + \Delta t \right) - u_i^{m,n}\left(  t \right)   \bigg|  u_i^{m,n}(t) \right]}{\Delta t} \\
&= \lim_{\Delta t \to 0} \left( \frac{1}{\Delta t}\right)  \left[ \frac{1}{m^2} P\left( u_i^{m,n}\left( t + \Delta t \right) - u_i^{m,n}\left(  t \right) = \frac{1}{m} \right) -  \frac{1}{m^2} P\left(  u_i^{m,n}\left( t + \Delta t \right) - u_i^{m,n}\left(  t \right)\right) + o\left(\Delta t \right) \right] \\
&=  \lim_{\Delta t \to 0} \left[ \left(\frac{1}{\Delta t} \right) \frac{1}{m^2} \left( O(m) \Delta t \right) \right]  \to 0 \: \: \mathrm{as} \: \: m \to \infty. 
\end{aligned}
\end{equation}
Because the infinitessimal variance vanishes in the limit as $m \to \infty$, we see that we expect $u^{m,n}(t)$ to be a constant equal to its mean in this limit. We then introduce the quantity $u_i^{n}(t) := \lim_{m \to \infty} u_i^{m,n}(t) = \lim_{m \to \infty} E\left[u_i^{m,n}(t)\right]$, and we can use an analogous argument to deduce that the number of doves should be given by the deterministic quantity $v_i^{n}(t) := \lim_{m \to \infty} v_i^{m,n}(t) = \lim_{m \to \infty} E\left[v_i^{m,n}(t)\right]$. Applying the limits as $m \to \infty$ to both sides of our expression for the infinitessimal mean and take the limit as $\Delta t \to 0$ to see that $u_i^n(t)$ will satisfy the following system of differential equations
\begin{subequations}
\begin{equation}
\begin{aligned}
& \dsddt{u_i^n(t)} \\ &= u_i^{n}(t)  \left[  p_{i,H}(t)- \kappa \left(u_i^{n}(t) + v_i^{n}(t) \right) \right] \\
&+ \mu_u u_{i-1}^{n}(t) \left( \frac{f_u\left(w_u p_{i,H}(t) \right)}{ f_u\left(p_{i,H}(t) \right) + f_u\left(w_u p_{i-2,H}(t)\right)} \right) +  \mu_u u_{i+1}^{n}(t) \left( \frac{f_u\left(w_u p_{i,H}(t) \right)}{ f_u\left(w_u p_{i,H}(t)\right) + f_u\left(w_u p_{i+2,H}(t)\right)} \right)   \\
&-  \mu_u u_{i}^{n}(t) \left( \frac{f_u\left(w_u p_{i-1,H}(t) \right)}{ f_u\left(w_u p_{i-1,H}(t) \right) + f_u\left(w_u p_{i+1,H}(t) \right)} \right) - \mu_u u_{i}^{n}(t) \left( \frac{f_u\left(w_u p_{i+1,H}(t) \right)}{ f_u\left(w_u p_{i-1,H}(t) \right) + f_u\left(w_u p_{i+1,H}(t) \right)} \right),
\end{aligned}
\end{equation}
and we can perform an analogous derivation for the number of doves to obtain the following ODE for the number of hawks $v_i^{n}(t)$ at patch $i$:
\begin{equation}
\label{eq:ODEsystem}
\begin{aligned}
& \dsddt{v_i^n(t)} \\ &= v_i^{n}(t)  \left[  p_{i,D}(t)- \kappa \left(u_i^{n}(t) + v_i^{n}(t) \right) \right] \\
&+ \mu_v v_{i-1}^{n}(t) \left( \frac{f_v\left(w_u p_{i,D}(t) \right)}{ f_v\left(p_{i,D}(t) \right) + f_v\left(w_v p_{i-2,D}(t)\right)} \right) +  \mu_v v_{i+1}^{n}(t) \left( \frac{f_v\left(w_v p_{i,D}(t) \right)}{ f_v\left(w_v p_{i,D}(t)\right) + f_v\left(w_v p_{i+2,D}(t)\right)} \right)   \\
&-  \mu_v v_{i}^{n}(t) \left( \frac{f_v\left(w_u p_{i-1,D}(t) \right)}{ f_v\left(w_v p_{i-1,D}(t) \right) + f_v\left(w_v p_{i+1,D}(t) \right)} \right) - \mu_v v_{i}^{n}(t) \left( \frac{f_v\left(w_v p_{i+1,D}(t) \right)}{ f_v\left(w_v p_{i-1,D}(t) \right) + f_v\left(w_v p_{i+1,D}(t) \right)} \right)
\end{aligned}
\end{equation}
\end{subequations} 

Considering the system of ODEs for $u_i^n(t)$ and $v_i^n(t)$ for each patch $i \in \{1,\cdots,n\}$ allows us to describe the spatial dynamics of hawks and doves with payoff-driven motion in terms of a continuum of population states distributed across finitely many spatial locations. We will then use this system of ODEs in the next section to derive our PDE model for payoff-driven motion in the limit of infinitely many patches and infinitessimal distance between nearest-neighbor patches.

\subsection{Derivation of PDE}
\label{sec:derivation_PDE}

To derive the PDE, we first think about directed motion from stochastic models in discrete space with biased random walks~\cite{gerlee2019persistence,painter2002volume,hillen2009user,bubba2020discrete,short2008statistical,alsenafi2018convection,alsenafi2021multispecies,lindstrom2020qualitative}. Our goal is to start with the ODE model derived in Section \ref{sec:ODEderivation} for evolutionary games with payoff-driven motion on a discrete spatial lattice, and to obtain a PDE limit by considering an expansion of the terms in the ODE model in the limit of infinitessimally small grid spacing. For the purpose of this derivation, we will use $x$ to denote the spatial location of a focal patch on our spatial lattice and we will denote $\Delta x$ as the distance between patches in our lattice. In the derivation, we will focus on the form of the spatial movement terms, and we will add the reaction terms once we have obtained our continuum limit. 

For this derivation, We will use $u$ and $v$ to denote population densities, with  $u(t,x)$ representing the population density of hawks at position $x$ and time $t$ and with $v(t,x)$ representing the population density of doves. For convenience, we will denote the weighting function for the biased random walks for hawks and doves as
\begin{subequations} \label{eq:fshorthand}
\begin{align}
f_u(t,x) &:= f_u\left(u(t,x),v(t,x) \right) \\
f_v(t,x) &:= f_v\left(u(t,x),v(t,x) \right),
\end{align}
\end{subequations}
and we will initially use this shorthand representation to simplify the process of obtaining our PDE limit. Using this new notation, we may write the system of ODEs from Equation \eqref{eq:ODEsystem} without the reaction terms as
\begin{subequations} \label{eq:uvdiscrete}
\begin{align}
\dsdel{u(t,x)}{t} &= \mu_u \left[ \frac{f_u(t,x) u\left( t,x-\Delta x \right) }{f_u(t,x) + f_u\left(t,x-2 \Delta x \right)} \right. \left. + \frac{f_u(t,x) u\left( t,x-\Delta x \right)}{f_u(t,x) + f_u\left(t,x+2 \Delta x \right)}   - u(t,x) \right] \label{eq:udiscreteu} \\
\dsdel{v(t,x)}{t} &= \mu_v \left[ \frac{f_v(t,x) v\left( t,x-\Delta x \right) }{f_v(t,x) + f_v\left(t,x-2 \Delta x \right)} \right. \left. + \frac{f_v(t,x) v\left( t,x-\Delta x \right)}{f_v(t,x) + f_v\left(t,x+2 \Delta x \right)}   - v(t,x) \right], \label{eq:vdiscretev}
\end{align}
\end{subequations}
where the first terms in each equation describe the movement of hawks or doves located at $x-\Delta x$ to $x$, the second terms describe the movement of hawks and doves located at $x+\Delta x$ moving to $x$, $\mu_u$ and $\mu_v$ are the movement rates of hawks and doves, and $\Delta x$ is the distance between neighboring patches. 

We will now derive the limiting PDE for the density of hawks $u(t,x)$ in the limit as the grid spacing satisfies $\Delta x \to 0$, and then apply an analogous approach to obtain a PDE for the density of doves $v(t,x)$. To further write our equation for the density of hawks in terms of our PDE limit, we perform the Taylor expansion of $f_u\left(t,x \pm 2 \Delta x\right)$ around $(t,x)$ to see that
\begin{equation}
    \begin{aligned}
    f_u(t,x-2\Delta x) = f_u(t,x) -  \dsdel{f_u(t,x)}{x} 2\Delta x + \doubledelsame{f_u(t,x)}{x} 2(\Delta x)^2 + O\left((\Delta x)^3\right),\\
    f_u(t,x+2\Delta x) = f_u(t,x) +  \dsdel{f_u(t,x)}{x} 2\Delta x  + \doubledelsame{f_u(t,x)}{x} 2(\Delta x)^2 + O\left((\Delta x)^3\right).
\end{aligned}
\end{equation}
This expansion allows us to rewrite \ref{eq:udiscreteu} in the following form
\begin{equation}
    \begin{aligned}
    & \dsdel{u(t,x)}{t}  \\ &= \mu_u \Biggl[\left(2-\left(f_u(t,x)\right)^{-1}\dsdel{f_u(t,x)}{x}2\Delta x+\left(f_u(t,x)\right)^{-1}\doubledelsame{f_u(t,x)}{x}2(\Delta x)^2 \right)^{-1}u(t,x-\Delta x)\\
    & +\left(2+\left(f_u(t,x)\right)^{-1}\dsdel{f_u(t,x)}{x}2\Delta x+\left(f_u(t,x)\right)^{-1}\doubledelsame{f_u(t,x)}{x}2(\Delta x)^2 \right)^{-1}u\left(t,x+\Delta x\right) -u(t,x)\Biggl].
\end{aligned}
\end{equation}
Because we are consider the limit of small $\Delta x$, %
we can use the Binomial series approximation $(2+y)^{-1}= \frac{1}{2} -\frac{1}{4}y + \frac{1}{8}y^2+ \mc{O}(y^3)$ to further expand our right-hand side as
\begin{equation}
    \begin{aligned}
    \dsdel{u(t,x)}{t}  &= \mu_u \Biggl\{\Biggl[\frac{1}{2} - \frac{1}{4} \left(-2\left(f_u(t,x)\right)^{-1}\dsdel{f_u(t,x)}{x}\Delta x+2\left(f_u(t,x)\right)^{-1}\doubledelsame{f_u(t,x)}{x}(\Delta x)^2\right) \\
    &+  \frac{1}{8} \left(-2\left(f_u(t,x)\right)^{-1}\dsdel{f_u(t,x)}{x}\Delta x+2\left(f_u(t,x)\right)^{-1}\doubledelsame{f_u(t,x)}{x}(\Delta x)^2\right)^2\Biggl]u(t,x-\Delta x) \\ 
    &+ \Biggl[\frac{1}{2} - \frac{1}{4} \left(2\left(f_u(t,x)\right)^{-1}\dsdel{f_u(t,x)}{x}\Delta x+2\left(f_u(t,x)\right)^{-1}\doubledelsame{f_u(t,x)}{x}(\Delta x)^2\right) \\
    &+  \frac{1}{8} \left(2\left(f_u(t,x)\right)^{-1}\dsdel{f_u(t,x)}{x}\Delta x+2\left(f_u(t,x)\right)^{-1}\doubledelsame{f_u(t,x)}{x}(\Delta x)^2\right)^2\Biggl]u(t,x+\Delta x) - u(t,x)\Biggl\} \\
    &+ \mc{O}\left( \left(\Delta x\right)^3 \right)
\end{aligned}
\end{equation}
as $\Delta x \to 0$. We can then further expand the terms on the right-hand side to write that
\begin{equation}
    \begin{aligned}
    \dsdel{u(t,x)}{t} &= \mu_u \Biggl\{\frac{1}{2} \left[u(t,x-\Delta x) + u(t,x+\Delta x) - 2u(t,x) \right]\\ 
    &-\frac{1}{4} \biggl[\left(-2\left(f_u(t,x)\right)^{-1}\dsdel{f_u(t,x)}{x}\Delta x+2\left(f_u(t,x)\right)^{-1}\doubledelsame{f_u(t,x)}{x}(\Delta x)^2\right)u(t,x-\Delta x) \\
    &+\left(2\left(f_u(t,x)\right)^{-1}\dsdel{f_u(t,x)}{x}\Delta x+2\left(f_u(t,x)\right)^{-1}\doubledelsame{f_u(t,x)}{x}(\Delta x)^2\right)u(t,x+\Delta x)\biggl] \\
    & + \frac{1}{8} \biggl[4 (\Delta x)^2 \left(\left(f_u(t,x)\right)^{-1}\dsdel{f_u(t,x)}{x}\right)^2 + 4(\Delta x)^4 \left(\left(f_u(t,x)\right)^{-1}\doubledelsame{f_u(t,x)}{x}\right)^2 \\
    & -8(\Delta x)^3 \left(\left(f_u(t,x)\right)^{-2}\dsdel{f_u(t,x)}{x}\doubledelsame{f_u(t,x)}{x}\right) \biggl] u(t,x-\Delta x) \\ 
    &+ \frac{1}{8} \biggl[ 4 (\Delta x)^2 \left(\left(f_u(t,x)\right)^{-1}\dsdel{f_u(t,x)}{x}\right)^2 + 4(\Delta x)^4 \left(\left(f_u(t,x)\right)^{-1}\doubledelsame{f_u(t,x)}{x}\right)^2 \\
    &+8(\Delta x)^3 \left(\left(f_u(t,x)\right)^{-2}\dsdel{f_u(t,x)}{x}\doubledelsame{f_u(t,x)}{x}\right) \biggl] u(t,x+\Delta x)  \Biggl\} +  \mc{O}\left( \left(\Delta x\right)^3 \right) \\
    &=  \mu_u \Biggl\{\frac{1}{2} \left[u(t,x-\Delta x) + u(t,x+\Delta x) - 2u(t,x) \right]\\
    &-\frac{1}{4} \biggl[\left(-2\left(f_u(t,x)\right)^{-1}\dsdel{f_u(t,x)}{x}\Delta x+2\left(f_u(t,x)\right)^{-1}\doubledelsame{f_u(t,x)}{x}(\Delta x)^2\right)u(t,x-\Delta x) \\
    &+\left(2\left(f_u(t,x)\right)^{-1}\dsdel{f_u(t,x)}{x}\Delta x+2\left(f_u(t,x)\right)^{-1}\doubledelsame{f_u(t,x)}{x}(\Delta x)^2\right)u(t,x+\Delta x)\biggl] \\
    &+ \frac{1}{2} \left(\Delta x\right)^2 \left(\left(f_u(t,x)\right)^{-1}\dsdel{f_u(t,x)}{x}\right)^2 \left[ u\left(t,x-\Delta x\right) + u\left(t,x+\Delta x \right) \right] \\
    &+ \mc{O}\left( \left( \Delta x \right)^3 \right) 
\end{aligned}
\end{equation}
as $\Delta x \to 0$. We can then perform expansions of the density of hawks, which are given by
\begin{equation}
u\left(t,x\pm \Delta x\right) = u(t,x) \pm  \dsdel{u(t,x)}{x} \left( \Delta x \right) + \frac{1}{2} \doubledelsame{u(t,x)}{x} \left( \Delta x \right)^2 + \mc{O}\left( \left(\Delta x \right)^3 \right) \: \: \mathrm{as} \: \: \Delta x \to 0,
\end{equation}
and we can apply this expansion to note that 
\begin{equation}
u(t,x-\Delta x) + u(t,x+\Delta x) - 2u(t,x) = \doubledelsame{u(t,x)}{x} \left( \Delta x \right)^2 + \mc{O}\left( \left(\Delta x \right)^3 \right) \: \: \mathrm{as} \: \: \Delta x \to 0.
\end{equation}
Using these expansions allows us to rewrite the right-hand side of our equation for $\dsdel{u(t,x)}{t}$ to see that
\begin{align}
\dsdel{u(t,x)}{t} &= \frac{1}{2} \mu_u \left( \Delta x \right)^2 \doubledelsame{u(t,x)}{x} -  \mu_u \left( \Delta x \right)^2 \left[ f_u(t,x)^{-1} \dsdel{f_u(t,x)}{x} \dsdel{u(t,x)}{x} \right] \\
&- \mu_u  \left( \Delta x \right)^2 \left[ f_u(t,x)^{-2} \doubledelsame{f_u(t,x)}{x} - f_u(t,x)^{-1} \left( \dsdel{f_u(t,x)}{x} \right)^2  \right] u(t,x) + \mc{O}\left( \left(\Delta x \right)^3 \right) \nonumber
\end{align}
as $\Delta x \to 0$, and we can further rearrange the right-hand side of this equation to write that
\begin{align}
\dsdel{u(t,x)}{t} &= \frac{1}{2} \mu_u \left( \Delta x \right)^2 \doubledelsame{u(t,x)}{x} -  \mu_u \left( \Delta x \right)^2  \dsdel{}{x} \left[u(t,x) \dsdel{}{x} \left\{ \log\left( f_u(t,x) \right) \right\} \right] +   \mc{O}\left( \left(\Delta x \right)^3 \right) %
\end{align}
as $\Delta x \to 0$. Finally, we can look to rescale time so that the right-hand side of our equation becomes finite in the limit as $\Delta x \to 0$. To do this, we introduce a characteristic timescale $t_c$ and the new time variable $\tau = \frac{t}{t_c}$, and this allows us to rewrite our equation for the density of hawks $u(\tau,x)$ as
\begin{equation} \label{eq:tauPDElimit}
\dsdel{u(\tau,x)}{\tau} =  \frac{1}{2} \mu_u \left[ \frac{\left(\Delta x \right)^2}{t_c} \right] \left[ \doubledelsame{u(\tau,x)}{x} -   \dsdel{}{x} \left[u(\tau,x) \dsdel{}{x} \left\{ \log\left( f_u(\tau,x) \right) \right\} \right] +   \mc{O}\left( \left(\Delta x \right)^3 \right) \right]
\end{equation}
as $\Delta x \to 0$. We can then choose to send $\Delta x \to 0$ and $t_c \to 0$ simultaneously at such a rate that the ratio of $\left(\Delta x\right)^2$ and $t_c$ satisfies
\begin{equation}
\lim_{\Delta x \to 0 \\ t_c \to 0} \frac{\left( \Delta x \right)^2}{t_c} = \alpha
\end{equation}
for a positive constant $\alpha$. If we then denote the diffusion coefficient for hawks by 
\begin{equation}
D_u = \frac{1}{2} \mu_u \alpha,
\end{equation}
we can evaluate the limit on both sides of Equation \eqref{eq:tauPDElimit} as $\Delta x \to 0$, $t_c \to 0$, and $\left( \Delta x \right)^2 / t_c \to \alpha$ to obtain the following partial differential equation for the density of hawks in terms of our rescaled time parameter $\tau$:
\begin{equation}
\dsdel{u(\tau,x)}{\tau} = D_u \doubledelsame{u(\tau,x)}{x} - 2 D_u \dsdel{}{x} \left( u(\tau,x) \dsdel{}{x} \left[ \log\left( f_u(t,x) \right) \right]\right).
\end{equation}

We can use similar expansions and the same rescaling of time to obtain the following equation for the density of doves
\begin{equation} \label{eq:tauPDElimitdove}
\dsdel{v(\tau,x)}{\tau} =  \frac{1}{2} \mu_v \left[ \frac{\left(\Delta x \right)^2}{t_c} \right] \left[ \doubledelsame{v(\tau,x)}{x} -    \dsdel{}{x} \left[v(\tau,x) \dsdel{}{x} \left\{ \log\left( f_v(\tau,x) \right) \right\} \right] +   \mc{O}\left( \left(\Delta x \right)^3 \right) \right].
\end{equation}
We can then take the limit as $\Delta x \to 0$, $t_c \to 0$, and $\left( \Delta x \right)^2 / t_c \to \alpha$ to obtain a PDE for the density of doves given by
\begin{equation}
\dsdel{v(\tau,x)}{\tau} = D_v \doubledelsame{v(\tau,x)}{x} - 2 D_v \dsdel{}{x} \left( v(\tau,x) \dsdel{}{x} \left[ \log\left( f_v(t,x) \right) \right]\right),
\end{equation}
where the diffusion coefficient $D_v$ for the doves is defined as
\begin{equation}
D_v := \frac{1}{2} \mu_v \alpha.
\end{equation}

Finally, we can incorporate the appropriate reaction terms in our equation by assuming that individuals of each strategy have a baseline per-capita birth or death rate proportional to their own payoffs, as well as a density-dependent per-capita death rate proportional their total population density at the point $x$. These assumptions on the reaction dynamics and the same rules for spatial motion allow us to obtain our PDE model for spatial evolutionary games with diffusive and payoff-driven motion:

\begin{equation}\label{eq:PDE-payoff-driven}
    \begin{aligned}
        \dsdel{u(\tau,x)}{\tau} &=D_u\doubledelsame{u(\tau,x)}{x} - 2D_u\dsdel{}{x}\left(u(\tau,x) \dsdel{}{x}  \log\left(f_u(\tau,x)\right)\right) + u\left(p_H(u,v) - k(u+v)\right), \\
       \dsdel{v(\tau,x)}{\tau} &=D_v\doubledelsame{v(\tau,x)}{x} - 2D_v\dsdel{}{x}\left(v(\tau,x) \dsdel{}{x} \log\left(f_v(\tau,x)\right)\right) + v\left(p_D(u,v) - k(u+v)\right). 
    \end{aligned}
\end{equation}
So far, we have considered the weights $f_u(\tau,x)$ and $f_v(\tau,x)$ for the bias in the random walks for hawks and doves as generic functions of the temporal variable $t$ and the spatial variable $x$. We can now incorporate the density of hawks $u(\tau,x)$, the density of doves $v(\tau,x)$ and the resulting payoff distributions $p_H\left( u(\tau,x),v(\tau,x)\right)$ and $p_D\left( u(\tau,x),v(\tau,x)\right)$ into our model of spatial movement. We can consider movement rules where $f_u(\tau,x)$ and $f_v(\tau,x)$ can be written in the form
\begin{equation}
\begin{aligned}
f_u(\tau,x) &= f_u\left( w_u p_H\left( u(\tau,x),v(\tau,x)\right) \right) \\
f_v(\tau,x) &= f_v\left( w_v p_D\left( u(\tau,x),v(\tau,x)\right)\right),
\end{aligned}
\end{equation}
so the movement of cooperators and defectors will depend on the payoff of their own strategy at the relevant spatial locations and a weight parameter $w_u$ and $w_v$ that the members of each strategy place on payoff differences in determining their movement probabilities. We can rewrite the PDE system by putting $w_u$ and $w_v$ as the coefficients of payoff-driven term
\begin{equation}
    \begin{aligned}
        \dsdel{u(\tau,x)}{\tau} &=D_u\doubledelsame{u(\tau,x)}{x} + u\left(p_H(u,v) - k(u+v)\right)\\
        &- 2D_uw_u\left(\dsdel{u}{x}\frac{f'( w_u p_H)}{f_u( w_u p_H)}\dsdel{p_H}{x}+u\dsdel{}{x}\left(\frac{f'( w_u p_H)}{f_u( w_u p_H)}\dsdel{p_H}{x}\right)\right)\\
       \dsdel{v(\tau,x)}{\tau} &=D_v\doubledelsame{v(\tau,x)}{x} + v\left(p_D(u,v) - k(u+v)\right)\\
       &- 2D_vw_v\left(\dsdel{v}{x}\frac{f_v'( w_v p_D)}{f_v( w_v p_D)}\dsdel{p_D}{x}+v\dsdel{}{x}\left(\frac{f_v'( w_v p_D)}{f_v( w_v p_D)}\dsdel{p_D}{x}\right)\right), \\
    \end{aligned}
\end{equation}
and we will use this form of our system of PDEs for stability analysis in the main paper.

\section{Additional Calculations for Linear Stability Analysis of PDE Model}
\label{sec:linearization_appendix}

In this section, we present the derivation of the linearized system of PDEs for our model of payoff-driven motion (Section \ref{sec:linearization_calculation}) and provide the expression for the threshold quantity $w_v^{\textnormal{III}b}$ required to satisfy the second necessary for a finite wavenumber pattern (Section \ref{sec:secondcondtion}).

\subsection{Linearization of PDE Model with Diffusive and Payoff-Driven Motion}
\label{sec:linearization_calculation}

We are looking to linearize the following nonlinear system of PDEs around the uniform coexistence state $(u_0,v_0)$ for the Hawk-Dove game with density-dependent regulation:
\begin{subequations}
    \begin{align}
     \dsdel{u(t,x)}{t} &=D_u\doubledelsame{u(t,x)}{x} - 2D_u\dsdel{}{x}\left(u(t,x) \dsdel{\log\left(w_u f_u(p_H(u,v))\right)}{x}\right) + u\left(p_H(u,v) - k(u+v)\right), \\
       \dsdel{v(\tau,x)}{\tau} &=D_v\doubledelsame{v(t,x)}{x} - 2D_v\dsdel{}{x}\left(v(t,x) \dsdel{\log\left(f_v(w_v p_D(u,v))\right)}{x}\right) + v\left(p_D(u,v) - k(u+v)\right).
    \end{align}
\end{subequations}
For this analysis, we are writing our nonlinear system of PDEs in terms of directed motion written in terms of $\log(f_u(w_u p_H(u,v)))$ and $\log(f_v(w_v p_D(u,v)))$ to highlight the dependence of directed motion on the spatial distributions of the payoffs for each strategy. To perform this linearization, we consider perturbations from the uniform state of the form
\begin{subequations}
\begin{align}
u(t,x)&=u_0+\epsilon u_1(t,x)\\
v(t,x)&=v_0+\epsilon v_1(t,x)
\end{align}
\end{subequations}
for sufficiently small $\epsilon$, and we look to plug this form of our solution into our nonlinear PDE system. To do this, we first perform a Taylor expansion around the uniform state $(u_0,v_0)$ under the assumption of small $\epsilon$, with the expansion of the logarithm the hawk's weight for directed motion taking the form
\begin{align}
    \log \left(f_u\left( w_u p_H\left( u,v\right) \right)\right) = \log\left( f_u\left( w_u p_H\left( u_0,v_0\right)\right)\right) + 
    \epsilon w_u c_1 u_1 + \epsilon w_v c_2 v_1 + \mc{O}(\epsilon^2) %
\end{align}
as $\epsilon \to 0$, with the constants $c_1$ and $c_2$ given by
\begin{align}\label{eq:cdefine-1}
\begin{split}
    c_1 &= \frac{f_u'\left( w_u p_H\left( u,v\right) \right)}{f_u\left(w_u \pi_H(u,v) \right)}\frac{\partial p_H\left( u,v\right)}{\partial u}\bigg\rvert_{u=u_0, v = v_0} \\
    c_2 &=  \frac{f_u'\left( w_u p_H\left( u,v\right) \right)}{f_u\left(w_u \pi_H(u,v) \right)}\frac{\partial p_H\left( u,v\right)}{\partial v}\bigg\rvert_{u=u_0, v = v_0} 
\end{split}
\end{align}
Substituting these expansions into the first PDE in our nonlinear system allows us to write that
\begin{align}
    \dsdel{u_0}{t}&+\epsilon\dsdel{u_1}{t} =D_u\left(\doubledelsame{u_0}{x}+\epsilon\doubledelsame{u_1}{x}\right)+ u_0\left(p_H(u_0,v_0) - k(u_0+v_0)\right) +\epsilon a_1 u_1 +\epsilon  a_2 v_1\\
        &- 2D_u\dsdel{}{x}\left[\left( u_0 + \epsilon u_1\right)\left(\log\left( f_u\left( w_u p_H\left( u_0,v_0\right)\right)\right) + \epsilon w_uc_1 u_1 + \epsilon w_uc_2 v_1\right)\right] + \mc{O}\left( \epsilon^2 \right), 
\end{align}
as $\epsilon \to 0$, where the constants $a_1$ and $b_1$ are given by
\begin{subequations}
    \begin{align}
    a_1 &= \frac{\partial}{\partial u} (u (p_H - \kappa (u+v))) \bigg\rvert_{u=u_0, v = v_0}= \frac{V (-C+V)(C+2V)}{2C^2 }< 0,\label{eq:coefficient-a1_appendix}  \\
    a_2& = \frac{\partial}{\partial v} (u (p_H - \kappa (u+v))) \bigg\rvert_{u=u_0, v = v_0}= \frac{V^3}{C^2} > 0.\label{eq:coefficient-b1_appendix} 
\end{align}
\end{subequations}
Since $u_0$ represents a spatially uniform equilibrium, we have conditions $\dsdel{u_0}{t}=0$, $\dsdel{u_0}{x}=0$, and $\doubledelsame{u_0}{x}=0$. Keeping terms only up to $\epsilon$, we arrive at the following linear PDE for the change in perturbation of the dove density $u_1(t,x)$:
\begin{align}
 \dsdel{u_1}{t} =D_u\doubledelsame{u_1}{x}+ a_1 u_1 +  b_1 v_1 - 2D_uw_u u_0 \left( c_1 \doubledelsame{u_1}{x}+ c_2 \doubledelsame{v_1}{x}\right).
\end{align}
With a similar approach, we can obtain the second PDE of our linearized system, which is given by
\begin{align}
 \dsdel{v_1}{t} =D_v\doubledelsame{v_1}{x}+ b_1 u_1 +  b_2 v_1 - 2D_vw_v v_0 \left(c_3 \doubledelsame{u_1}{x}+c_4 \doubledelsame{v_1}{x}\right),
\end{align}
where the constants $a_2$, $b_2$, $c_3$, and $c_4$ take the form
\begin{subequations}
\label{eq:appendixconstants}
    \begin{align}
    b_1 &= \frac{\partial}{\partial u} (v (p_D - \kappa (u+v))) \bigg\rvert_{u=u_0, v = v_0}=  \frac{-V (V-C)^2}{C^2}< 0, \label{eq:coefficient-a2_appendix} \\
    b_2 &= \frac{\partial}{\partial v} (v (p_D - \kappa (u+v))) \bigg\rvert_{u=u_0, v = v_0}= \frac{V (V-C)(C-2V)}{2 C^2} \begin{cases} >0 ~~\text{if}~~C < 2V\\ <0 ~~\text{if}~~ C > 2V \end{cases}\label{eq:coefficient-b2_appendix} \\
    c_3 &= \frac{f_v'\left( w_v p_D\left( u,v\right) \right)}{f_v\left( w_v p_D\left( u,v\right) \right)} \frac{\partial p_D\left( u,v\right)}{\partial u}\bigg\rvert_{u=u_0, v = v_0}\\
    c_4 &=\frac{f_v'\left( w_v p_D\left( u,v\right) \right)}{f_v\left( w_v p_D\left( u,v\right) \right)}\frac{\partial p_D\left( u,v\right)}{\partial v}\bigg\rvert_{u=u_0, v = v_0}.
\end{align}
\end{subequations}

\subsection{Second Condition for Finite Wavenumber Instability with Payoff-Driven Motion}
\label{sec:secondcondtion}

We now present the derivation of the second necessary condition required for the emergence of finite wavenumber patterns in our PDE model with payoff-driven motion. Our goal is to provide conditions on the values of $w_v$ for which it is possible to achieve a negative determinant $\det(A(m))$ of the linearization matrix for finitely many wavenumbers $m$. In Section \ref{sec:PDE-fullypayoff-driven}, we showed that this can be written as a polynomial of the form
\begin{equation}
\det(A(m)) = \alpha \left( \frac{m \pi}{l} \right)^4 + \beta \left( \frac{m \pi}{l} \right)^2 + \gamma,
\end{equation}
where the coefficients $\alpha$, $\beta$, and $\gamma$ are given by
\begin{subequations}
\label{eq:alphabetagammarewritten}
\begin{align}
\alpha &= A w_v + B w_u + D \\
\beta &= E w_v + F w_u + G \\
\gamma &= a_1 b_2 - a_2 b_1
\end{align}
\end{subequations}
with the constants $A$, $B$, $C$, $D$, $E$, and $F$ in Equation \eqref{eq:alphabetagammarewritten} taking the form
\begin{subequations} \label{eq:capitalletterconstants}
\begin{align}
A &= - \frac{D_u D_v u_0 v_0 V}{\left( u_0 + v_0 \right)^2} \\
B &=  \frac{D_u D_v u_0 v_0 \left( V + C \right)}{\left( u_0 + v_0 \right)^2}  \\
D &= D_u D_V \\ 
E &= \frac{\left( a_1 u_0 + a_2 v_0 \right) v_0 D_u \left( V + C \right)}{\left( u_0 + v_0 \right)^2}\\
F &= - \frac{\left( b_1 u_0 + b_2 v_0 \right) D_u u_0 \left( V + C \right)}{\left( u_0 + v_0 \right)^2} \\
G &= -\left( a_1 D_v + b_2 D_u \right).
\end{align}
\end{subequations}

We would like to derive a condition on $w_v$ in order to guarantee that $\beta^2 - 4 \alpha \gamma > 0$. Using the expressions from Equation \eqref{eq:alphabetagammarewritten}, we can rewrite this inequality as
\begin{equation}
E^2 w_v^2 + \left[2 E \left(F w_u + G \right) - 4 A \gamma \right] w_v + \left[ \left( F w_u + G\right)^2 - 4 \gamma \left( B w_u + D \right) \right] > 0.
\end{equation}
We can then solve this inequality to see that finite wavenumber patterns can only be achieved if the payoff sensitivity $w_v$ for doves satisfies
\begin{equation}
\begin{aligned}
w_v > w_v^{\textnormal{III}b} &:= \frac{4 A \gamma - 2 E \left( Fw_u + G \right)}{2 E^2} \\ &+ \frac{1}{2E^2} \sqrt{\left[4 A \gamma - 2 E \left( Fw_u + G \right)\right]^2 - 4 E^2 \left[\left( F w_u + G\right)^2 - 4 \gamma \left( B w_u + D \right)  \right] }
\end{aligned}
\end{equation}
We use this expression for $w_v^{\textnormal{III}b}$ and the constants from Equation \eqref{eq:capitalletterconstants} to help show the possibility of finite wavenumber patterns in Section \ref{sec:mixedeffectsresults} when we consider faster hawk diffusion $D_v > D_u$ and greater payoff sensitivity for doves $w_v > w_u$. 

\section{Linearization of Model of Payoff-Driven Motion with Nonlocal Payoff Sensing}
\label{sec:nonlocal-linearization}
In this section, we will perform the linearization of our nonlocal PDE model for payoff-driven motion around the uniform coexistence equilibrium $(u_0,v_0)$ for the Hawk-Dove game. We recall that the nonlinear, nonlocal PDE system is given by
\begin{align}
\begin{split}
    \frac{\partial u(t,x)}{\partial t} &= D_u \frac{\partial^2 u(t,x)}{\partial x^2}  + u \left(p_H(u(t,x),v(t,x)) - \kappa \left(u(t,x)+v(t,x)\right)\right) \\
    &- 2D_u \frac{\partial}{\partial x} \left( u(t,x) \left[ \frac{1}{2 \rho} \left( \log(f_u(u(t,x+\rho), v(t,x+\rho))) - \log(f_u(u(t,x-\rho), v(t,x-\rho))) \right) \right] \right)  \\
    \frac{\partial v(t,x)}{\partial t} &= D_v \frac{\partial^2 v(t,x)}{\partial x^2}  + v \left(p_D(u(t,x),v(t,x)) - \kappa \left(u(t,x)+v(t,x)\right)\right)    \\
    &- 2D_v \frac{\partial}{\partial x} \left( v(t,x) \left[ \frac{1}{2 \rho} \left( \log(f_v(u(t,x+\rho), v(t,x+\rho))) - \log(f_v(u(t,x-\rho), v(t,x-\rho))) \right) \right] \right),
\end{split}
\end{align}
and we aim to linearize the system by considering a solution of the PDE system featuring small perturbations of the uniform state, which take the form
\begin{subequations}
    \begin{align}
    u(t,x) &= u_0 + \epsilon u_1(t,x), \\
    v(t,x) &= v_0 + \epsilon v_1(t,x).
\end{align}
\end{subequations}
for a small value of $\epsilon$. To plug in this ansatz and evaluate the linearization, we need to expand the nonlocal term $ \log(f_u(u(t,x+\rho), v(t,x+\rho))) - \log(f_u(u(t,x-\rho), v(t,x-\rho)))$ in terms of the small parameter $\epsilon$ and the perturbation functions $u_1$ and $v_1$. Using a Taylor expansion of $\log(f_u(u, v))$ around the point $(u_0, v_0)$ up to second-order terms allows us to see that

\begin{subequations}
    \begin{align}
    &\log(f_u(u(t,x), v(t,x)) \\ &= \log(f_u(u_0, v_0))  + \epsilon\left(\frac{\partial \log f}{\partial u}\bigg\rvert_{u_0, v_0}  u_1(t,x) + \frac{\partial \log f}{\partial v}\bigg\rvert_{u_0, v_0} v_1(t,x)\right) \\
    &+ \epsilon^2\left(\frac{1}{2} \frac{\partial^2 \log f}{\partial u^2} \bigg\rvert_{u_0, v_0} u_1(t,x)^2 + \frac{1}{2} \frac{\partial^2 \log f}{\partial v^2} \bigg\rvert_{u_0, v_0} v_1(t,x)^2 + \frac{\partial^2 \log f}{\partial u \partial v} \bigg\rvert_{u_0, v_0} u_1(t,x)v_1(t,x)\right) \\
    &+ \mc{O}(\epsilon^3) \: \: \mathrm{as} \: \: \epsilon \to 0.
\end{align}
\end{subequations}

Retaining terms up to first order in $\epsilon$, we see that the contribution of the term $\log(f_u(u,c))$ up to order $\epsilon$ will take the form

\begin{align}
    \log(f_u(u(t,x), v(t,x)))\approx \log(f_u(u_0, v_0)) + \epsilon w_u c_1 u_1(t,x) +\epsilon w_uc_2 v_1(t,x),
\end{align}
and we can evaluate this expression at points $x+\rho$ and $x-\rho$ gives to see that
\begin{subequations}
\begin{align}
    \log(f_u(u(t,x+\rho), v(t,x+\rho))) &\approx \log(f_u(u_0, v_0)) + \epsilon  w_u c_1  u_1(t,x+\rho) +\epsilon w_u  c_2 v_1(t,x+\rho), \\
    \log(f_u(u(t,x-\rho), v(t,x-\rho))) &\approx \log(f_u(u_0, v_0)) + \epsilon w_u  c_1 u_1(t,x-\rho) +\epsilon w_u c_2 v_1(t,x-\rho),
\end{align}
\end{subequations}
where $c_1$ and $c_2$ are the same constants from Equation \eqref{eq:cdefine-1} that arose in the linearization of the local PDE model for payoff-driven motion. We can then combine these expanded expressions to see that the nonlocal gradient of the payoff weight function $f_u(u,v)$ for hawks will take the form
\begin{align}
    &\log(f_u(u(t,x+\rho),v(t,x+\rho))) - \log(f_u(u(t,x-\rho),v(t,x-\rho))) \\
    &=\epsilon w_u \left(c_1(u_1(t,x+\rho)-u_1(t,x-\rho)+c_2(v_1(t,x+\rho)-v_1(t,x-\rho)\right).
\end{align}

We can use a similar approach to see that the nonlocal gradient for the payoff weight function $f_v(u,v)$ for doves will be given by
\begin{align}
    &\log(f_v( u(t,x+\rho), v(t,x+\rho) )) - \log(f_v(u(t,x-\rho), v(t,x-\rho)))\\
    &=\epsilon w_v\left(c_3(u_1(t,x+\rho)-u_1(t,x-\rho)+c_4(v_1(t,x+\rho)-v_1(t,x-\rho)\right),
\end{align}
where the constants $c_3$ and $c_4$ are the constants from Equation \eqref{eq:appendixconstants} from the local PDE model.

To linearize the diffusion terms, we substitute our perturbation ansatz $u = u_0 + \epsilon u_1(t,x)$ and $v = v_0 + \epsilon v_1(t,x)$ into our PDE and note that $\frac{\partial^2 u_0}{\partial x^2} = 0$ and $\frac{\partial^2 v_0}{\partial x^2} = 0$ for uniform solutions. This allows us to write the diffusive term in our linearized PDE as

\begin{subequations}
    \begin{align}
        D_u \frac{\partial^2}{\partial x^2}(u_0 + \epsilon u_1(t,x))&=\epsilon D_u \frac{\partial^2 u_1(t,x)}{\partial x^2} \\
        D_v \frac{\partial^2}{\partial x^2}(v_0 + \epsilon v_1(t,x))&=\epsilon D_v \frac{\partial^2 v_1(t,x)}{\partial x^2},
    \end{align}
\end{subequations}

and we can similarly obtain the lineraization of our reaction terms around  $(u_0, v_0)$ to see that
\begin{subequations}
    \begin{align}
      u \left(p_H(u(t,x),v(t,x)) - \kappa (u(t,x)+v(t,x))\right) &\approx \epsilon \left(a_1 u_1(t,x) + b_1 v_1(t,x)\right), \\
     v \left(p_D(u(t,x),v(t,x)) - \kappa (u(t,x)+v(t,x))\right)&\approx \epsilon \left(a_2 u_1(t,x) + b_2 v_1(t,x)\right),
\end{align}
\end{subequations}
where $a_1,a_2,b_1,$ and $b_2$ are the entries of the Jacobian matrix for the reaction dynamics described in Section \ref{sec:ODEstabilityreview}.

Putting together all of the terms we have obtained so far, we see that the perturbation functions $u_1(t,x)$ and $v_1(t,x)$ will satisfy the following linearized system of nonlocal PDEs
\begin{align}
\begin{split}
      \frac{\partial u_1(t,x)}{\partial t} &= D_u \frac{\partial^2 u_1(t,x)}{\partial x^2}  +a_1 u_1(t,x) + b_1 v_1(t,x) \\
    &- \frac{w_u u_0}{\rho} \frac{\partial}{\partial x} \left( c_1(u_1(t,x+\rho)-u_1(t,x-\rho)) + c_2(v_1(t,x+\rho)-v_1(t,x-\rho))\right), \\
    \frac{\partial v_1(t,x)}{\partial t} &= D_v \frac{\partial^2 v_1(t,x)}{\partial x^2}  + a_2 u_1(t,x) + b_2 v_1(t,x) \\
    &- \frac{w_v v_0}{ \rho} \frac{\partial}{\partial x} \left( c_3(u_1(t,x+\rho)-u_1(t,x-\rho)) + c_4(v_1(t,x+\rho)-v_1(t,x-\rho))\right),
\end{split}
\end{align}
which we can use to study the stability of the coexistence state under the nonlocal PDE model.

\end{document}